\documentclass[a4paper]{cas-sc}

\usepackage[numbers,sort&compress]{natbib}
\usepackage{nomencl}
\usepackage{framed}
\usepackage{amsmath}
\usepackage{tikz}

\makenomenclature

\newcommand\ie{i.e.}

\newcommand\eg{e.g.}

\usepackage{caption}
\usepackage[section]{placeins}
\usepackage{appendix}
\usepackage{csquotes}
\usepackage{subcaption}
\usepackage{etoolbox}
\renewcommand\nomgroup[1]{%
	\item[\bfseries
	\ifstrequal{#1}{P}{Physics constants}{%
		\ifstrequal{#1}{N}{Number sets}{%
			\ifstrequal{#1}{O}{Other symbols}{}}}%
	]}
\def\tsc#1{\csdef{#1}{\textsc{\lowercase{#1}}\xspace}}
\tsc{WGM}
\tsc{QE}
\begin{document}

	\let\WriteBookmarks\relax
	\def\floatpagepagefraction{1}
	\def\textpagefraction{.001}
	\let\printorcid\relax % 可去掉页面下方的ORCID(s)
	
	% \shorttitle{<short title of the paper for running head>}    
	\shorttitle{Overshoot-resolved transition modeling}   
	
	% \shortauthors{<short author list for running head>} 
	\shortauthors{L. Wu and Z. Xiao}
	
	%\title[mode = title]{Self-similar solution based $\gamma$-$\widetilde{Re}_{\theta \rm{t}}$ transition model for high-speed flows: Improvement of transition prediction and detailed physical investigations}
	\title[mode = title]{Overshoot-resolved transition modeling based on field inversion and symbolic regression}
	
	% \author[<aff no>]{<author name>}[<options>]
	% 填写如下
	\author[1]{Lei Wu}
	
	\author[1,2,3]{Zuoli Xiao}
	\ead{z.xiao@pku.edu.cn}
	\cormark[1]
	\cortext[1]{corresponding author}
	
	\affiliation[1]{organization={State Key Laboratory for Turbulence and Complex Systems, School of Mechanics and Engineering Science, Peking University},
		city={Beijing},
		postcode={100871}, 
		country={China}}
	
	\affiliation[2]{organization={HEDPS and Center for Applied Physics and Technology, School of Mechanics and Engineering Science, Peking University},
		city={Beijing},
		postcode={100871}, 
		country={China}}
	
	\affiliation[3]{organization={Nanchang Innovation Institute, Peking University},
		city={Nanchang},
		postcode={330008}, 
		country={China}}

	\begin{abstract}
		 Overshoot of high-speed transitional skin-friction and heat-transfer values over their fully turbulent levels is well documented by numerous direct numerical simulations (DNS) and experimental studies. However, this high-speed-specific overshoot phenomenon remains a longstanding challenge in Reynolds-averaged Navier-Stokes (RANS) transition models. In this paper, field inversion and symbolic regression (FISR) methodologies are adopted to explore a generalizable and interpretable augmentation for resolving the missing overshoot characteristic. Specifically, field inversion is implemented on our previous high-speed-improved $k$-$\omega$-$\gamma$-$\widetilde{Re}_{\theta \rm{t}}$ transition-turbulence model. Then symbolic regression is employed to derive an analytical map from RANS mean flow variables to the pre-defined and inferred corrective field $\beta(\mathbf{x})$. Results manifest that the excavated expression faithfully reproduces the overshoot phenomena of transition region over various test cases while does not corrupt model behavior in transition location and length. Based on its transparent functional form, mechanistic investigations are conducted to illustrate the underlying logic for accurate capture of overshoot phenomenon. In addition, importance of protect function in $\beta(\mathbf{x})$, feasibility of a more concise expression for $\beta(\mathbf{x})$, and reliable performance of $\beta(\mathbf{x})$ in low-speed transitional flows are emphasized.
	\end{abstract}
	
	
	\begin{highlights}
		\item The first generalizable and interpretable expression is derived to capture overshoot phenomenon.
		\item Current model accurately reproduces the overshoot magnitudes while avoiding spurious behavior.
		\item A clear and physical pathway for overshoot resolution is identified.
	\end{highlights}
	
	% Keywords
	\begin{keywords}
		Overshoot phenomenon \sep
		Boundary layer transition \sep 
		Ensemble Kalman filter method \sep
		Symbolic regression \sep
		High-speed flows \sep
	\end{keywords}
	
	\maketitle
	
	% Main text
	\section{Introduction}\label{sec:intro}
	Accurate laminar-turbulent transition prediction is paramount for design of high-speed configurations. Firstly, location and length of transition determines the spatial scope of laminar and turbulent regions. Given the stark contrast in aerodynamic and thermodynamic characteristics between laminar and turbulent flow states, even small errors in transition location and length can lead to significant discrepancies in performance prediction. Secondly, in high-speed transitional flows, wall skin-friction and heat-transfer in transition region commonly exhibit a peak that overshoot their fully turbulent state. The magnitude of this overshoot phenomenon directly governs the peak aerodynamic and thermodynamic loads over the vehicle. Consequently, distinct from low-speed circumstance, reliable transition prediction in high-speed transitional flow must accounts not only for the appropriate placement of transition region but also for the occurrence and degree of overshoot behavior within it. These two factors collectively govern critical design metrics, including aerodynamic forces, aeroheating, propulsive efficiency, and thermal protection system requirements \cite{Anderson, Vogel-overshoot}.
	
	Although the overshoot phenomena have been extensively observed by direct numerical simulations (DNS) and experimental investigations across a range of high-speed transitional flows (\eg, flat plates \cite{Flat-Plate-DNS2.25, Mayer-2011-JFM, Franko-2013-JFM, Unnikrishnan-2019-CAF, XuDehao-2022-JFM, GuoPeixu-2022-POF, LiuMeikuan-2022-POF1}, sharp/blunt straight cones \cite{Sivasubramanian-2015-JFM, LiXinliang-2023-POF, GuoPeixu-2024-JFM, Schuabb-2025-JFM, Horvath, Paredes-2019-JSR, LiuMeikuan-2022-POF2}, flare cones \cite{Hader-2018-JFM, Meersman-2018-conference}, HIFiRE-1 \cite{Wadhams-2008-JSR}, and MF-1 \cite{TuGuohua-2021-AMS}), the physical mechanisms of overshoot are still unclear \cite{XuDehao-2022-JFM}. Leveraging the capability of DNS to illuminate potential transition scenarios \cite{Franko-2013-JFM}, a growing body of studies has sought to unravel the specific origins of this behavior. However, a consensus on its fundamental cause has yet to emerge. To foster a deeper understanding of the overshoot and to substantiate the physical rationality of current overshoot-resolved transition model in this paper, several representative mechanistic interpretations are reviewed below from those DNS studies.
	
	From the perspectives of flow field, \citet{Franko-2013-JFM} attributed the overshoot of skin-friction and heat-transfer coefficients to the enhanced wall-normal transport of momentum and thermal energy driven by the streamwise vorticity. This enhanced transport is physically manifested through elevated Reynolds shear stress and Reynolds heat flux in the transitional boundary layer. Based on the twofold repeated integration for decomposing skin-friction and heat-transfer values, the overshoot in skin-friction coefficient is shown to arise primarily from the drastic change of mean velocity profiles, while the overshoot of heat-transfer coefficient is predominantly attributed to the viscous dissipation \cite{XuDehao-2022-JFM}. Consistent with this conclusion, \citet{GuoPeixu-2022-POF} conducted an energy budget analysis and identified that the heat-transfer overshoot originates from a significant amplification of viscous dissipation in the near-wall region. And this amplification, can be further traced back to the pronounced mean flow distortion.
	
	In parallel with flow field-based interpretations, another class of studies adopted the instability waves perspective. The common approach is to selectively force one or two dominant instability waves at set frequencies and spanwise wave numbers to investigate specific transition scenarios \cite{Franko-2013-JFM}. Although this method does not replicate the broad spectrum of instabilities present in actual flight or wind-tunnel environments, it is valuable for elucidating specific transition mechanisms, particularly those responsible for the overshoot. To determine potential mechanisms for both boundary layer transition and overshoot of heat-transfer, \citet{Franko-2011-AIAAconference} exerted four different types of forcing (\ie, impulse, oblique, Mack 2D, and Mack 3D) and concluded that the oblique breakdown scenario offers a underlying explanation for the heat-transfer overshoot. Similarly, \citet{Franko-2013-JFM} introduced three different types of disturbances, namely, first mode oblique breakdown, second mode oblique breakdown, and second mode fundamental resonance, through blowing and suction, and found that only the first mode oblique breakdown produced a distinct peak in wall skin-friction and heat-transfer. These findings were further corroborated by \citet{Unnikrishnan-2019-CAF, Unnikrishnan-2020-JFM, GuoPeixu-2022-POF, ZhouTeng-2022-POF}. Therefore, of all the instability waves examined, first mode oblique breakdown (a canonical form of oblique-wave transition) exhibits the strongest association with the skin-friction and heat-transfer overshoot phenomena. Given that oblique-wave transition is the predominant pathway in high-speed transitional flows \cite{Chang-1994-JFM, Mayer-2011-JFM, Malik-2011-JFM, Laible-2016-JFM, Fezer-2000-IUTAM, Leinemann-2021-AIAA, Hader-2021-AIAA}, so it is natural that DNS \cite{Flat-Plate-DNS2.25, Mayer-2011-JFM, Franko-2013-JFM, Unnikrishnan-2019-CAF, XuDehao-2022-JFM, GuoPeixu-2022-POF, Sivasubramanian-2015-JFM, LiXinliang-2023-POF, GuoPeixu-2024-JFM, Schuabb-2025-JFM, Hader-2018-JFM, Meersman-2018-conference} and experiments \cite{LiuMeikuan-2022-POF1, Horvath, Paredes-2019-JSR, LiuMeikuan-2022-POF2, Wadhams-2008-JSR, TuGuohua-2021-AMS} consistently exhibit overshoot phenomena, albeit with varying magnitude, across a wide range of high-speed flow conditions.
	
	% 尽管过冲现象产生的最底层原因可能需要追溯到不稳定波的种类，但由于我们发现过冲现象在高速流动中或多或少都存在，因此从工程转捩预测角度，我们可以不需要在RANS方法中考虑不稳定波这一科学上都还未有统一定论的事情(事实上RANS方法也无法考虑不稳定波)，而是从更为上层的平均流场角度对RANS方法进行过冲现象的建模，正如诸多从流场角度进行过冲现象产生机制的DNS研究一样。因此，基于RANS方法的过冲解析转捩预测其实是可能的。但很遗憾地是，截至目前，尽管RANS研究工作浩如烟海，但真正具有过冲解析能力的RANS模型屈指可数
	Although the most fundamental cause of overshoot phenomenon may need to be traced back to the type of instability waves, the fact that overshoot is observed in virtually all high-speed flows implies that, from the perspective of engineering transition prediction (\ie, Reynolds-averaged Navier-Stokes (RANS) method), it is unnecessary to account for the instability waves--particularly given that their roles in transition have not yet reached a scientific consensus. Instead, RANS modeling of the overshoot phenomenon should be approached from a more macroscopic and well-defined mean flow field perspective, just as those DNS studies that investigated the generation mechanism of overshoot from the flow field viewpoint \cite{Franko-2013-JFM, XuDehao-2022-JFM, GuoPeixu-2022-POF}.
	
	In this scenario, RANS-based prediction of overshoot is indeed feasible. Regrettably, however, despite the vast body of RANS transition research, only a handful of models to date possess the capability to capture the overshoot phenomenon. Based on the structural ensemble dynamics (SED) theory of wall turbulence, \citet{XiaoMengjuan-2019-ScienceChina} presented a multi-layer stress length (SL) function $\ell_{12}$ (which defines the eddy viscosity $\nu_{\rm{T}}$ in RANS equations) to characterize the laminar-turbulent transition. This SED-SL model was later extended to hypersonic straight cone cases by \citet{BiWeitao-2022-AIA}, who introduced a multi-regime structure that supplemented the original parameters (transition center $x^*$ and after-transition near-wall eddy length $\ell_{0\infty}^*$) with a transition-width parameter $\beta_l$ to quantify the overshoot behavior. Larger $\beta_l$ contributes to a wider transition regime and stronger transition overshoot. Those model parameters are determined by a prediction-correction procedure that iteratively sets initial parameter values, solves the RANS equations with the SED-SL model, and updates the parameters by comparing the numerical predictions with experimental data until satisfactory agreement is achieved. On this basis, \citet{BiWeitao-2023-CJA} further correlated the transition center parameter with five influential factors $Re_{x*}(Tu, Ma_{\rm{e}}, T_{\rm{w}}/T_{\rm{aw}}, \theta, Re_{\rm{N}})$ based on available experimental data. These two works validated the ability of SED-SL model to capture the characteristic overshoot in hypersonic transitional flows \cite{BiWeitao-2022-AIA, BiWeitao-2023-CJA}. Proceeding from the widely used concept of intermittency $\gamma$ in RANS-based transition modeling, \citet{Vogel-overshoot} modified the parameterized $\gamma$ distribution of \citet{Narasimha-1985} to accommodate the possibility of overshoot phenomenon:
	\begin{eqnarray}
		\gamma=\left(1+C_{o v}\right)\left(1-\exp \left(-\eta_1 \zeta_1^2\right)\right)-C_{o v}\left(1-\exp \left(-\eta_2 \zeta_2^2\right)\right),\label{eqn:gamma-Vogel}
	\end{eqnarray}
	here, $\zeta_1$ and $\zeta_2$ are nondimensionalized transition zone coordinates that depend on the local streamwise position, onset and end locations of transition, and position of overshoot. The intermittency $\gamma$ is then multiplied by the eddy viscosity $\mu_{\rm{T}}$, yielding a transition-predictive RANS equation. $C_{ov}$, $\eta_1$, and $\eta_2$ are freestream condition-dependent correlations formulated using ground experiments or flight tests. Combined with their stability analysis tool, valid estimate of the transition overshoot was exhibited.
	
	Despite their preliminary success in overshoot-resolved RANS methods \cite{BiWeitao-2022-AIA, BiWeitao-2023-CJA, Vogel-overshoot}, both approaches rely on several case-specific model parameters. Establishing the required correlations for them necessitates an extensive set of experimental results. Moreover, in practical transition simulations, the SED-SL model \cite{BiWeitao-2022-AIA} still requires an iterative prediction-correction procedure to adjust its remaining parameters. While the intermittency-based approach \cite{Vogel-overshoot} depends on an external stability tool to close the formulation Eq.~\eqref{eqn:gamma-Vogel}. These dependencies may limit the generalizability and autonomous predictive capability of the current overshoot-resolved RANS models.
	
	It is worth noting that the aforementioned overshoot-capable RANS approaches are formulated without a transport-based transition model. Yet, in industrial transition prediction, the coupled solution of turbulence and transition models remains the most widely adopted strategy for closing the RANS equations. However, as \citet{Vogel-overshoot} noted, in transport-based transition-turbulence model framework, the characteristics of transition zone, such as its shape and extent, are inherently coupled with the prediction of transition onset through the same set of governing equations. This tight coupling makes it challenging to modify the transition zone independently without affecting other aspects of the model's behavior. Perhaps for this reason, almost all high-speed transition models succeed only in accurate prediction of transition location and length, while remaining incapable of capturing the overshoot phenomenon. This limitation has been repeatedly acknowledged across numerous studies in high-speed transition models \cite{XuJiakuan-2017-AIAA, LiuZaijie-2020-AST, LiuZaijie-2022-AIAA, Vogel-2025-IJHMT, WuLei-2026-IJHMT}. To the best of our knowledge, the only exception to current situation is the study of \citet{QinYaping-2018-IJHMT}, in which a piecewise amplification function was introduced to scale the intermittency factor $\gamma_1$ computed from transport-based transition model to an enhanced counterpart $\gamma_2$, thus accelerating the development process of turbulence and bringing about the appearance of overshoot. Nevertheless, the approach is a purely numerical adjustment that does not address the underlying physics of overshoot phenomenon. Even so, the core idea that amplifying the intermittency factor beyond unity to reproduce the overshoot is conceptually insightful and has also been mentioned in recent studies \cite{Vogel-overshoot, Vogel-2025-IJHMT}.
	
	Against this backdrop, sophisticated and physical augmentation that intrinsically captures the overshoot through mean-flow physics rather than ad hoc numerical adjustments is the only viable path toward a truly overshoot-resolved transport-based transition model. Such an expectation was presumed to be difficult through conventional modeling approaches as evidenced by the high-speed transition model community \cite{XuJiakuan-2017-AIAA, LiuZaijie-2020-AST, LiuZaijie-2022-AIAA, Vogel-2025-IJHMT, WuLei-2026-IJHMT}. In recent years, data-driven paradigm has emerged as a powerful strategy to overcome the long-standing limitations of traditional transition model \cite{Duraisamy-2021-PRF, TangZhigong-2023}. Based on the hierarchy of modeling objectives, existing machine learning-based transition models can be broadly categorized into three classes. The first class refers to correcting and modeling individual terms of the transition model equations using machine learning (ML) technology. Following the field inversion and machine learning (FIML) framework \cite{Duraisamy-2016-JCP, Singh-2017-AIAA}, \citet{YangMuchen-2020-POF} obtained the distribution of correction terms in four-equation $k-\omega-\gamma-A_r$ transition model through the ensemble Kalman filtering (EnKF) method, and mapped the relation between mean flow variables and correction terms using random forest (RF) and artificial neural network (ANN) algorithms. A similar study was conducted by \citet{Srivastava-2021-PRF}. Recognizing the well-known deficiency of original $\gamma-\widetilde{Re}_{\theta \rm{t}}$ model \cite{Langtry-2009-AIAA} in high-speed transition flows, \citet{ZhangTianxin-2023-POF} also employed the FIML strategy to infer the spatial perturbation correction term $\beta$ for transition onset function $F_{\rm{onset}}$ and subsequently developed a RF model to represent it. Results revealed that the transitional location and length can be correctly obtained. Another school of data-driven transition models pertain to the direct modeling of specific term in traditional transition model. The first attempt is made by \citet{Duraisamy-2015-AIAAconference}, who introduced their FIML framework \cite{Duraisamy-2016-JCP, Singh-2017-AIAA} to bypass transitional flow, in which the source term of intermittency factor transport equation is modeled to improve the accuracy of traditional transition model. Inheriting from the gene expression programming (GEP) \cite{Weatheritt-2016-JCP} and computational fluid dynamics (CFD)-driven framework \cite{ZhaoYaomin-2020-JCP}, \citet{Akolekar-2021-energies} developed a more accurate expression for laminar kinetic energy (LKE) production and transfer terms of LKE transition model. Similar work can also be found in study of \citet{FangYuan-2024-JOT}. To estimate the onset of crossflow instability, \citet{Barraza-2025-JSR} adopted an ANN to predict the production term of amplification factor transport equation. The third category involves the direct modeling and substitution of equation variables in traditional transition model. Learning from flow field generated by the partial differential equation (PDE)-based transition-turbulence model, ML models for intermittency factor $\gamma$ were established, thus replacing the original PDE-constrained $\gamma$ \cite{WuLei-2022-POF, WuLei-2022-TAML, ChenYanjun-2022-Aerospace, WuLei-2024-POF}. Different from those models that trained on RANS-based flow fields, \citet{LiZhen-2023-POF} artificially constructed a turbulence intermittency field from the fluctuation velocity of large eddy simulation (LES) data and employed an ANN to model this quantity, yielding a data-driven-substituted transition model with improved accuracy. Similar to the aforementioned FIML studies \cite{YangMuchen-2020-POF, Srivastava-2021-PRF, ZhangTianxin-2023-POF, Duraisamy-2015-AIAAconference}, \citet{Ahmed-2025-AIAAconference} adopted the discrete adjoint method to derive a LES-consistent intermediate field for $\gamma$, which was then used to scale the production term of turbulent kinetic energy equation.
	
	The above data-driven efforts underscore the effectiveness of FIML framework for enhancing conventional transition models. During FIML process, data assimilation (DA) techniques, such as EnKF and adjoint-based methods, are employed to infer the predefined, spatially varying multiplicative correction field, a procedure commonly referred to as field inversion (FI). Then, state-of-the-art machine learning (ML) algorithms including ANN, RF, and symbolic regression (SR) are adopted to learn the inferred correction fields. FIML-augmented transition modeling provides a paradigm shift that can overcome the well-known bottleneck of traditional physics and intuition-based modeling creations. In fact, the FIML methodology has achieved considerable success in a wide range of turbulence model studies, such as separated flows \cite{WuChenyu-2023-PRF, WuChenyu-2025-AIAA, ChenLong-2025-PRF, WuChutian-2025-JCP}, transonic regimes \cite{LiuYi-2023-AIAA}, and flows with shock-wave/turbulent boundary-layer interactions \cite{TangDenggao-2023-POF, TangDenggao-2024-AA}. Among these achievements, SR has recently gained increasing attention within the FIML community due to its white-box nature and strong interpretability \cite{WuChenyu-2023-PRF, WuChenyu-2025-AIAA, WuChutian-2025-JCP, TangDenggao-2024-AA}. To some extent, the FISR framework not only serves as a strategy to augment conventional transition or turbulence models like other black-box ANN, RF, and Gaussian process regressor (GPR) models, but also functions as a tool for scientific discovery. For example, \citet{WuChenyu-2023-PRF} discovered a generalizable expression that significantly enhanced the accuracy of shear-stress-transport (SST) model in separated flows.
	
	Motivated by this, the present paper leverages the powerful FISR framework to develop the first truly overshoot-resolved RANS transition model. Specifically, combined with sparse Hi-Fi data, EnKF-based DA technique is employed to derive the overshoot-resolved RANS flow field. Subsequently, symbolic regression is applied to discover an interpretable and white-box corrective expression that endows the baseline transition model with intrinsic overshoot prediction capability. To the best of our knowledge, this work represents not only the first real sense of RANS-based transition model capable of faithfully capturing the overshoot phenomenon, but also the first scientific discovery of a generalizable and physical mechanism for overshoot entirely from the perspective of RANS mean flow field.

	\section{Field inversion}\label{sec:FI}	
	\subsection{Improved $k$-$\omega$-$\gamma$-$\widetilde{Re}_{\theta \rm{t}}$ model for high-speed flows}\label{subsec:improved-model}
	The original $k$-$\omega$-$\gamma$-$\widetilde{Re}_{\theta \rm{t}}$ transition-turbulence model \cite{Langtry-2009-AIAA} exhibits well-documented deficiency in high-speed transitional flows. To address this limitation, we have developed a comprehensive high-speed extension with the full set of governing equations \cite{WuLei-2026-IJHMT} provided as
	\begin{eqnarray}
		\frac{\partial (\rho k)}{\partial t} + \frac{\partial (\rho u_j k)}{\partial x_{j}} = \frac{\partial}{\partial x_{j}}\left[\left(\mu+\sigma_k \mu_{\rm{T}}\right) \frac{\partial k}{\partial x_{j}}\right]+\gamma_{\rm{eff}}P_{k} - \min(\max(\gamma_{\rm{eff}},0.1),1.0)D_k + \Pi_{cc},
		\label{equ:k-equ}
	\end{eqnarray}
	\begin{eqnarray}
		\frac{\partial (\rho \omega)}{\partial t} + \frac{\partial (\rho u_j \omega)}{\partial x_{j}} = \frac{\partial}{\partial x_{j}}\left[\left(\mu+\sigma_{\omega} \mu_{\rm{T}}\right) \frac{\partial \omega}{\partial x_{j}}\right]+\frac{C_\omega \rho}{\mu_{\rm{T}}}P_{k} - D_{\omega} + 2(1-f_1)\frac{\rho\sigma_{\omega 2}}{\omega}\frac{\partial k}{\partial x_j}\frac{\partial \omega}{\partial x_j},
		\label{equ:omega-equ}
	\end{eqnarray}
	\begin{eqnarray}
		\frac{\partial (\rho \gamma)}{\partial t} + \frac{\partial (\rho u_j \gamma)}{\partial x_{j}} = \frac{\partial}{\partial x_{j}}\left[\left(\mu+\frac{\mu_{\rm{T}}}{\sigma_{\gamma}}\right) \frac{\partial \gamma}{\partial x_{j}}\right]+P_{\gamma}-E_{\gamma},
		\label{equ:gamma-equ}
	\end{eqnarray}
	\begin{eqnarray}
		\frac{\partial ( \rho \widetilde{Re}_{\theta \rm{t}} )}{\partial t} + \frac{\partial (\rho u_j \widetilde{Re}_{\theta \rm{t}} )}{\partial x_{j}} = 
		P_{\theta \rm{t}} + \frac{\partial}{\partial x_j}\left[ \sigma_{\theta \rm{t}}(\mu + \mu_{\rm{T}}) \frac{\partial \widetilde{Re}_{\theta \rm{t}}}{\partial x_j} \right].
		\label{equ:Re_theta-equ}
	\end{eqnarray}
	Here, $P_\gamma$ and $E_\gamma$ denote the production and destruction terms of $\gamma$ equation. $P_{\theta \rm{t}}$ is the source term in the $\widetilde{Re}_{\theta \rm{t}}$ equation defined as follows
	\begin{subequations}
		\begin{gather}\label{equ:P_gamma-E_gamma-P_thetagamma}
			P_\gamma = F_{\rm{length}} c_{a1} \rho S (\gamma F_{\rm{onset}})^{0.5} (1-c_{e1}\gamma), \\
			E_\gamma = c_{a2} \rho \Omega \gamma  F_{\rm{turb}} (c_{e2}\gamma-1), \\
			P_{\theta \rm{t}} = c_{\theta \rm{t}} \frac{\rho}{t}(Re_{\theta \rm{t}} - \widetilde{Re}_{\theta \rm{t}}) (1.0-F_{\theta \rm{t}}),
		\end{gather}
	\end{subequations}
	in which the transition onset function $F_{\rm{onset}}$ is core of the whole transition-turbulence model and is formulated as
	\begin{equation}\label{equ:F-onset}
		F_{\rm{onset}} = \max(F_{\rm{onset2}}-F_{\rm{onset3}},0),
	\end{equation}
	with
	\begin{subequations}\label{equ:F-onset123}
		\begin{gather}
			F_{\rm{onset1}} = \frac{Re_\nu}{f(Ma_{\rm{local}},\frac{T_{\rm{w}}}{T_{\rm{local}}}) Re_{\theta \rm{c,new}}},\label{equ:F-onset1} \\
			F_{\rm{onset2}} = \min(\max(F_{\rm{onset1}},F_{\rm{onset1}}^4), 2.0),\\
			F_{\rm{onset3}} = \max\left[1-\left( \frac{R_{\rm{T}}}{2.5} \right)^3,0\right].
		\end{gather}
	\end{subequations}
	Compared to the previous low-speed model, both improvements are focused on the core subfunction $F_{\rm{onset1}}$ (see Eq.~\eqref{equ:F-onset1}) of $F_{\rm{onset}}$. One is the compressible self-similar solutions (CSS)-based fitting correlation $f(Ma_{\rm{local}},\frac{T_{\rm{w}}}{T_{\rm{local}}})$ consist of local Mach number $Ma_{\rm{local}}$ and ratio of wall temperature to local temperature $T_{\rm{w}}/T_{\rm{local}}$ to replace the constant $2.193$ that derived from Blasius boundary layer. Another one is the revised empirical correlation $Re_{\theta \rm{c,new}}$ for $Re_{\theta \rm{c}}$ that considers compressibility and nose bluntness effects. For brevity and simplicity, readers can referred to our original paper \cite{WuLei-2026-IJHMT} for formulation of both modifications and other details about the improved $k$-$\omega$-$\gamma$-$\widetilde{Re}_{\theta \rm{t}}$ model. In such endeavors, it is revealed that the transition location and length of the high-speed model is basically consistent with the Hi-Fi results, considerably overcoming the disability of its low-speed counterpart. Nevertheless, similar to all previous transition models appropriate for high-speed flows \cite{LiuZaijie-2020-AST, LiuZaijie-2022-AIAA, XuJiakuan-2017-AIAA}, our improved model completely failed to reproduce the overshoot phenomenon that observed in Hi-Fi results.

	\subsection{Model-form uncertainty for overshoot}\label{subsec:model-form-uncertainty}
	In FI-based correction for transition/turbulence model, the critical first step is to identify the appropriate location within the traditional model equation where model-form uncertainty should be introduced. In the majority of existing FIML-based works, the introduction of model-form uncertainty is often straightforward without explicit assessment of its physical or structural relevance to the phenomenon of interest (\eg, multiplicative correction to the Spalart-Allmaras (SA) production terms \cite{Yan-2022-IJHFF, Fang-2024-POF, Nishi-2024-Aerospace}, SST production \cite{Ho-2021-AIAAconference} or destruction terms \cite{ChenLong-2025-PRF, WuChenyu-2023-PRF, WuChenyu-2025-AIAA}, and $F_{\rm{onset1}}$ in $\gamma$ model \cite{ZhangTianxin-2023-POF}). This is probably because the target physical phenomena are well understood and unambiguous such as separated flows. However, as highlighted in Sec.~\ref{sec:intro}, the physical origin of overshoot phenomenon is still an open scientific question without consensus explanation, and the mechanistic understanding from RANS mean flow field perspective is largely undeveloped. Consequently, a systematic identification of the model-form uncertainty associated with overshoot is essential.
	
\begin{figure}[h]
		\centerline{\includegraphics[scale=0.3]{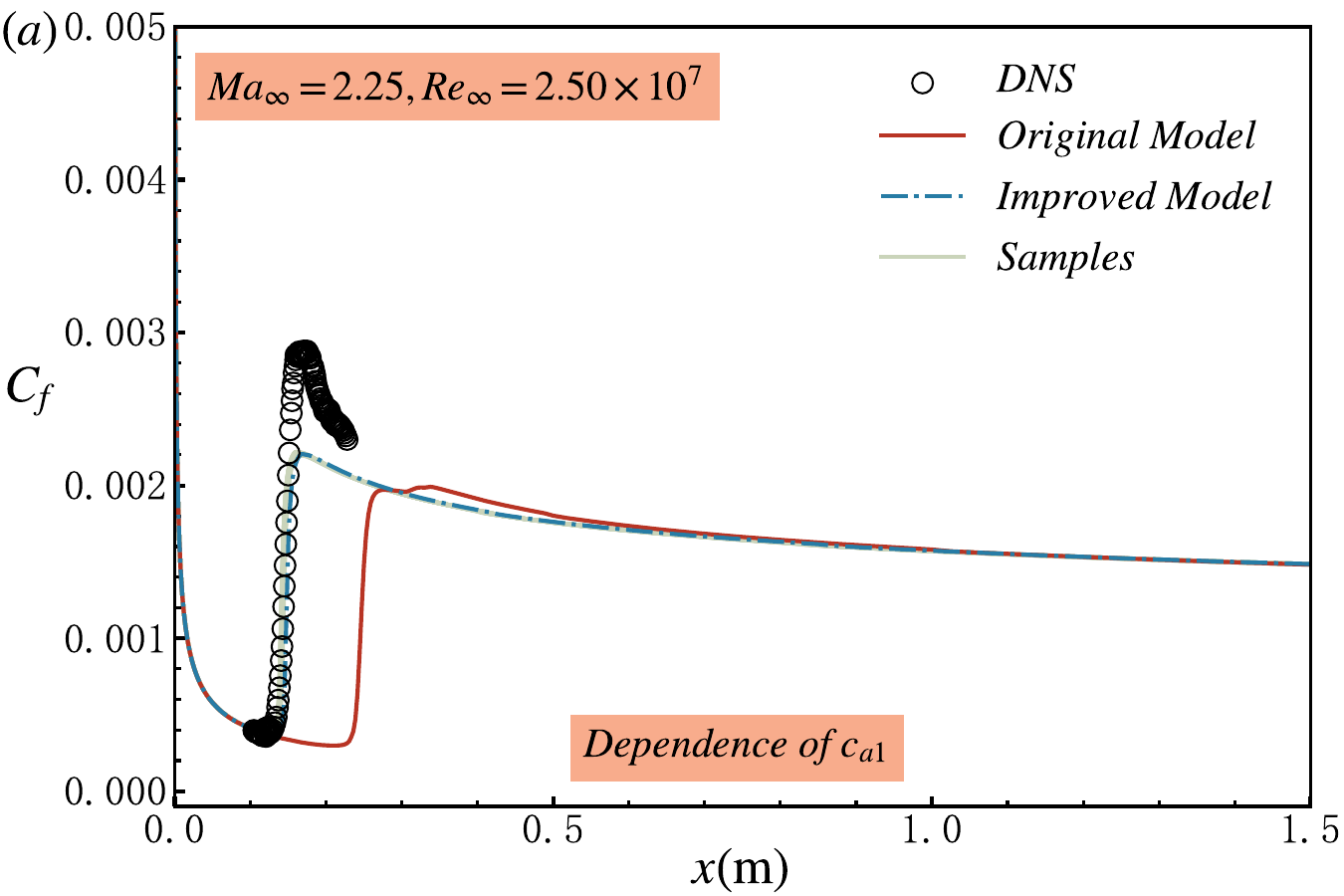}~\includegraphics[scale=0.3]{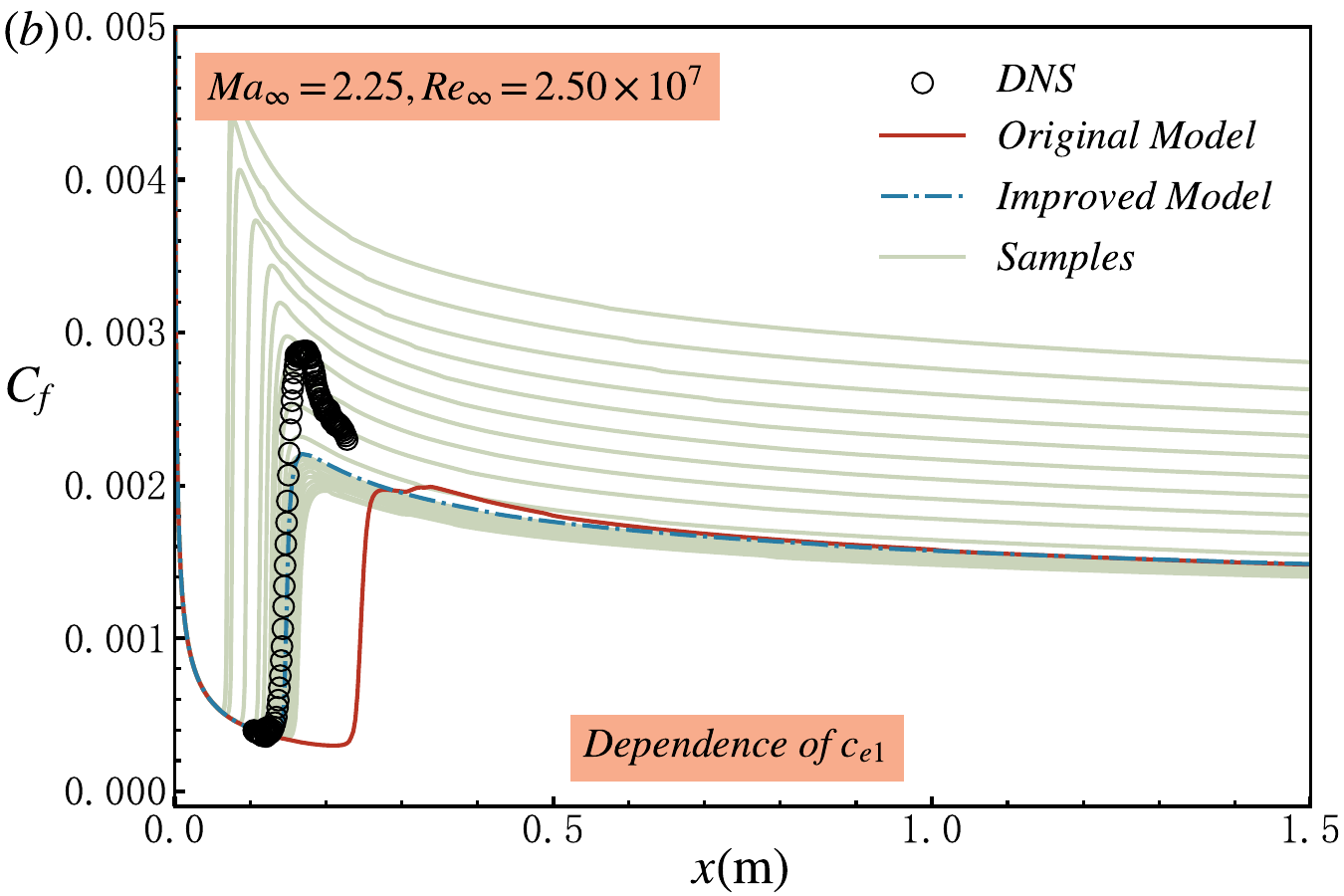}}
		\centerline{\includegraphics[scale=0.3]{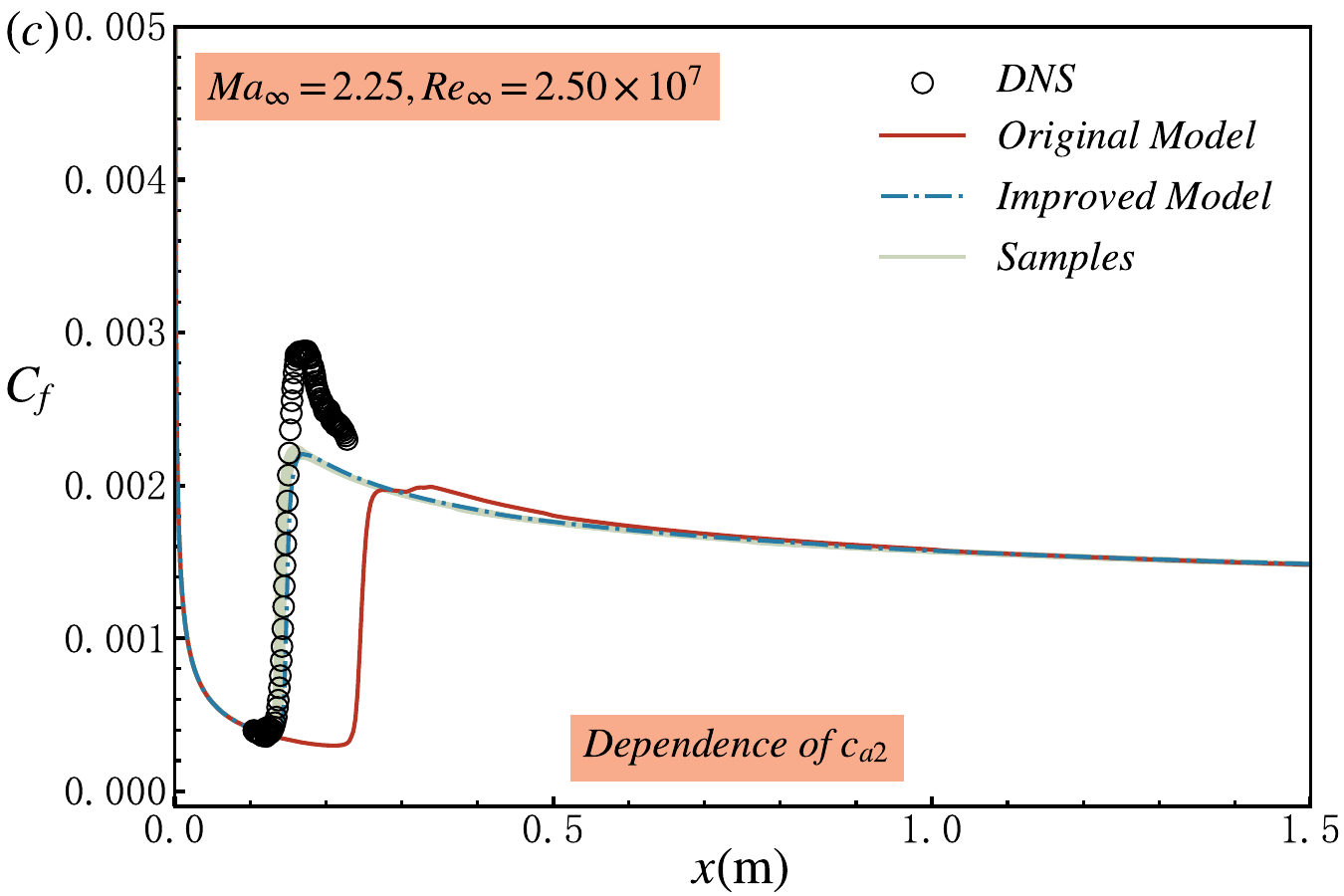}~\includegraphics[scale=0.3]{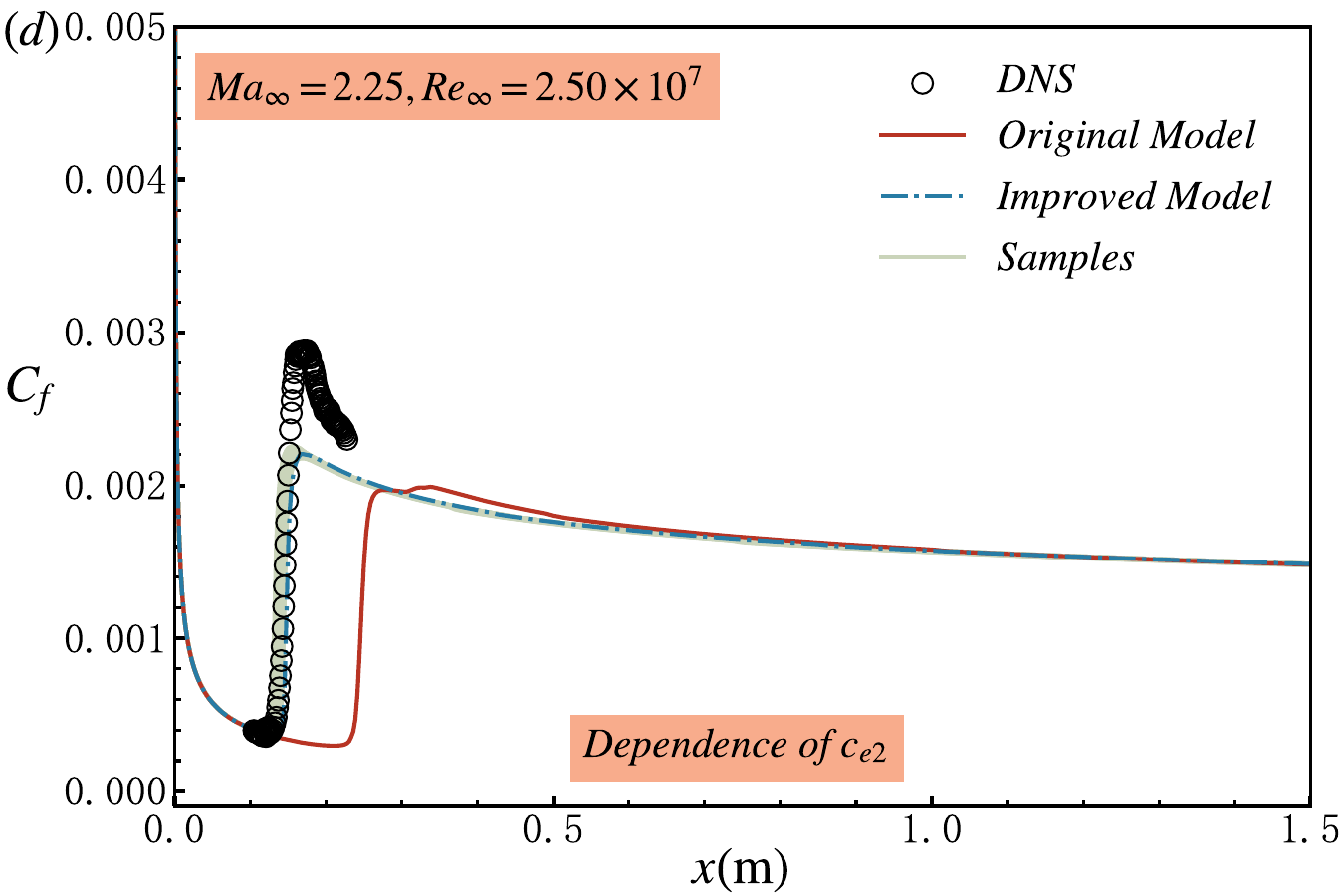}}
		\centerline{\includegraphics[scale=0.3]{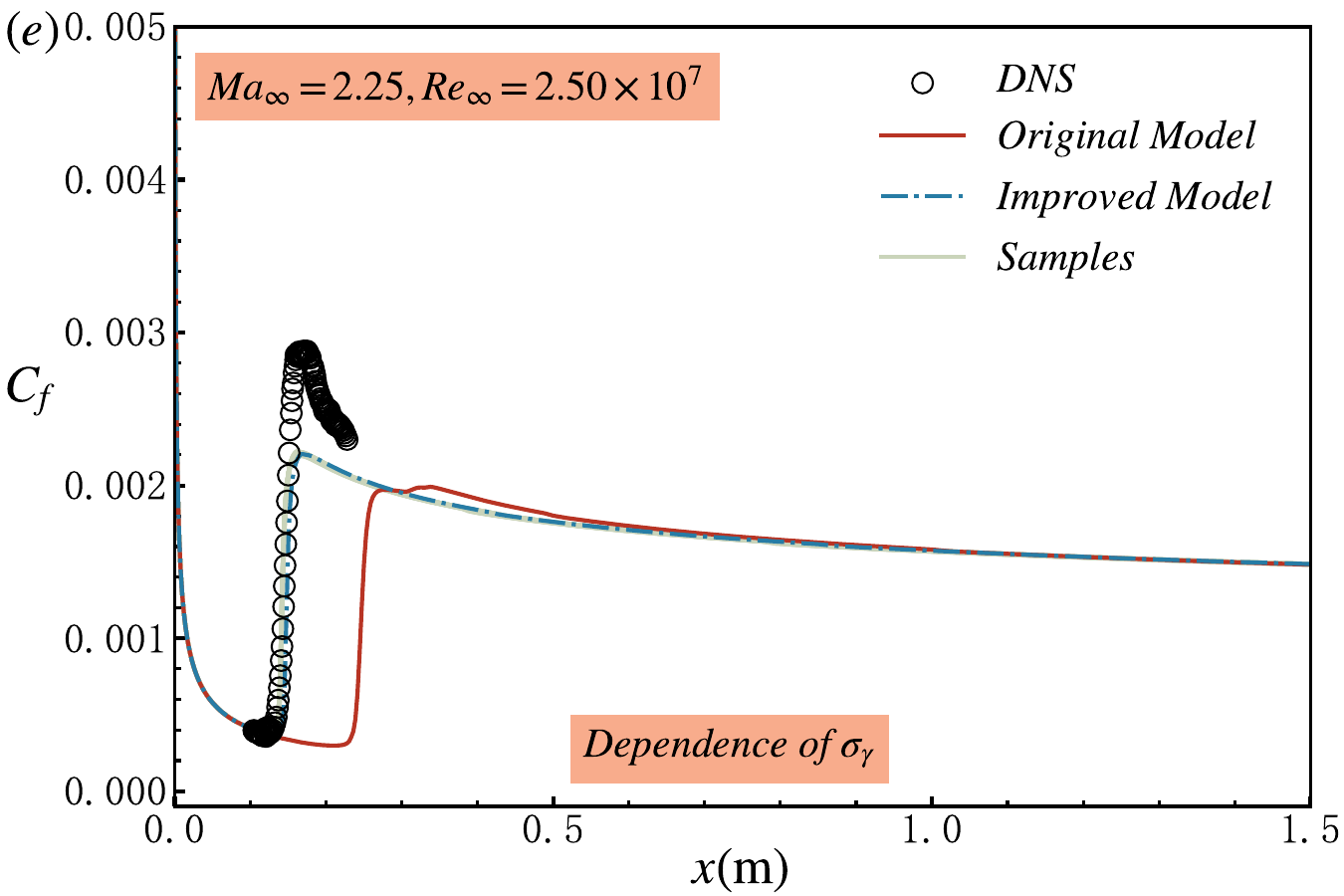}~\includegraphics[scale=0.3]{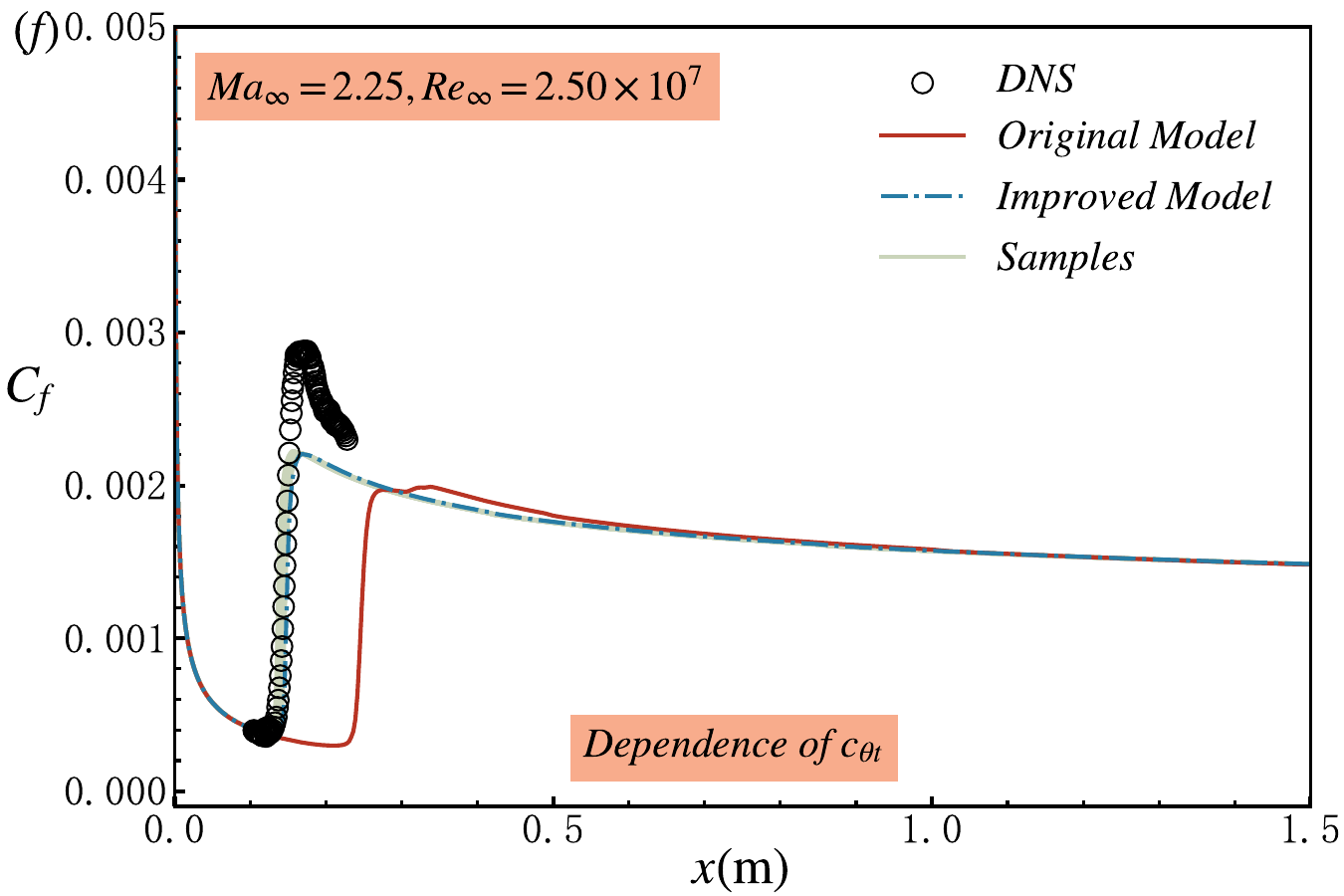}}
		\centerline{\includegraphics[scale=0.3]{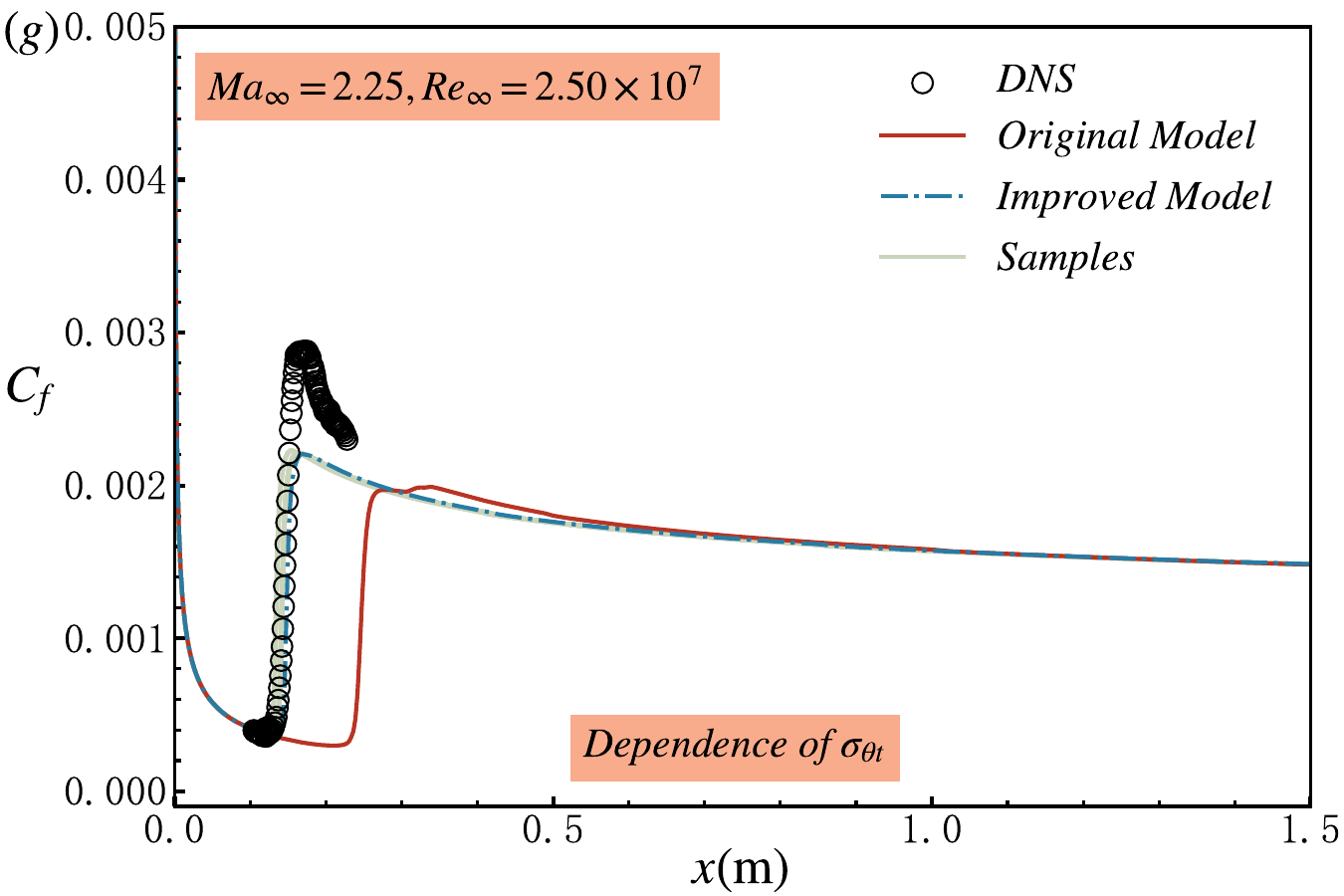}}
		\caption{Model parameter sensitivity analysis for adiabatic flat plate at $Ma_\infty=2.25, Re_\infty=2.50\times10^7$: (a) $c_{a1}$, (b) $c_{e1}$, (c) $c_{a2}$, (d) $c_{e2}$, (e) $\sigma_\gamma$, (f) $c_{\theta \rm{t}}$, and (g) $\sigma_{\theta \rm{t}}$. Here, legends ``Original Model" and ``Improved Model" refer to the results of low-speed model \cite{Langtry-2009-AIAA} and our high-speed improved transition-turbulence model \cite{WuLei-2026-IJHMT}, respectively, while ``Samples" stands for results from individual ensemble members.}
		\label{fig:Dependence-parameter}
	\end{figure}

	Given that the seven parameters ($c_{a1}$, $c_{e1}$, $c_{a2}$, $c_{e2}$, $\sigma_{\gamma}$, $c_{\theta \rm{t}}$, and $\sigma_{\theta \rm{t}}$) in the $\gamma$-$\widetilde{Re}_{\theta \rm{t}}$ model equations collectively represent a near-complete set of model-form uncertainties associated with transition, a sensitivity analysis is performed by perturbing each parameter individually. Specifically, each parameter is varied over $20$ uniform values ranging from $0.5$ to $1.5$ times its original value (\ie, $\pm 50 \%$  perturbation), while all other parameters keep unchanged. The perturbed parameter sets are then fed into the CFD solver, yielding a suite of converged flow field samples. Fig.~\ref{fig:Dependence-parameter} depicts the results of model parameter sensitivity analysis for adiabatic flat plate at $Ma_\infty=2.25, Re_\infty=2.50\times10^7$. Among all seven parameters, only the uncertainty encoded by $c_{e1}$ exerts a decisive influence on both the emergence and disappearance of the skin-friction overshoot.

\begin{figure}[h]
		\centerline{\includegraphics[scale=0.16]{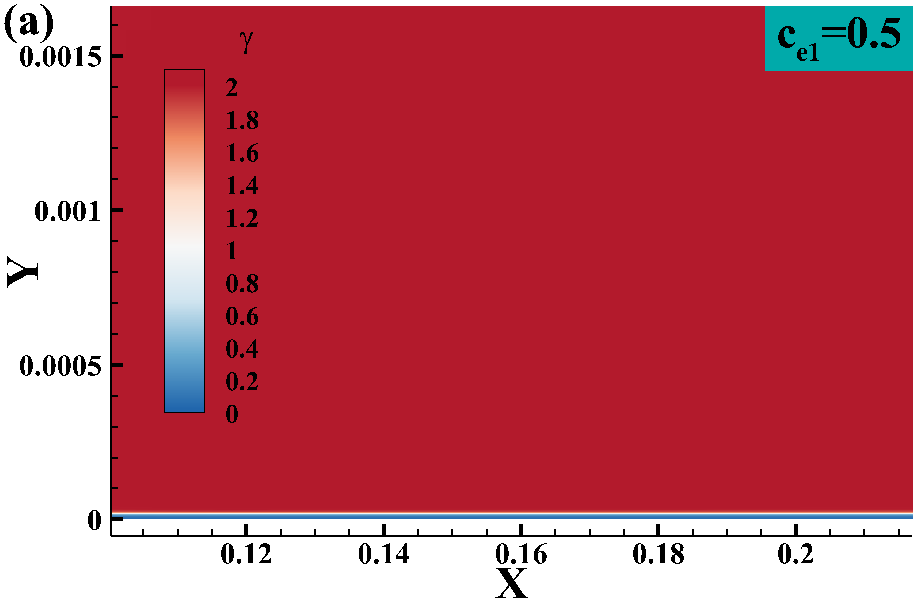}~\includegraphics[scale=0.16]{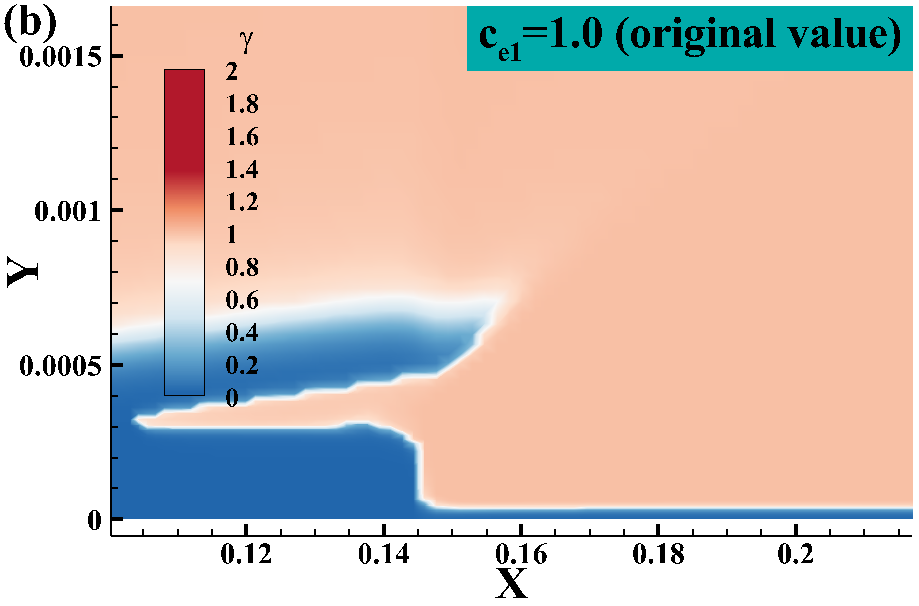}~\includegraphics[scale=0.16]{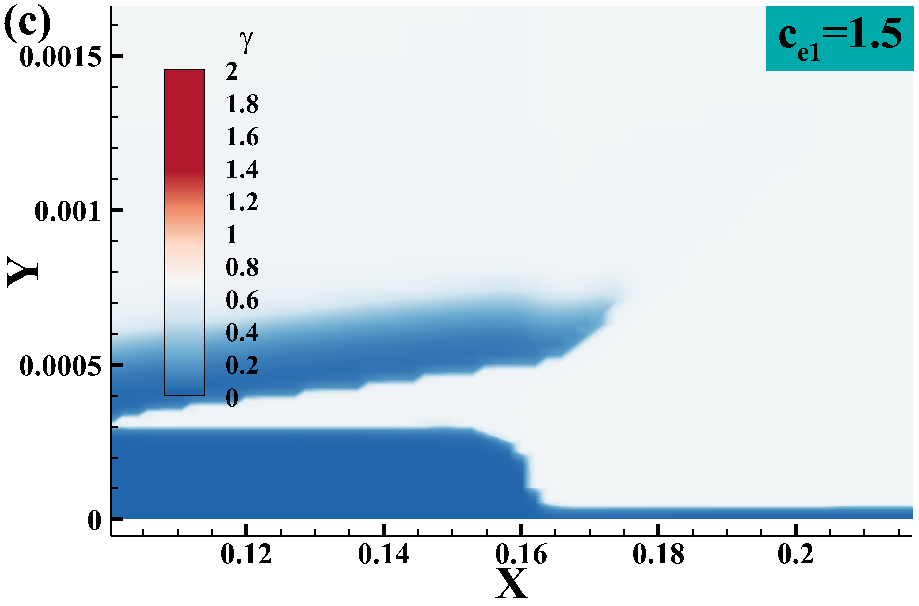}}
		\caption{Zoom-in view of intermittency factor $\gamma$ contours with (a) $c_{e1}=0.5$, (b) $c_{e1}=1.0$, and (c) $c_{e1}=1.5$ for adiabatic flat plate at $Ma_\infty=2.25, Re_\infty=2.50\times10^7$.}
		\label{fig:Yt-ce1}
	\end{figure}

	Three representative $\gamma$ field contours with $c_{e1}=0.5$, $1.0$, and $1.5$ are shown in Fig.~\ref{fig:Yt-ce1}, which correspond to the strongest overshoot (also earliest transition), baseline improved model, and no overshoot with the latest transition, respectively in Fig.~\ref{fig:Dependence-parameter} (b). It is revealed that the presence or absence of overshoot correlates directly with the magnitude of $\gamma$. This observation also explains why \citet{QinYaping-2018-IJHMT} were able to induce overshoot by artificially amplifying $\gamma$ in their simulations. Contour of the production term in $\gamma$ model equation (not shown for brevity) further shows that $c_{e1}$ directly controls the magnitude of $\gamma$ in regions outside the laminar boundary layer and turbulent viscous sublayer. In those regions, $\gamma=1/c_{e1}$ due to the vanished production term. For $c_{e1}=0.5$, $\gamma$ remains uniformly large across the flow field, leading not only to an overshoot but also to an overprediction of skin-friction in the fully turbulent region compared to Hi-Fi data.

	\subsection{EnKF for field inversion}\label{subsec:EnKF}
	It is indicated in Sec.~\ref{subsec:model-form-uncertainty} that model-form uncertainty (denoted as $\beta(\mathbf{x})$) should be introduced at the structural location with parameter $c_{e1}$, which acts as a spatially varying correction to original parameter $c_{e1}=1$
	\begin{eqnarray}
		F_{\rm{length}} c_{a1} \rho S (\gamma F_{\rm{onset}})^{0.5} (1-c_{e1}\gamma) \rightarrow F_{\rm{length}} c_{a1} \rho S (\gamma F_{\rm{onset}})^{0.5} (1-c_{e1}\beta(\mathbf{x})\gamma).
		\label{equ:pro-with-beta}
	\end{eqnarray}
	Leveraging its Bayesian and derivative-free framework, field inversion based on ensemble Kalman filtering (EnKF) \cite{Evensen-EnKF} is employed to quantify the model-form uncertainty with sparse Hi-Fi data and is thus ideally suited for complex, nonlinear inverse problems in engineering.

	EnKF is a Monte Carlo implementation of the classical Kalman filter (KF), designed to address the limitations of KF in handling nonlinear systems and high-dimensional state spaces. Unlike KF, which requires the explicit propagation and storage of the full covariance matrix (which becomes computationally prohibitive for transition-turbulence flow problems where the state dimension scales with the number of grid points), the EnKF represents the state probability distribution through an ensemble of state samples. The covariance matrix is then estimated from the ensemble spread, thereby avoiding the need to store or compute large covariance matrices directly.
	
	Standard EnKF algorithm consists of forecast (prediction) step and analysis (update) step. Consider an ensemble of $N$ independent members. In forecast step, the analysis state vector of $i$-th ensemble member at $k$-th epoch $\mathbf{x}_{i,k}^a$ is propagated forward to form the forecast state vector of next epoch $\mathbf{x}_{i,k+1}^f$
	\begin{eqnarray}
		\mathbf{x}_{i,k+1}^f = \mathcal{G}(\mathbf{x}_{i,k}^a) + \mathbf{q}_{i,k},
		\label{equ:EnKF-forecast}
	\end{eqnarray}
	where $\mathcal{G}(\cdot)$ represents the equivalent nonlinear forward operator for RANS and current high-speed transition-turbulence model equations. $\mathbf{q}_{i,k} \sim \mathcal{N}(\mathbf{0}, \mathbf{Q})$ is the Gaussian white noise with zero mean and covariance matrix $\mathbf{Q}$. The ensemble-based forecast covariance matrix $\mathbf{P}_{k+1}^{f}$ is then estimated from the ensemble deviations
	\begin{eqnarray}
		\mathbf{P}_{k+1}^f = \frac{1}{N-1} \sum_{i=1}^N \left(\mathbf{x}_{i,k+1}^f-\overline{\mathbf{x}_{i,k+1}^f}\right) \left(\mathbf{x}_{i,k+1}^f-\overline{\mathbf{x}_{i,k+1}^f}\right)^T, 
		\label{equ:EnKF-covariance}
	\end{eqnarray}
	where $\overline{\mathbf{x}_{i,k+1}^f}=\frac{1}{N} \sum_{i=1}^N \mathbf{x}_{i,k+1}^f$ is the ensemble mean.
	
	In the analysis step, the forecast ensemble is updated using available observations $\mathbf{y}_{i,k+1}$, which is related to the forecast state vector through the observation operator $\mathcal{H}$
	\begin{eqnarray}
		\mathbf{y}_{i,k+1} = \mathcal{H}\mathbf{x}_{i,k+1}^f + \mathbf{e}_{i,k+1},
		\label{equ:EnKF-analysis}
	\end{eqnarray}
	where $\mathbf{e}_{i,k+1} \sim \mathcal{N}(\mathbf{0}, \mathbf{R})$ is the observation error with covariance matrix $\mathbf{R}$. The Kalman gain $\mathbf{K}_{k+1}$ is computed using the sample covariance
	\begin{eqnarray}
		\mathbf{K}_{k+1} = \mathbf{P}_{k+1}^f \mathcal{H}^T (\mathcal{H} \mathbf{P}_{k+1}^f \mathcal{H}^T + \mathbf{R})^{-1}.
		\label{equ:EnKF-Kalman-gain}
	\end{eqnarray}
	Each ensemble member is then updated independently as
	\begin{eqnarray}
		\mathbf{x}_{i,k+1}^a = \mathbf{x}_{i,k+1}^f + \mathbf{K}_{k+1}(\mathbf{y}_{i,k+1}-\mathcal{H}\mathbf{x}_{i,k+1}^f),
		\label{equ:EnKF-updata}
	\end{eqnarray}
	here, $\mathbf{x}_{i,k+1}^a$ denotes the analysis state vector of $j$-th member at $k+1$ EnKF epoch.
	
	A critical prerequisite for the success of EnKF is the construction of an informative initial ensemble that adequately represents the prior uncertainty in the state to be inferred. Without sufficient variability among ensemble samples, the filter will lack the necessary degrees of freedom to effectively assimilate observational information and update the state estimate. According to our preliminary attempts, it is challenging to generate a physically plausible and sufficiently diverse initial sample for $\beta(\mathbf{x})$ primarily due to the lack of prior knowledge about its spatial structure and magnitude. To circumvent this difficulty, variable of model equation (\ie, intermittency factor $\gamma$) is taken as the field inversion variable. Therefore, in current EnKF framework, the state vector $\mathbf{x}$ corresponds to the discrete field of $\gamma$ defined over the computational domain. Sparse Hi-Fi skin-friction or heat-transfer coefficient results serve as observations $\mathbf{y}$. Through iterative application of the EnKF forecast and analysis steps, the ensemble of $\gamma$ fields converges toward a spatial distribution that minimizes the discrepancy between RANS predictions and Hi-Fi reference data.
	%, which comprises the analysis state vector of all ensemble members ($\mathbf{x}_{i,k}^a$, $i=1...N$)

	\subsection{Field inversion results}\label{subsec:EnKF-res}
\begin{figure}[h]
		\centerline{\includegraphics[scale=0.3]{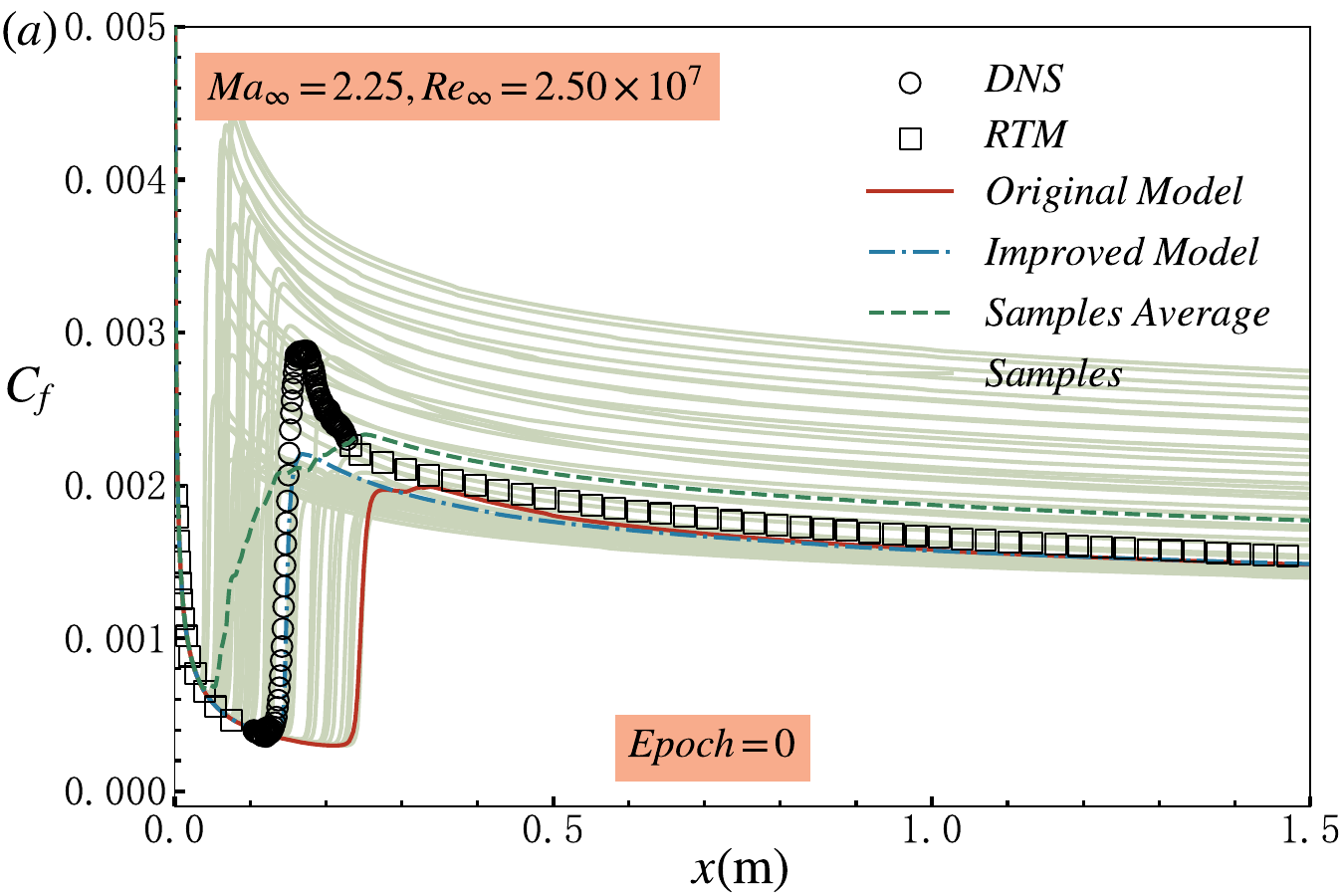}~\includegraphics[scale=0.3]{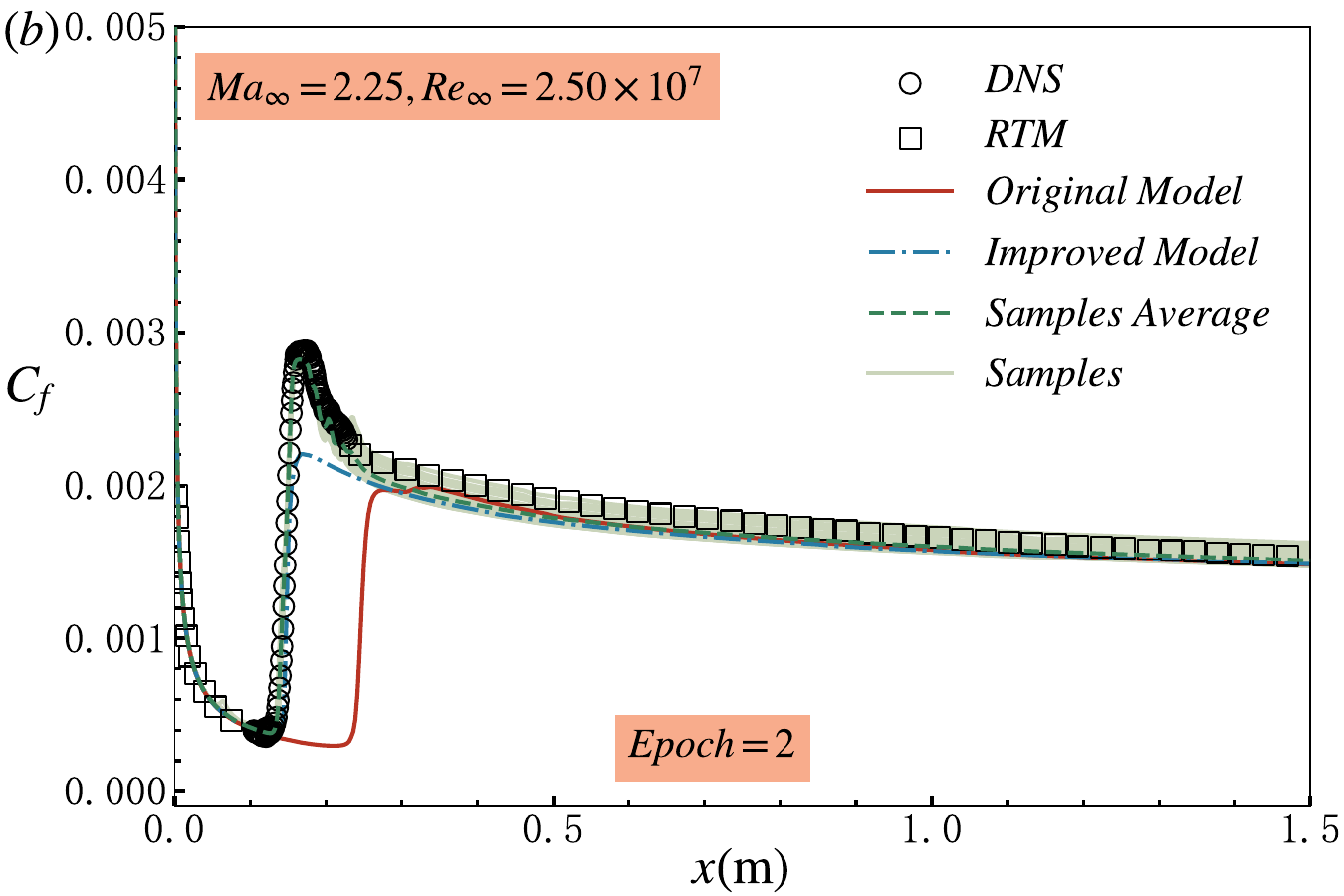}}
		\centerline{\includegraphics[scale=0.3]{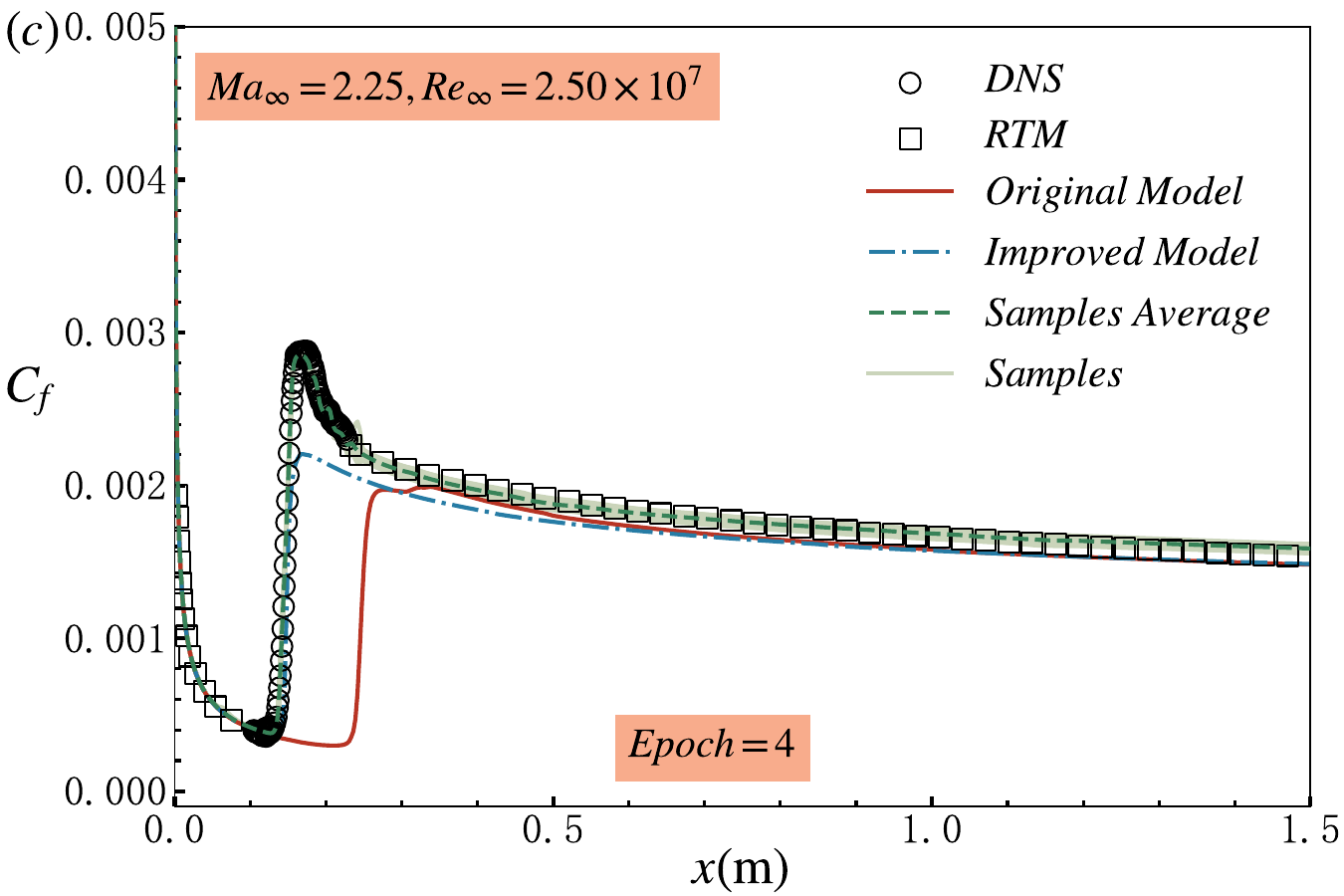}~\includegraphics[scale=0.3]{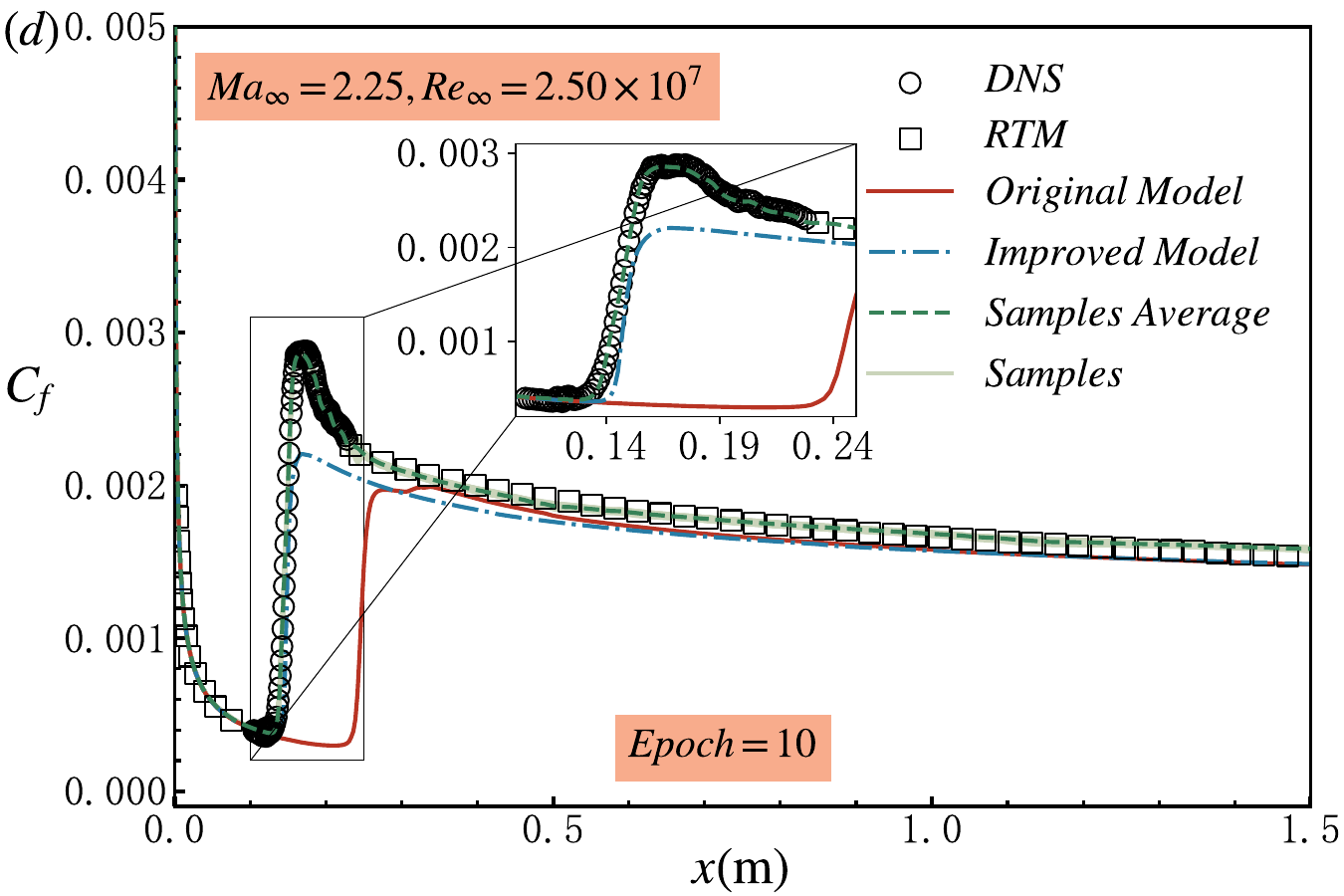}}
		\caption{Distributions of skin-friction coefficient $C_f$ for an adiabatic flat plate at $Ma_{\infty}=2.25, Re_{\infty}=2.50\times10^7$ obtained at different EnKF epochs. Legend ``Samples Average" denotes the mean among all ensemble samples. DNS \cite{Flat-Plate-DNS2.25} and turbulent skin-friction estimated via the reference temperature method (RTM) \cite{WuLei-2026-IJHMT} are shown for comparison. All other legends are consistent with those in Fig.~\ref{fig:Dependence-parameter}.}
		\label{fig:F1-EnKF-Cftau}
	\end{figure}

	EnKF-based field inversion is first implemented for an adiabatic flat plate at $Ma_{\infty}=2.25, Re_{\infty}=2.50\times10^7$. Since DNS data \cite{Flat-Plate-DNS2.25} are unavailable in the turbulent region, skin-friction there is estimated using the reference temperature method (RTM) and included as part of the observation data. Details of RTM can be found in our previous research on the high-speed improved model \cite{WuLei-2026-IJHMT}.
	Fig.~\ref{fig:F1-EnKF-Cftau} presents the distributions of skin-friction coefficient $C_f$ from EnKF at different iterative epochs. As shown, modified by the two improvements, the high-speed improved model yields a significant improvement over the original model in transition location and length. Unfortunately, the overshoot phenomenon around the end of transition process has yet to be resolved. From the EnKF ensemble results, it is evident that the initial samples are sufficiently dispersed, providing a informative prior distribution that underpins the success for subsequent field inversion. Remarkably, the ensemble members converge closely to the Hi-Fi data by the second epoch. After ten EnKF iterations, full convergence is achieved. The resulting RANS flow fields faithfully resolves the skin-friction overshoot phenomenon and also exhibit excellent agreement with Hi-Fi data across transition onset and length.

\begin{figure}[h]
		\centerline{\includegraphics[scale=0.2]{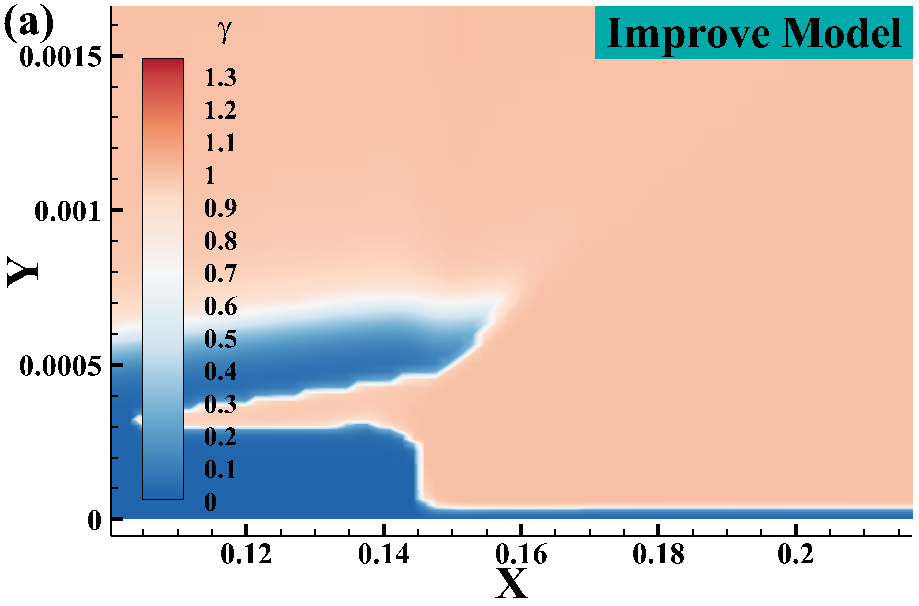}~\includegraphics[scale=0.2]{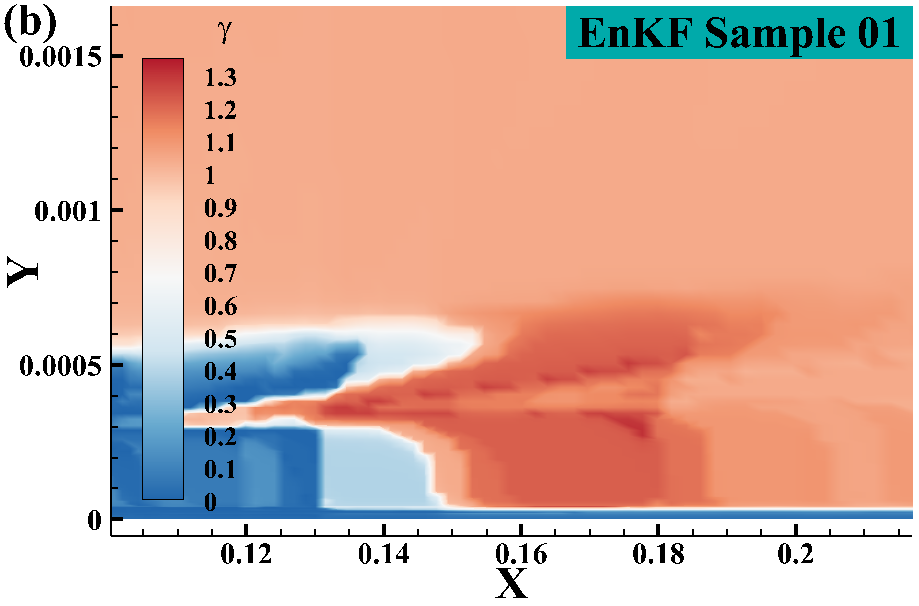}}
		\caption{Intermittency factor fields of an adiabatic flat plate at $Ma_{\infty}=2.25, Re_{\infty}=2.50\times10^7$ near the transition region from (a) improved model and (b) first ensemble member of EnKF at tenth epoch.}
		\label{fig:F1-Yt-contour}
	\end{figure}

	As a representative case, Fig.~\ref{fig:F1-Yt-contour} compares the intermittency factor fields near the transition region for the improved model and first ensemble member of EnKF at tenth epoch. In the improved model, $\gamma$ is constrained to unity throughout both transitional and turbulent region due to the fixed value $c_{e1}=1.0$, which inherently suppresses the overshoot phenomenon. In contrast, the EnKF-based $\gamma$ exhibits a larger value than unity in the transitional region (approximately $x=0.15-0.19$ in Fig.~\ref{fig:F1-EnKF-Cftau} (d)), which coincides precisely with the location of skin-friction overshoot. As the flow fully transitions into turbulence, $\gamma$ gradually decays back toward unity, and the corresponding level of skin-friction retreats to its canonical turbulent level. Therefore, appropriate spatial evolution of $\gamma$ not only triggers the overshoot of skin-friction but also governs its natural damping. 
		
\begin{figure}[h]
		\centerline{\includegraphics[scale=0.3]{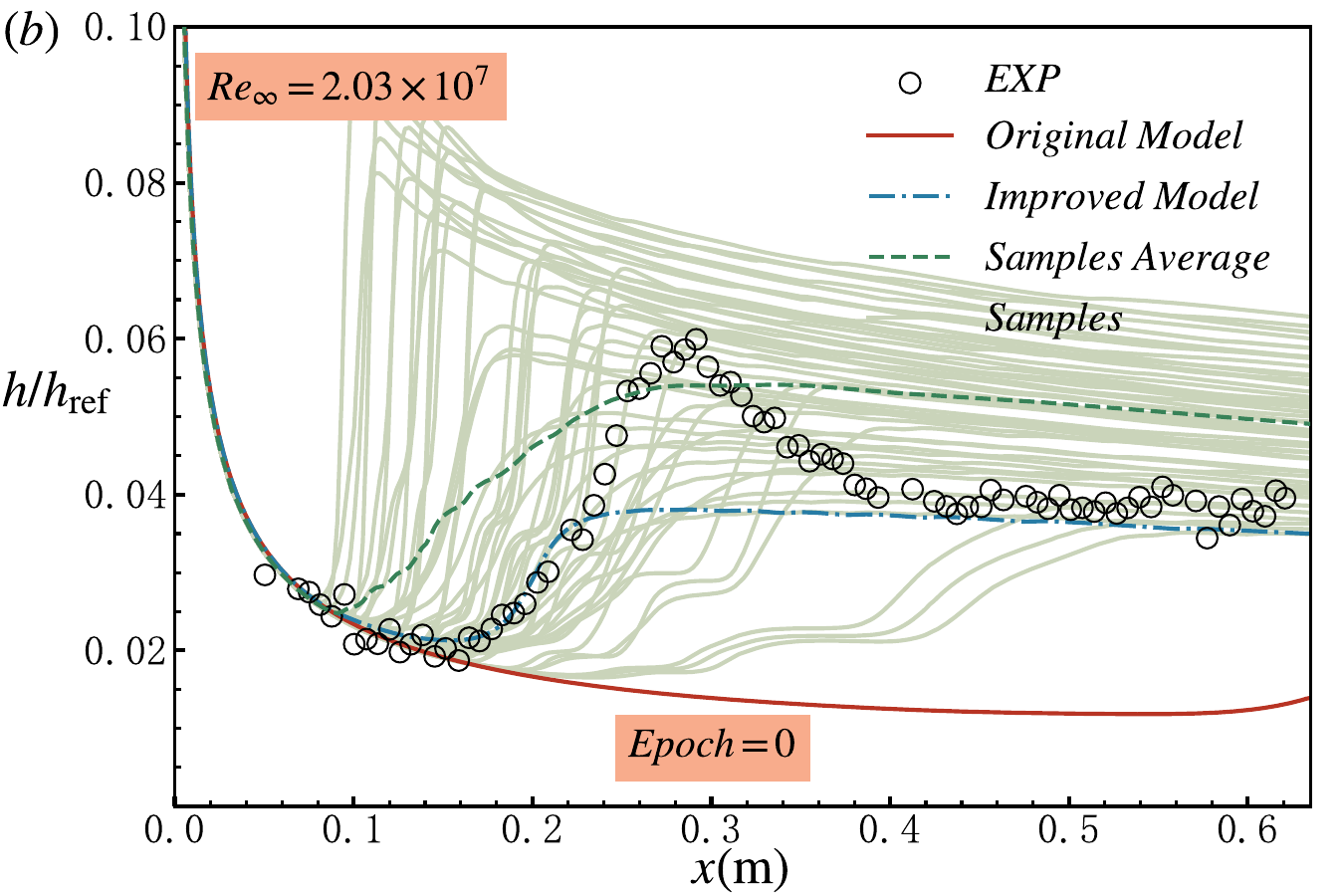}~\includegraphics[scale=0.3]{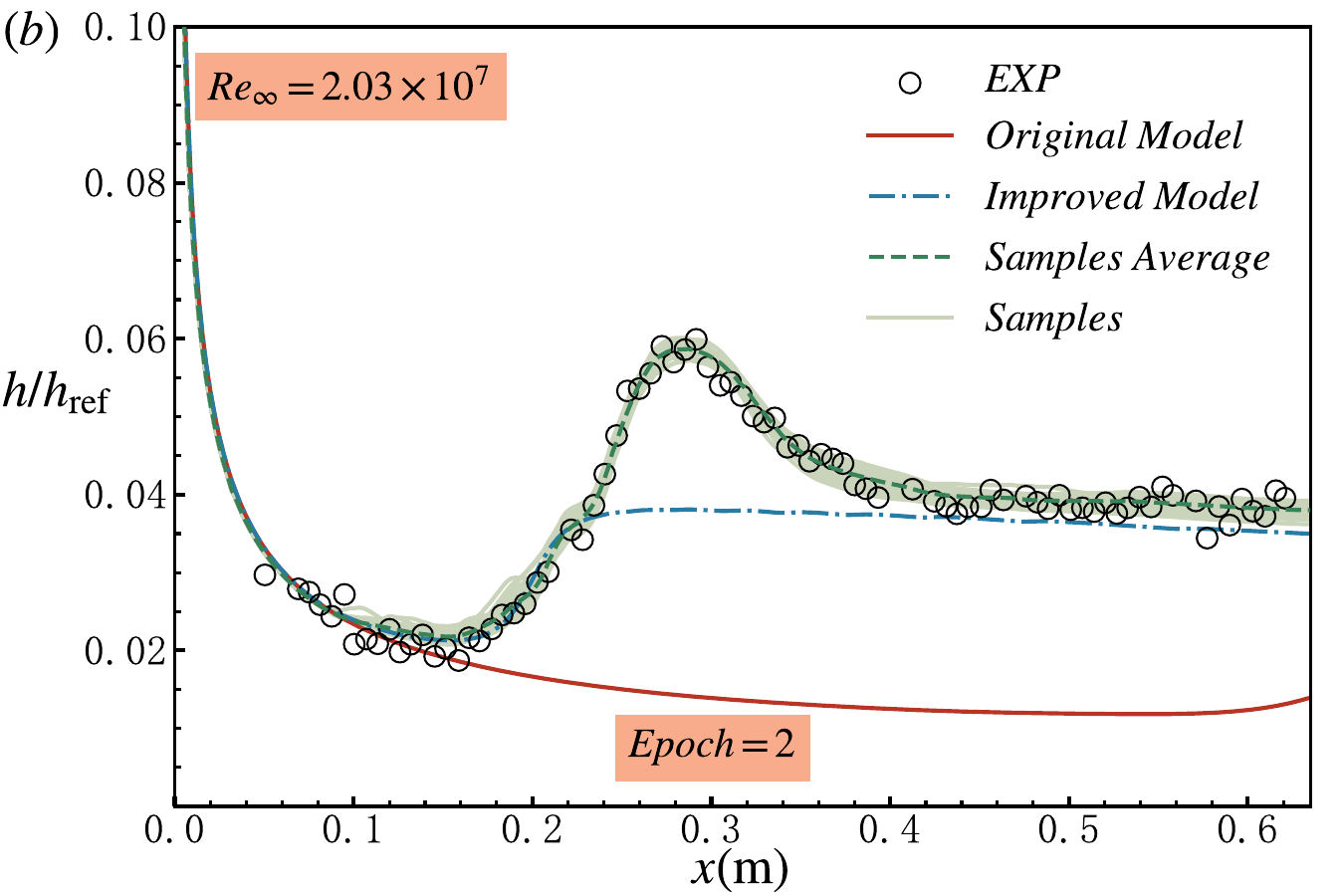}}
		\centerline{\includegraphics[scale=0.3]{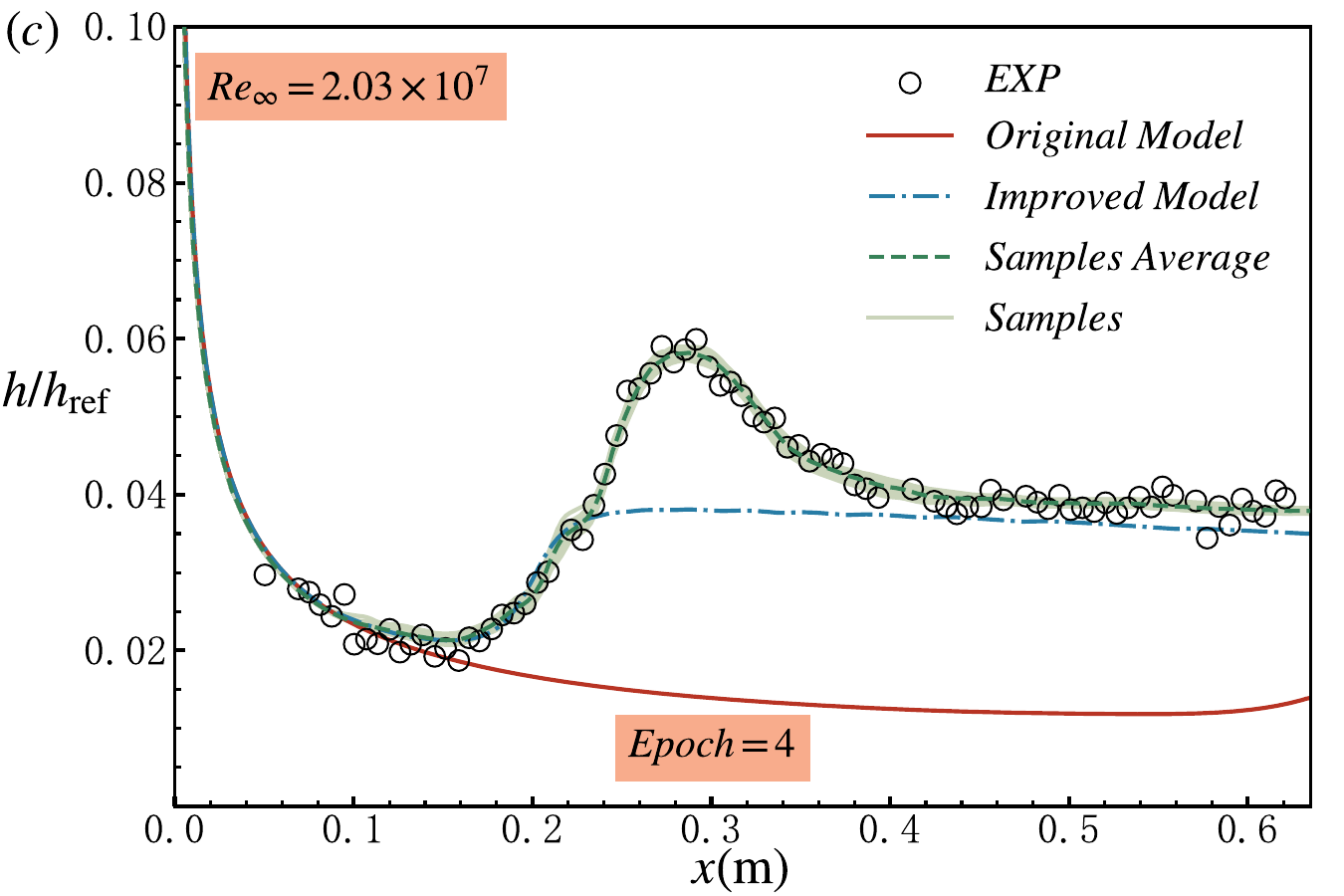}~\includegraphics[scale=0.3]{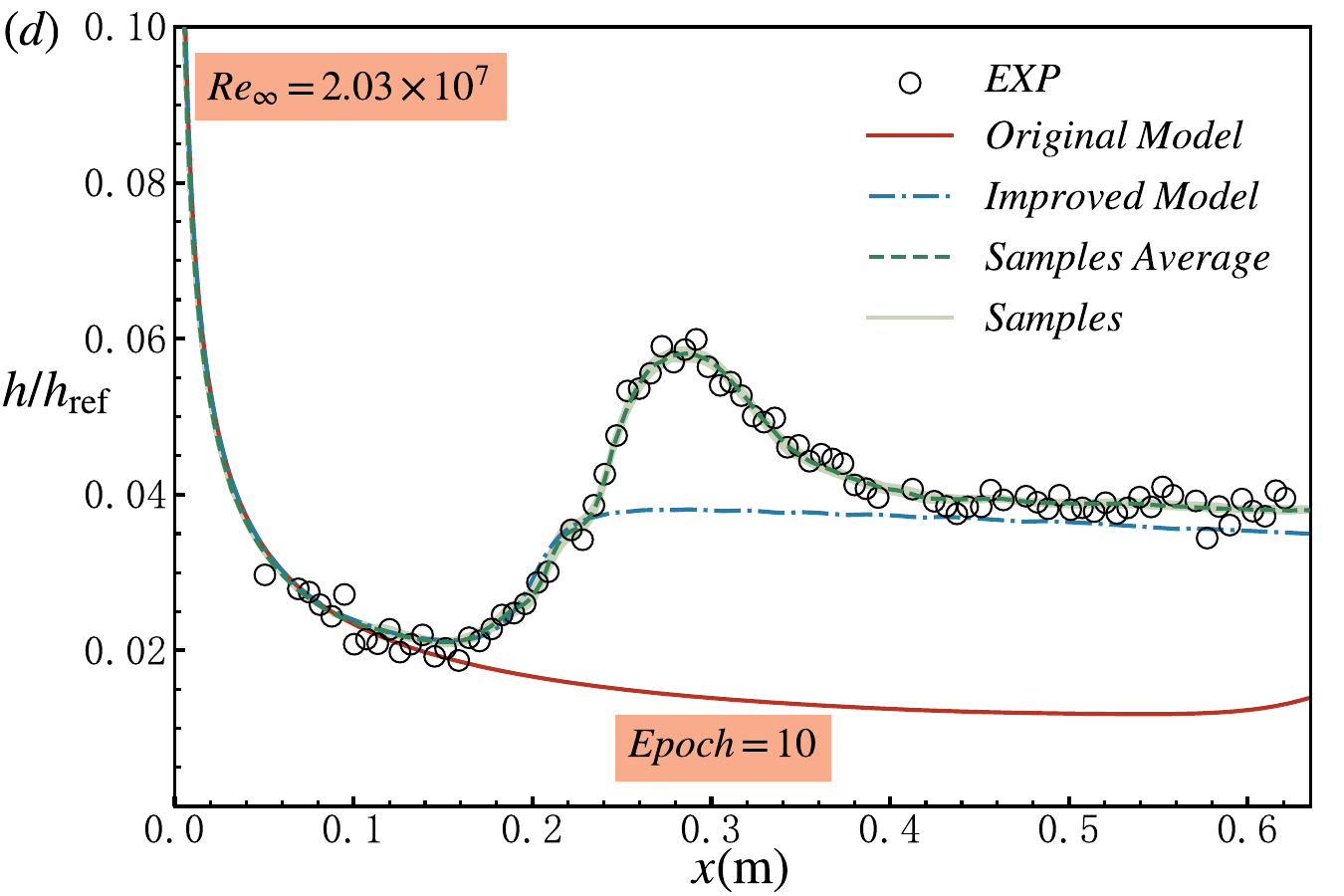}}
		\caption{Distributions of heat-transfer coefficient $h/h_{\rm{ref}}$ for an isothermal sharp cone at $Ma_{\infty}=6.00, Re_{\infty}=2.03\times10^7$ obtained at different EnKF epochs. Legend ``EXP" refers to the experimental results of \citet{Horvath}. All other legends are consistent with those in Fig.~\ref{fig:F1-EnKF-Cftau}.}
		\label{fig:S4-EnKF-h}
	\end{figure}

	A similar EnKF-based field inversion process is also conducted for an isothermal sharp cone at $Ma_{\infty}=6.00, Re_{\infty}=2.03\times10^7$. Fig.~\ref{fig:S4-EnKF-h} presents the distributions of heat-transfer coefficient $h/h_{\rm{ref}}$ at different EnKF iterative epochs. The results exhibit the same trend as observed in the flat plate case: the initial ensemble is widely dispersed, rapid convergence toward Hi-Fi data occurs within a few epochs, and the final RANS flow fields accurately captures the heat-transfer overshoot. These outcomes confirm the effectiveness and generalizability of the EnKF-based field inversion in producing overshoot-resolved RANS flow fields. This achievement overcomes the long-standing bottleneck of existing high-speed transition models and paves the way for discovering interpretable and white-box model expression that imbue the baseline model with a native capacity for overshoot prediction.

	\section{Symbolic regression}\label{sec:SR}
	\subsection{Derivation of model-form uncertainty dataset}\label{subsec:derivation-beta}
\begin{figure}[h]
		\centerline{\includegraphics[scale=0.2]{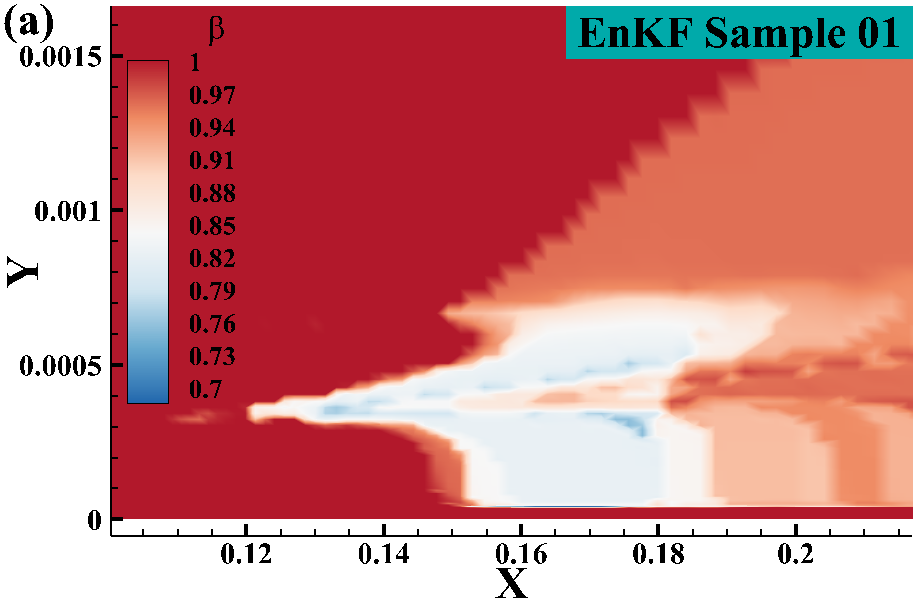}~\includegraphics[scale=0.2]{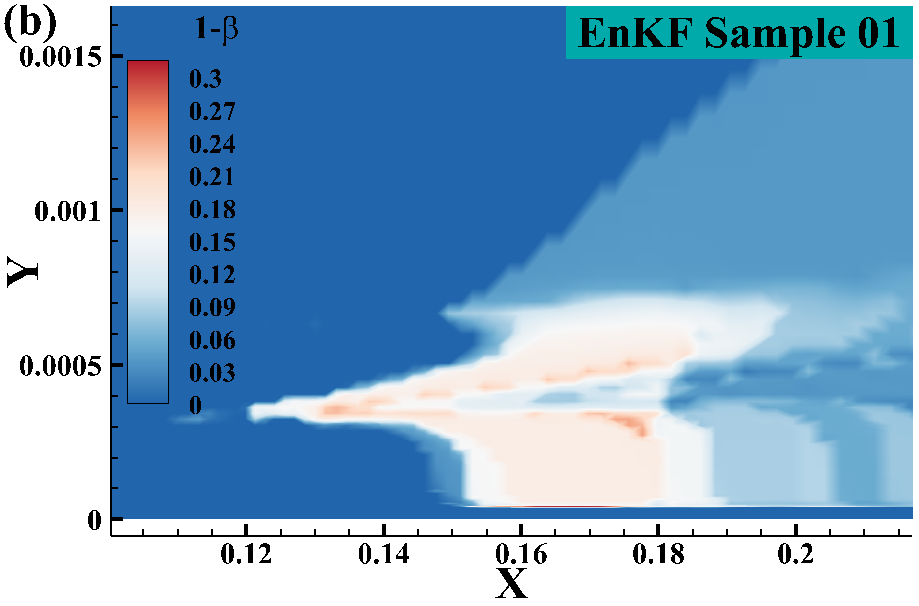}}
		\caption{Model-form uncertainty of (a) $\beta(\mathbf{x})$ and (b) $1-\beta(\mathbf{x})$ fields of an adiabatic flat plate at $Ma_{\infty}=2.25, Re_{\infty}=2.50\times10^7$ near the transition region from first ensemble member of EnKF at tenth epoch.}
		\label{fig:beta_inverse_1_beta_inverse}
	\end{figure}

	Based on the overshoot-resolved field inversion RANS solution in Sec.~\ref{subsec:EnKF-res}, the spatially varying model-form uncertainty $\beta(\mathbf{x})$ that we have introduced into the $c_{e1}$ term of $\gamma$ (see Eq.~\ref{equ:pro-with-beta}) can be algebraically inverted as a function of RANS local variables
	\begin{eqnarray}
		\beta(\mathbf{x})=\left(1-\frac{\partial_j\left(\rho u_j \gamma\right)+E_\gamma-\partial_j\left(\left(\mu_L+\mu_T / \sigma_\gamma\right) \partial_j \gamma\right)}{F_{\text {length }} c_{a 1} \rho S\left(\gamma F_{\rm {onset }}\right)^{0.5}}\right) / c_{e 1} \gamma,
		\label{equ:beta}
	\end{eqnarray}
	in which the temporal term $\partial(\rho \gamma)/\partial t$ is omitted due to the steady characteristic. When $\rho$, $S$, $\gamma$, and $F_{\rm{onset}}$ approach zero, a regularization condition that $\beta(\mathbf{x})=1$ is imposed to eliminate the algebraic singularities. Plotted in Fig.~\ref{fig:beta_inverse_1_beta_inverse} is the derived $\beta(\mathbf{x})$ and $1-\beta(\mathbf{x})$ fields from first ensemble member of EnKF at tenth epoch. Note that the previously discussed relationship between $\gamma$ and $c_{e1}$ in Sec.~\ref{subsec:model-form-uncertainty}, we are not puzzled by these results. Specifically, $\beta(\mathbf{x})$ and $\gamma$ move away from or approach unity in opposite trends. In transition region, $\beta$ deviates significantly from the nominal value of $1$, corresponding to a larger extent of $\gamma > 1$, thus exhibiting a certain degree of overshoot. As $\beta(\mathbf{x})$ gradually retreats to value near $1$, $\gamma$ also recovers to $1$, at which point the transition model reverts to our original high-speed model. This behavior carries clear physical meaning that the deviation of $\beta$ from $1$ serves as a quantitative measure of the local correction exerted to the baseline transition model. Thus, the quantity $1-\beta(\mathbf{x})$ provides a natural metric for the intensity of model-form uncertainty introduced at each spatial location, which naturally serves as the learning target for symbolic regression.
	%Therefore, $1-\beta(\mathbf{x})$ represents the degree of correction exerted to the previous model.

	\subsection{Input features}\label{subsec:inputs}
	Model performance hinges critically on the appropriate selection of input features. As detailed in Table~\ref{tab:inputs}, the candidate inputs for symbolic regression are chosen based on the following physical rationale. First, flow transition is characterized by the production of turbulent kinetic energy (TKE), and TKE level of transitional region typically remains at a relatively high level. So the turbulent Mach number $Ma_k = \sqrt{2k}/c$ is adopted as a candidate feature. However, for high inlet turbulence intensity (e.g., $Tu_\infty = 0.4\%$ in the experiments of sharp and blunt straight cones \cite{Horvath}), $Ma_k$ remains non-negligible across the freestream, so the raw turbulent Mach number may inevitably induce a certain degree of correction in the freestream if it is discovered by SR as an indicator for quantifying the model-form uncertainty $\beta(\mathbf{x})$. This spurious behavior has been repeatedly emphasized and explicitly prohibited by \citet{Srivastava-2021-PRF} in ML-augmented RANS models. To comply with this principle, a background-subtracted turbulent Mach number which subtracts the freestream $Ma_{k,\infty}$ is employed.
	
	Second, wall-distance and turbulent kinetic energy-based Reynolds number $Re_k$ and model function in production term of $\gamma$ equation $F_{\rm{onset}}$ are canonical inputs that have repeatedly proven effective in data-driven transition modeling \cite{WuLei-2022-POF, WuLei-2022-TAML, WuLei-2024-POF}. Dimensionless entropy increase $\Delta s$ and high-speed correlation for $Re_{\nu,\rm{max}}/Re_\theta$, to some extent, serve as a proxies for distance to the nearest wall. Furthermore, as reviewed in Sec.~\ref{sec:intro}, \citet{Franko-2013-JFM} have attributed the overshoot phenomenon to intense Reynolds shear stress $\tau_{12}$, so it is included as an additional input feature. Together, these physically motivated candidates form a comprehensive and representative input space for discovering an interpretable white-box expression for the transitional overshoot.
	\begin{table*}[width=.9\textwidth,cols=4,pos=h]
		\caption{\label{tab:inputs}Candidate input parameters $p_1-p_6$ for symbolic regression. Here, $k$ denotes the turbulent kinetic energy, $c$ is the sound velocity, and $Ma_{k,\infty}=\sqrt{2k_{\infty}}/c_{\infty}$ is the freestream turbulent Mach number. $d$, $\nu$, and $\mu_{\rm{T}}$ are distance to the nearest wall, kinematic viscosity, and eddy viscosity.}
		\begin{tabular*}{\tblwidth}{@{} LLL@{}}
			\toprule
			Input features & Description & Formula \\
			\midrule
			$p_1=Ma_k$   & Background-subtracted turbulent Mach number       & $\frac{\sqrt{2k}}{c}-Ma_{k,\infty}$          \\
			$p_2=Re_k$   & Wall-distance and turbulent kinetic energy-based Reynolds number       & $\min\left(\frac{\sqrt{k}d}{50\nu},2\right)$          \\
			$p_3=\Delta s$   & Dimensionless entropy increase       & $\frac{C_p}{R}\ln\left(\frac{T}{T_{\infty}}\right) + \ln\left(\frac{p_{\infty}}{p}\right)$          \\
			$p_4=F_{\rm{onset}}$   & Model function in production term of $\gamma$ equation      & see Eqs.~\eqref{equ:F-onset} and \eqref{equ:F-onset123}         \\
			$p_5=f\left(Ma_{\rm{local}},\frac{T_{\rm{w}}}{T_{\rm{local}}}\right)$   & High-speed correlation for $\frac{Re_{\nu,\rm{max}}}{Re_\theta}$      & Ref.~\cite{WuLei-2026-IJHMT}        \\
			$p_6=\tau_{12}$   & Reynolds shear stress      & $\mu_{\rm{T}}\left(\frac{\partial u}{\partial y} + \frac{\partial v}{\partial x}\right)$         \\
			\bottomrule
		\end{tabular*}
	\end{table*}

	\subsection{Description of symbolic regression}\label{subsec:descr-sr}
	An open-source symbolic regression tool PySR \cite{Cranmer-2023-PySR} is employed to discover a compact and interpretable analytical expression for the variant of model-form uncertainty $1-\beta(\mathbf{x})$. In PySR, genetic programming is utilized to evolve mathematical expressions through an iterative process of selection and variation. Specifically, the algorithm initiates with a population of randomly generated expression trees, where internal nodes represent mathematical operators and leaf nodes correspond to variables and constants. Through successive generations, these expressions undergo genetic operations including crossover (subtree exchange) and mutation (random subtree modification). This evolutionary mechanism enables the systematic exploration of functional forms without requiring prior assumptions about the model structure.
	
	During the training process, the expression accuracy is quantified using the sum of squared error (SSE)
	\begin{eqnarray}
		\mathrm{SSE}(\mathbf{p})=\sum_{i=1}^N\left[\left(1-\beta_i(\mathbf{x})\right)- f_i(\mathbf{p}) \right]^2,
		\label{equ:SSE}
	\end{eqnarray}
	where $f_i(\mathbf{p})$ denotes the prediction of PySR by input feature vector $\mathbf{p}$ and $N$ is the total number of training samples. To maintain the model interpretability, the complexity of every expression $\mathrm{Comp}(f(\mathbf{p}))$ in which number of variables $N_{\rm{var}}$, operators $N_{\rm{op}}$, and constants $N_{\rm{const}}$ contribute equally with unit weights is calculated
	\begin{eqnarray}
		\mathrm{Comp}(f(\mathbf{p}))=N_{\rm{var}} + N_{\rm{op}} + N_{\rm{const}}.
		\label{equ:complexity}
	\end{eqnarray}
	The loss function used during evolution is designed to balance accuracy and simplicity
	\begin{eqnarray}
		\mathrm{loss}(f(\mathbf{p})) = \mathrm{SSE}(f(\mathbf{p})) \cdot \exp(\mathrm{frecency}[\mathrm{Comp}(f(\mathbf{p}))]),
		\label{equ:loss}
	\end{eqnarray}
	here, the ``frecency" of $\mathrm{Comp}(f(\mathbf{p}))$ is some combined measure of the frequency and recency of expressions occurring at complexity $\mathrm{Comp}(f(\mathbf{p}))$ in the population.
	
	A set of unary and binary operators are predefined as summarized in Table~\ref{tab:operators}. Two EnKF flow fields of first ensemble member at tenth epoch (\ie, Figs.~\ref{fig:F1-EnKF-Cftau} and \ref{fig:S4-EnKF-h}) are used as the training dataset. To balance the proportion of trivial sample (\ie, $1-\beta(\mathbf{x})>0$) and nontrivial samples (\ie, $1-\beta(\mathbf{x}) \le 0$), the downsampling technique is applied. 
	
	Throughout the training process, PySR preserves the best-performing expression (elite) at each complexity level. Upon algorithm termination, it returns a Pareto frontier which contains the best-performing expression at each complexity level, thereby offering a spectrum of candidate models ranging from simple expressions to more accurate but complex formulations. By weighting the predictive accuracy, physical plausibility and domain-specific interpretability requirements, the final expression are selected as
	\begin{eqnarray}
		\beta(\mathbf{x}) = 1 - \frac{Re_k (\exp(0.944Ma_k)-1)} {1+\tanh(F_{\rm{onset}})(1.686 \cdot f\left(Ma_{\rm{local}},\frac{T_{\rm{w}}}{T_{\rm{local}}}\right) \cdot \tanh(\Delta s) - 5.844f\left(Ma_{\rm{local}},\frac{T_{\rm{w}}}{T_{\rm{local}}}\right) + 19.58)  }.
		\label{equ:SR-expression1}
	\end{eqnarray}
	\begin{table*}[width=.9\textwidth,cols=4,pos=h]
		\caption{\label{tab:operators}Predefined unary and binary operators in PySR. Here, $p_i$ and $p_j$ denote the base inputs listed in Table~\ref{tab:inputs}. During evolution, PySR will recursively combines these inputs using the following operators to construct complex expressions.}
		\begin{tabular*}{\tblwidth}{@{} LL@{}}
			\toprule
			Operator type & Operators \\
			\midrule
			Unary operators   & $\exp(p_i),\tanh(p_i),\frac{1}{1+p_i},\frac{1}{p_i}$          \\
			Binary operators  & $p_i+p_j, p_i-p_j, p_i*p_j, \frac{p_i}{p_j}, p_i^{p_j}$       \\
			\bottomrule
		\end{tabular*}
	\end{table*}

	\subsection{Elucidation about the discovered expression}\label{subsec:elucidation}
\begin{figure}[h]
		\centerline{\includegraphics[scale=0.2]{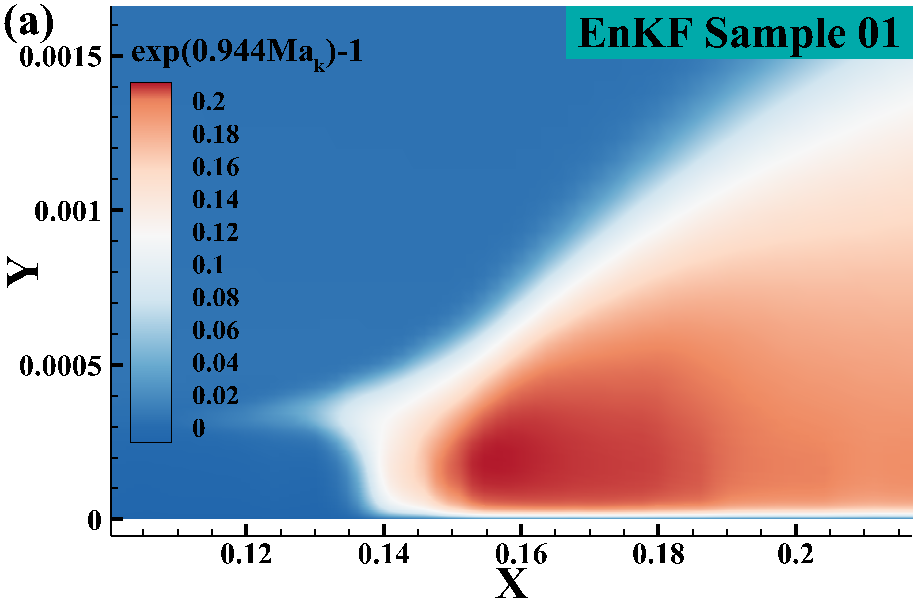}~\includegraphics[scale=0.2]{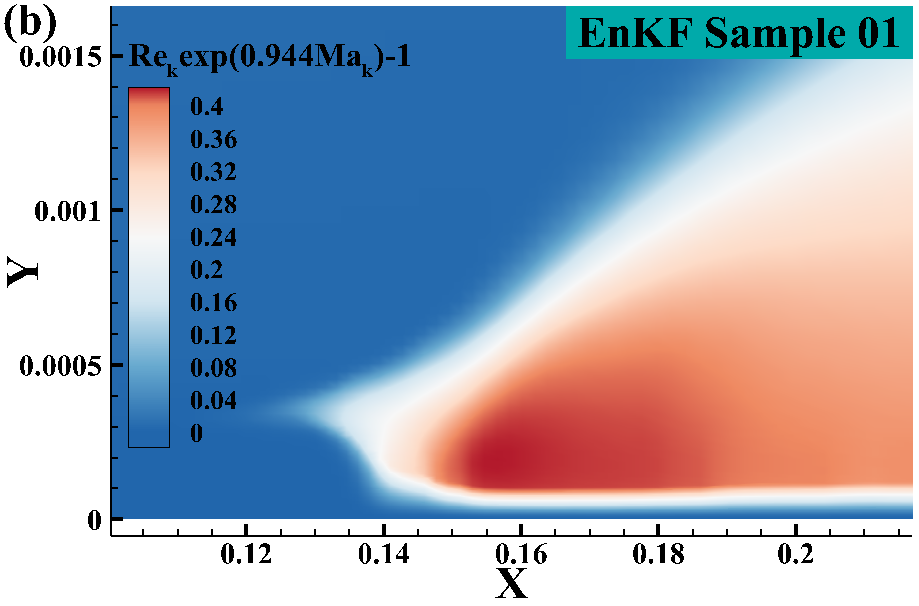}}
		\centerline{\includegraphics[scale=0.2]{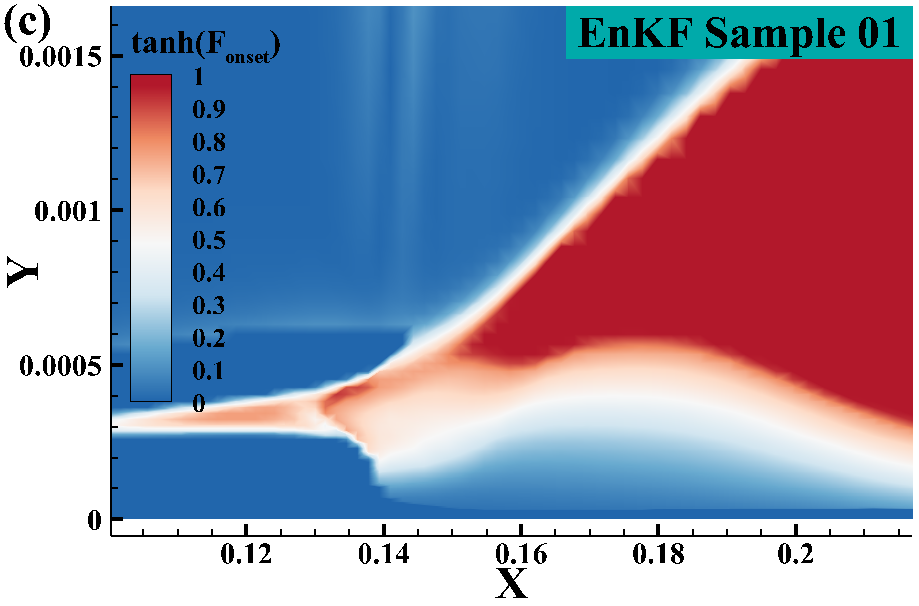}~\includegraphics[scale=0.2]{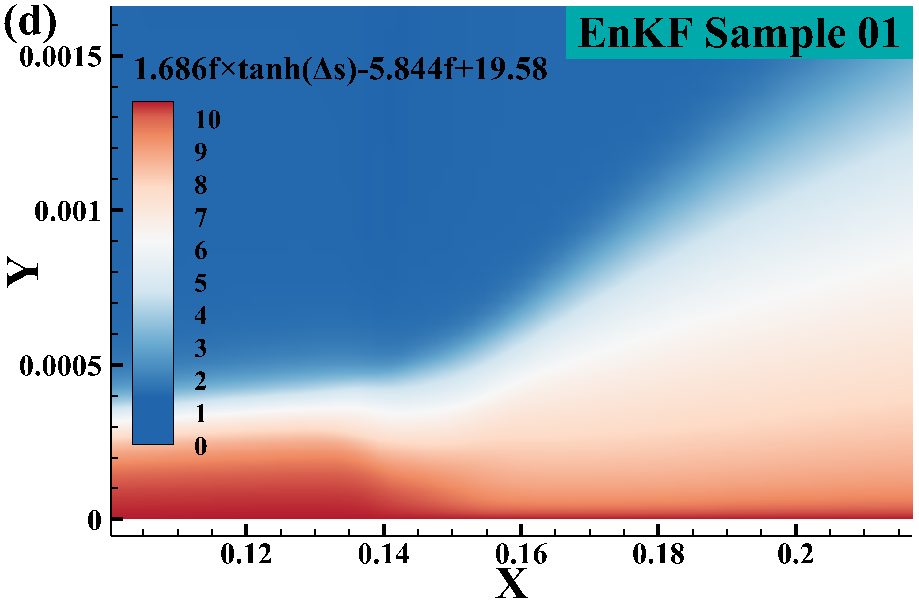}}
		\caption{Contours of each component in analytical expression Eq.~\eqref{equ:SR-expression1} of an adiabatic flat plate at $Ma_{\infty}=2.25, Re_{\infty}=2.50\times10^7$ near the transition region from first ensemble member of EnKF at tenth epoch. Meanings of each panel are labeled in title of legend in which $f$ denotes $f\left(Ma_{\rm{local}},\frac{T_{\rm{w}}}{T_{\rm{local}}}\right)$.}
		\label{fig:SR-expression1-each-term}
	\end{figure}

	Structure and meaning of analytical expression Eq.~\eqref{equ:SR-expression1} discovered by PySR are mined to reveal the physical rationale behind each component. Also using the adiabatic flat plate at $Ma_{\infty}=2.25, Re_{\infty}=2.50\times10^7$ as the representative case, Fig.~\ref{fig:SR-expression1-each-term} shows the contours of each term in Eq.~\eqref{equ:SR-expression1}. It is observed that transition region (see Fig.~\ref{fig:F1-EnKF-Cftau} (d)) where the overshoot occurs is characterized by a pronounced peak in $\exp(0.944Ma_k)-1$ (also $Ma_k$). This elevated $\exp(0.944Ma_k)-1$ in turn reduces $\beta(\mathbf{x})$ and thereby enhances $\gamma$ to trigger the overshoot. The subtraction of unity $-1$ in $\exp(0.944Ma_k)-1$ guarantees that the correction vanishes when $Ma_k=0$, which is essential for preserving the baseline model behavior in laminar or freestream regions. Moreover, since $Ma_k$ scales inversely with the speed of sound $c$, this formulation provides protection for low-speed transitional flows that the correction term $\exp(0.944Ma_k)-1$ remains negligible in subsonic regimes and prevents spurious overshoot phenomenon.
	
	In the viscous sublayer of turbulent region, the $\gamma$ is effectively zero, so no correction to $\beta(\mathbf{x})$ is exhibited (see Fig.~\ref{fig:beta_inverse_1_beta_inverse}). However, the term $\exp(0.944Ma_k)-1$, which depends on TKE, begins to take non-negligible values within the viscous sublayer and increases rapidly with distance from the wall. To prevent the near-wall spurious correction, this term is premultiplied by a damping-like quantity $Re_k=\min(\sqrt{k}d/50\nu,2)$ that is inspired by the near-wall damping functions commonly used in RANS models \cite{SST1994}. This quantity remains very small in both the viscous sublayer and buffer layer, thereby acting as a near-wall protection mechanism. As a result, the composite correction term $Re_k(\exp(0.944Ma_k)-1)$ (see Fig.~\ref{fig:SR-expression1-each-term} (b)) vanishes and thus ensures that $\beta(\mathbf{x})$ approaches unity in turbulent viscous sublayer.
	
	As the flow transitions into fully turbulent state, it is shown in Fig.~\ref{fig:beta_inverse_1_beta_inverse} that the correction to $\beta(\mathbf{x})$ gradually diminishes. In this regime, model function $F_{\rm{onset}}$ reaches its upper bound of $2$, and the term $\tanh(F_{\rm{onset}})$ in denominator of expression effectively attenuates the overall correction magnitude. Conversely, within the transitional region, $F_{\rm{onset}}$ lies between its lower bound $0$ and upper bound $2$, yielding in a smaller $\tanh(F_{\rm{onset}})$ and thus a larger correction to $\beta(\mathbf{x})$. This behavior is physically consistent with the expectation that model exhibits stronger augmentation in transition region and reduces correction once the flow becomes turbulent, which aligns with the physical role of $F_{\rm{onset}}$ as a transition indicator.
	
	The final term of expression Eq.~\eqref{equ:SR-expression1}, $1.686 \cdot f(Ma_{\rm{local}},T_{\rm{w}}/T_{\rm{local}}) \cdot \tanh(\Delta s) - 5.844f(Ma_{\rm{local}},T_{\rm{w}}/T_{\rm{local}}) + 19.58$, is discovered by PySR to depict the law of wall. As illustrated in Fig.~\ref{fig:SR-expression1-each-term} (d), this term increases monotonically as the wall-normal distance decreases. Together with the preceding terms, it forms a physically coherent correction mechanism that collectively adjusts $\beta(\mathbf{x})$.

\begin{figure}[h]
		\centerline{\includegraphics[scale=0.3]{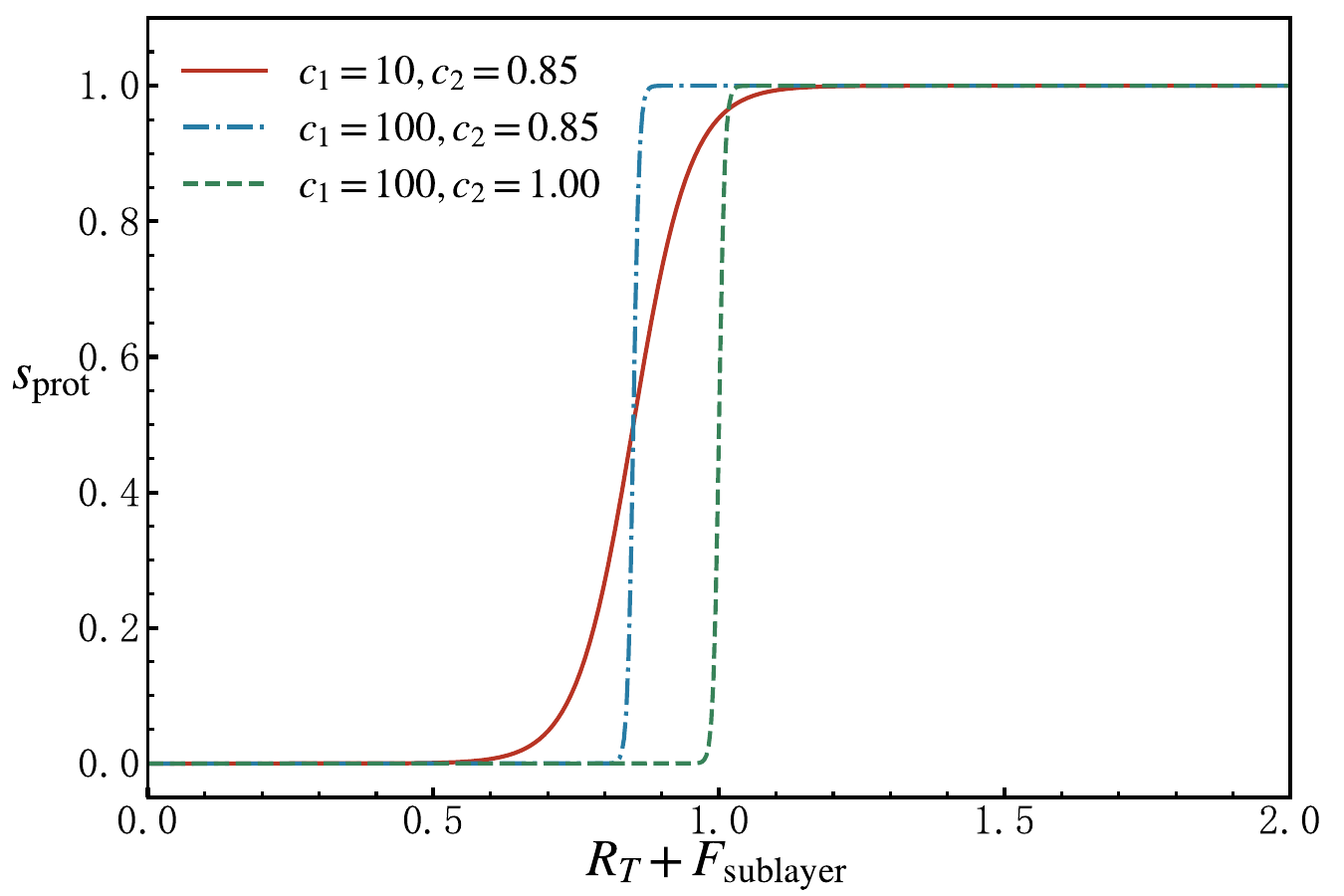}}
		\caption{Profiles of $s_{\rm{prot}}$ under different $c_1$ and $c_2$.}
		\label{fig:s_prot-profile}
	\end{figure}

	Although the analytical expression derived by PySR reflects clear and rational pathway for resolving the overshoot phenomenon, a potential limitation of Eq.~\eqref{equ:SR-expression1} still exists. As depicted in Fig.~\ref{fig:SR-expression1-each-term} (a), turbulent Mach number $Ma_k$ exhibits a certain level in the laminar region close to the onset of transition. This causes the correction term $\exp(0.944Ma_k)-1$ to activate prematurely, leading to an unexpected advancement of transition location. To address this, a protection function $s_{\rm{prot}}$ structurally similar to that of \citet{WuChenyu-2023-PRF} is proposed and multiplicatively applied to the numerator of original expression Eq.~\eqref{equ:SR-expression1}
	\begin{eqnarray}
		\beta(\mathbf{x}) = 1 - \frac{s_{\rm{prot}} \cdot Re_k (\exp(0.944Ma_k)-1)} {1+\tanh(F_{\rm{onset}})(1.686 \cdot f\left(Ma_{\rm{local}},\frac{T_{\rm{w}}}{T_{\rm{local}}}\right) \cdot \tanh(\Delta s) - 5.844f\left(Ma_{\rm{local}},\frac{T_{\rm{w}}}{T_{\rm{local}}}\right) + 19.58)  },
		\label{equ:SR-expression2}
	\end{eqnarray}
	in which
	\begin{eqnarray}
		s_{\rm{prot}} = 0.5\tanh(c_1(R_{\rm{T}}+F_{\rm{sublayer}}-c_2)) + 0.5.
		\label{equ:s_prot}
	\end{eqnarray}
	Here, $R_{\rm{T}}=\mu_{\rm{T}}/(\mu_{\rm{T}}+\mu_{\rm{L}})$ is the turbulent viscosity ratio that widely used in ML transition modeling \cite{WuLei-2022-POF, WuLei-2022-TAML, WuLei-2024-POF}. $F_{\rm{sublayer}}=\exp(-R_{\omega}^2)$ is a variant of turbulent viscous sublayer damping function employed in transition model of \citet{Langtry-2009-AIAA} in which $R_{\omega}=\rho d^2 \omega / 500\mu$. The parameter $c_2$ is threshold for the sum $R_{\rm{T}}+F_{\rm{sublayer}}$, while $c_1$ controls the steepness of hyperbolic tangent function (see Fig.~\ref{fig:s_prot-profile}).
	
\begin{figure}[h]
		\centerline{\includegraphics[scale=0.17]{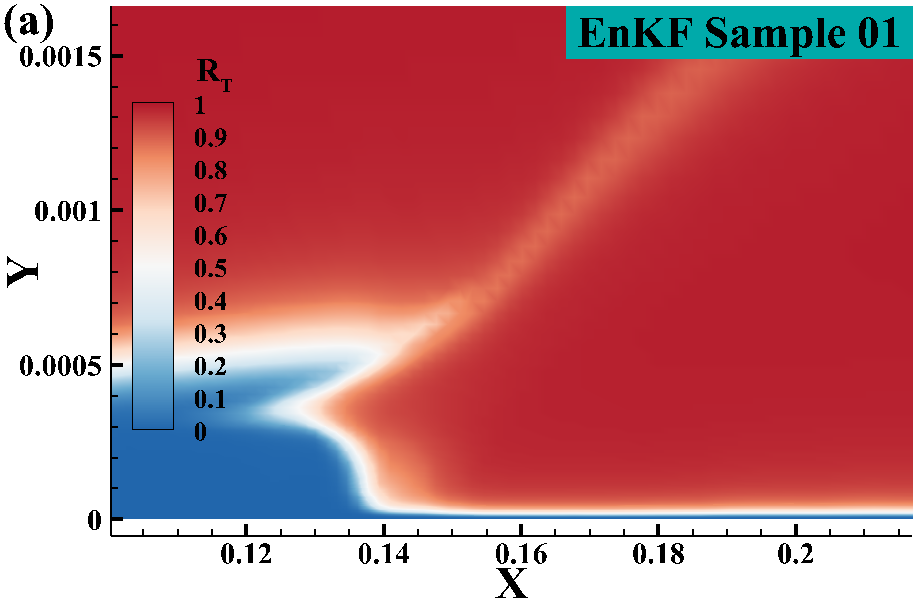}~\includegraphics[scale=0.17]{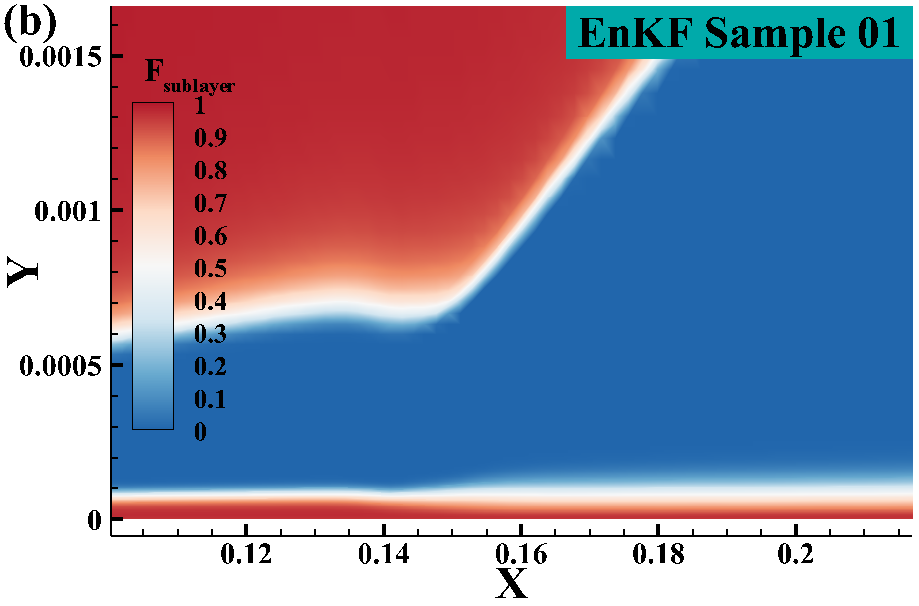}~\includegraphics[scale=0.17]{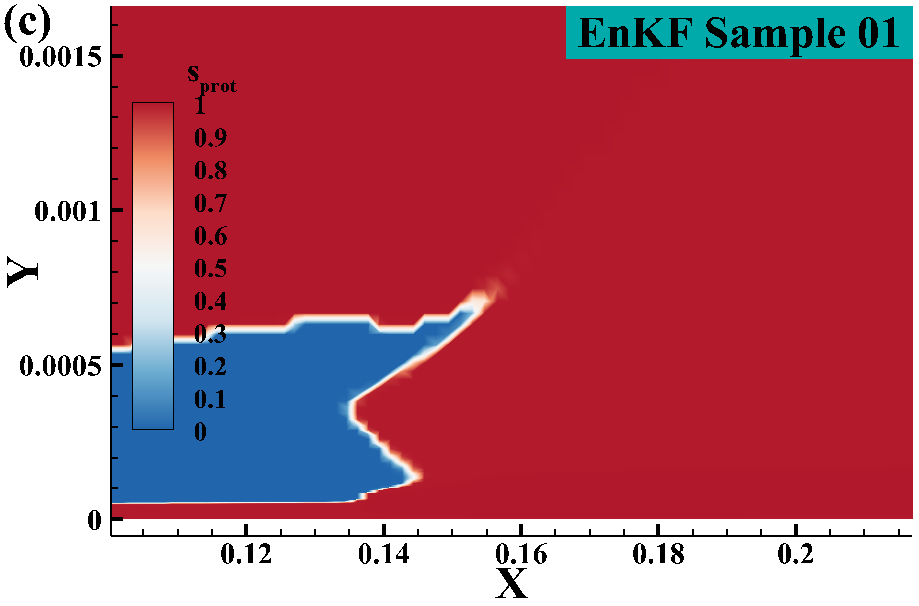}}
		\caption{Contours of (a) $R_{\rm{T}}$, (b) $F_{\rm{sublayer}}$, and (c) $s_{\rm{prot}}$ of an adiabatic flat plate at $Ma_{\infty}=2.25, Re_{\infty}=2.50\times10^7$ near the transition region from first ensemble member of EnKF at tenth epoch.}
		\label{fig:s_prot-contour}
	\end{figure}

	As defined in the $s_{\rm{prot}}$ expression Eq.~\ref{equ:s_prot}, the correction to $\beta(\mathbf{x})$ is triggered whenever $R_{\rm{T}}>c_2$, $F_{\rm{sublayer}}>c_2$, or their sum exceeds $c_2$. It is shown in Fig.~\ref{fig:s_prot-contour} (c) that $s_{\rm{prot}}$ activates precisely after the onset of transition, thereby preventing premature correction in the pre-transitional zone. Furthermore, $s_{\rm{prot}}$ does not interfere with the flow field downstream of transition onset, including the viscous sublayer near the wall. In fact, as can be seen in Fig.~\ref{fig:s_prot-contour} (a), if $s_{\rm{prot}}$ is a function of $R_{\rm{T}}$ alone, it will incorrectly suppress the correction in turbulent viscous sublayer. Despite the original $\beta(\mathbf{x})$ field imposes no correction there, such interference is still undesirable as it destroys the structural independence of previous correction. In the present model, parameters $c_1$ and $c_2$ are calibrated to be $100$ and $0.85$, based on the adiabatic flat plate case.

	\section{Results}\label{sec:results}
	To validate the ability of newly developed SR-augmented transition model, three configurations including flat plate, sharp straight cones at various Reynolds numbers, and blunt straight cones with several nose bluntness are employed. All simulations are performed using an in-house structured, finite-volume solver. Within this framework, the inviscid fluxes are computed based on Roe's flux-difference splitting scheme, while the viscous terms are discretized by a standard second-order central-difference method. Temporal integration is achieved via the Lower-Upper Symmetric Gauss-Seidel (LU-SGS) method, which is applied without sub-iterations.

	\subsection{Flat plate}\label{subsec:FP}
	High-speed flow over the flat plate is first considered, with a computational domain $1.5\rm{m} \times 0.2\rm{m}$ discretized by $553 \times 181$ grid points in the streamwise and wall-normal directions, respectively. The simulation is conducted under canonical DNS \cite{Flat-Plate-DNS2.25, Flat-Plate-DNS4.50} and experimental \cite{Mee-Exp} cases with its flow regimes summarized in Table~\ref{tab:Flat-plate-cases}.
	\begin{table*}[width=.9\textwidth,cols=4,pos=h]
		\caption{\label{tab:Flat-plate-cases}Flow regimes over the flat plate.}
		\begin{tabular*}{\tblwidth}{@{} LLLLLLL@{}}
			\toprule
			Case & $Ma_\infty$ & $Re_\infty(\rm{m}^{-1})$ & $T_{\rm{w}}(\rm{K})$ & $T_\infty(\rm{K})$ & $Tu_\infty(\%)$ & Reference  \\
			\midrule
			F1   & $2.25$      & $2.50\times10^7$         & $T_{\rm{aw}}$        & $169.44$           & $0.10$            & DNS, Ref.~\cite{Flat-Plate-DNS2.25} \\
			F2   & $4.50$      & $6.43\times10^6$         & $T_{\rm{aw}}$        & $61.11$            & $0.10$            & DNS, Ref.~\cite{Flat-Plate-DNS4.50} \\
			F3   & $5.5$       & $1.6\times10^6$          & $300$                & $1560$             & $0.32$            & EXP, Ref.~\cite{Mee-Exp} \\
			F4   & $6.1$       & $4.9\times10^6$          & $300$                & $800$              & $0.32$            & EXP, Ref.~\cite{Mee-Exp} \\
			F5   & $6.2$       & $2.6\times10^6$          & $300$                & $690$              & $0.32$            & EXP, Ref.~\cite{Mee-Exp} \\
			F6   & $6.3$       & $1.7\times10^6$          & $300$                & $570$              & $0.32$            & EXP, Ref.~\cite{Mee-Exp} \\
			\bottomrule
		\end{tabular*}
	\end{table*}
	
\begin{figure}[h]
		\centerline{\includegraphics[scale=0.3]{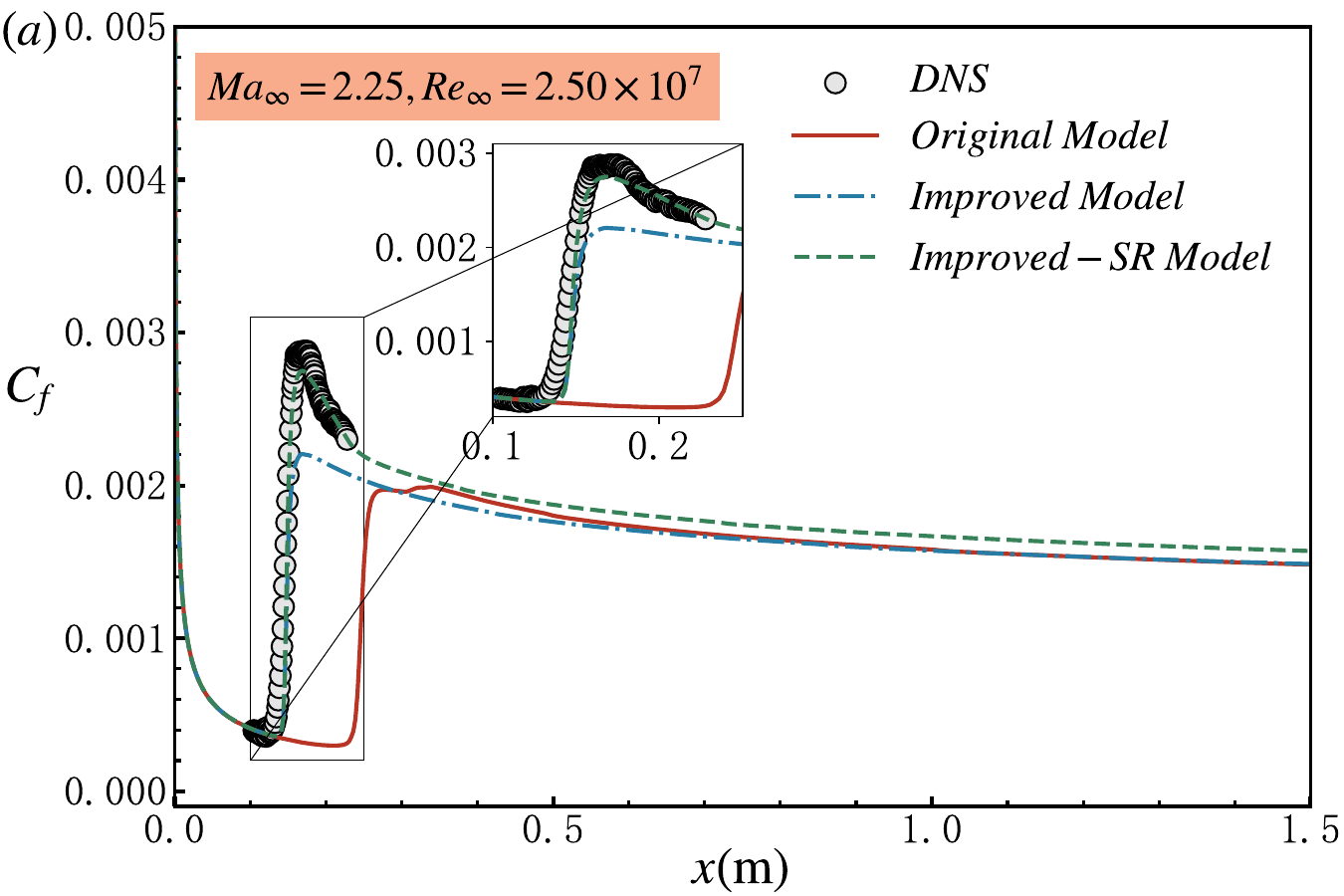}~\includegraphics[scale=0.3]{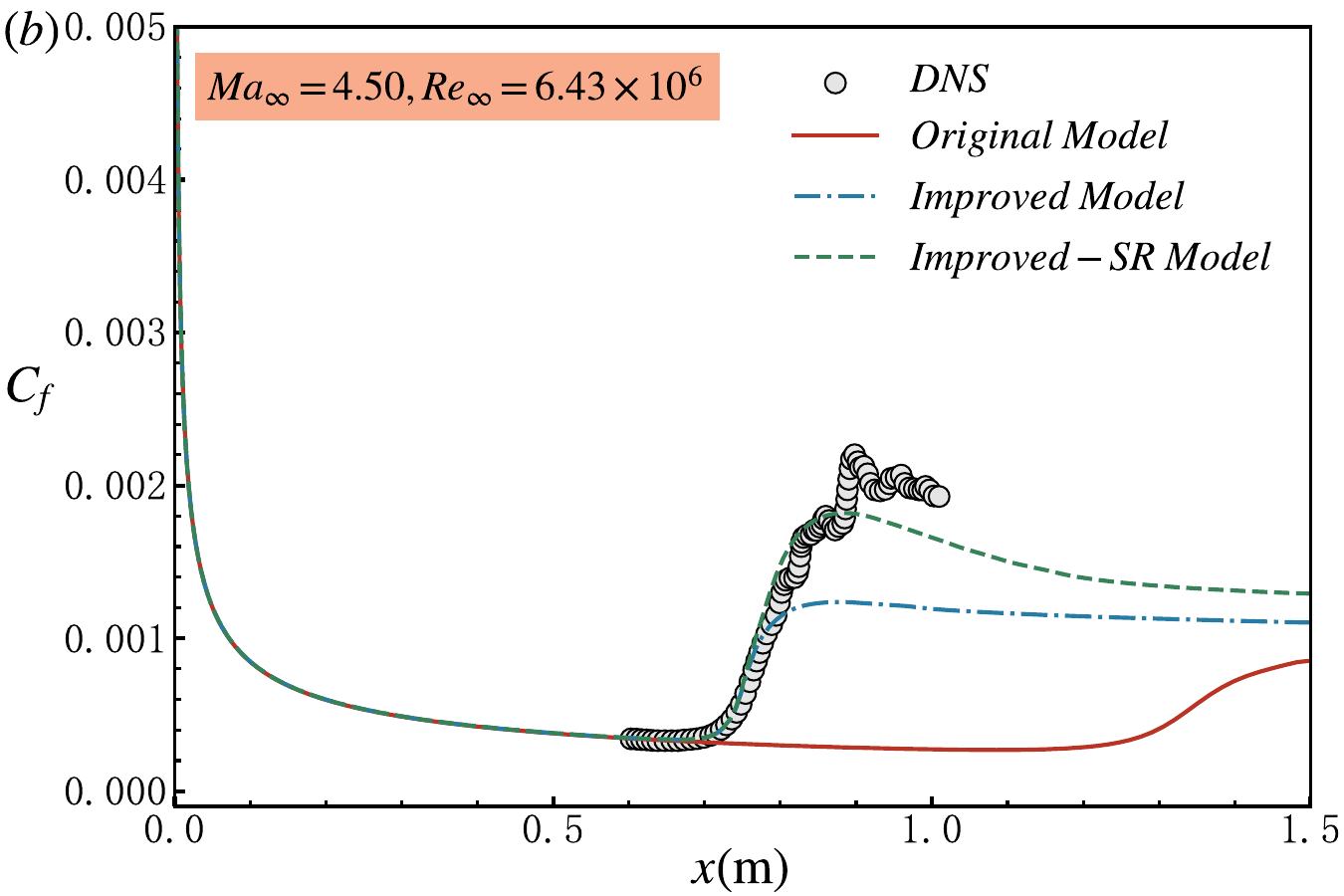}}
		\caption{Distributions of skin-friction coefficient $C_f$ for adiabatic flat plate predicted by different models in comparison with DNS data \cite{Flat-Plate-DNS2.25, Flat-Plate-DNS4.50}: (a) case F1 and (b) case F2 as listed in Table~\ref{tab:Flat-plate-cases}.}
		\label{fig:F12-Cftau-Menter-LeiWu-Exp}
	\end{figure}

	Fig.~\ref{fig:F12-Cftau-Menter-LeiWu-Exp} compares distributions of skin-friction coefficient $C_f$ for cases F1-F2. As can be seen, the original low-speed model significantly fails to predict the transition location and overshoot phenomenon. The high-speed improved model corrects the transition onset and length but still cannot reproduce the skin-friction overshoot. In contrast, the SR-modified improved model accurately predicts the magnitude and spatial extent of overshoot while also maintains the fidelity of baseline model in transition onset and length, achieving remarkable agreement with DNS data. This targeted enhancement, which improves overshoot prediction without interfering with the baseline model's accuracy, is a key requirement for any RANS model augmentation. In addition, although the peak magnitude of the overshoot in Fig.~\ref{fig:F12-Cftau-Menter-LeiWu-Exp} (b) exhibits a noticeable discrepancy with DNS, SR model demonstrates a substantial improvement over the baseline model, confirming the strong generalization capability of Eq.~\eqref{equ:SR-expression2} across different flow regimes.

\begin{figure}[h]
		\centerline{\includegraphics[scale=0.3]{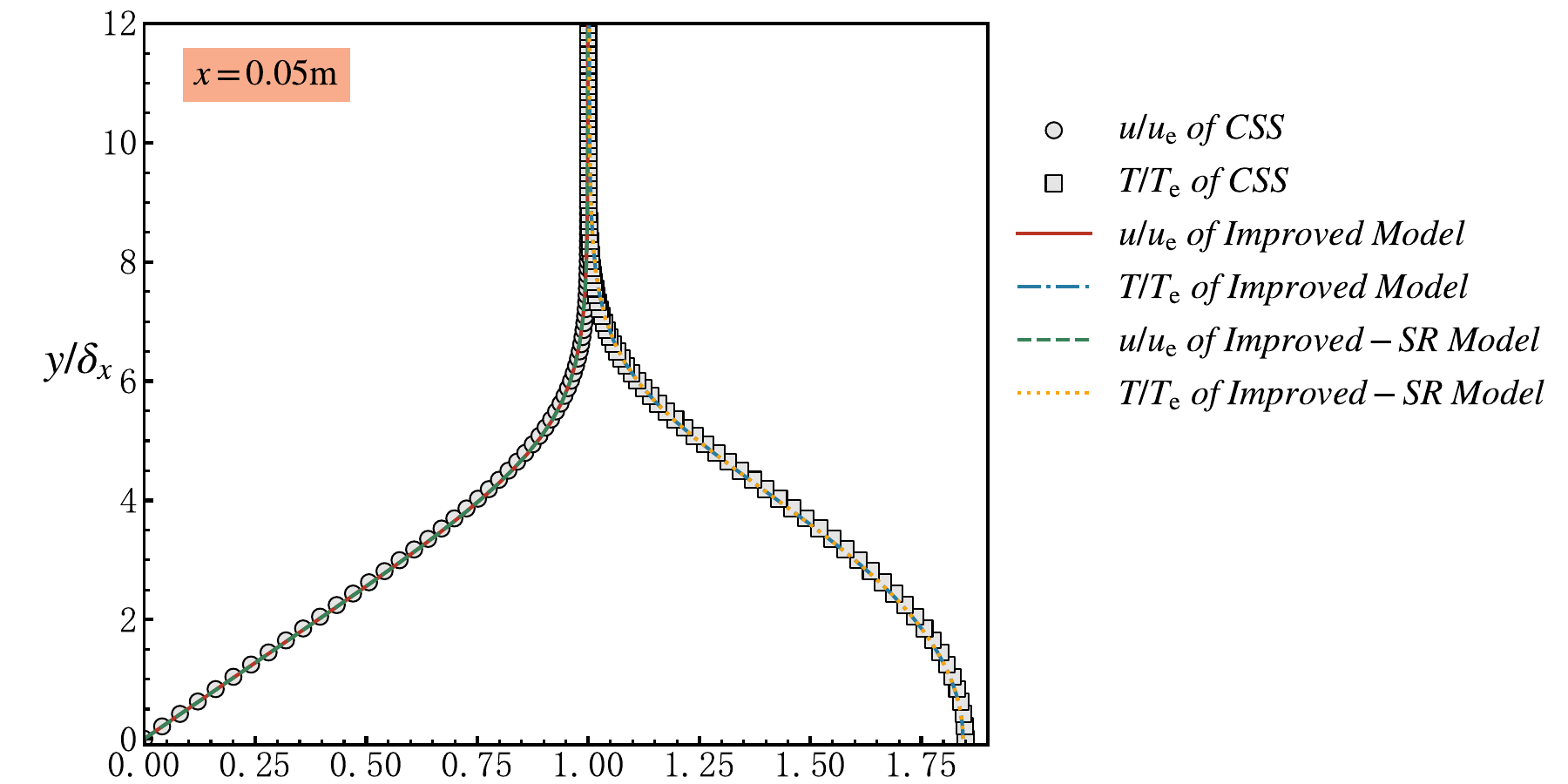}}
		\caption{Streamwise velocity and temperature profiles given by the improved model, SR-enhanced model, and CSS at $x=0.05\rm{m}$ (laminar state) for case F1 (see Table~\ref{tab:Flat-plate-cases}). The vertical coordinate represents the dimensionless distance to the wall $y/\delta_x=y/\sqrt{{\mu_{\rm{e}}x}/\rho_{\rm{e}}u_{\rm{e}}}$.}
		\label{fig:F1-uTrhou-profile-SST-G-CSS}
	\end{figure}

	To further assess the impact of $\beta(\mathbf{x})$ correction, streamwise velocity and temperature profiles in the laminar region of case F1 are compared between the improved model and SR-augmented model. The compressible self-similar solution (CSS) serves as a benchmark for laminar flow \cite{WuLei-2026-IJHMT}. As shown in Fig.~\ref{fig:F1-uTrhou-profile-SST-G-CSS}, both models yield identical profiles that coincide with CSS results, confirming that the symbolic regression augmentation introduces no spurious modification to pre-transitional laminar field.

\begin{figure}[h]
		\centerline{\includegraphics[scale=0.3]{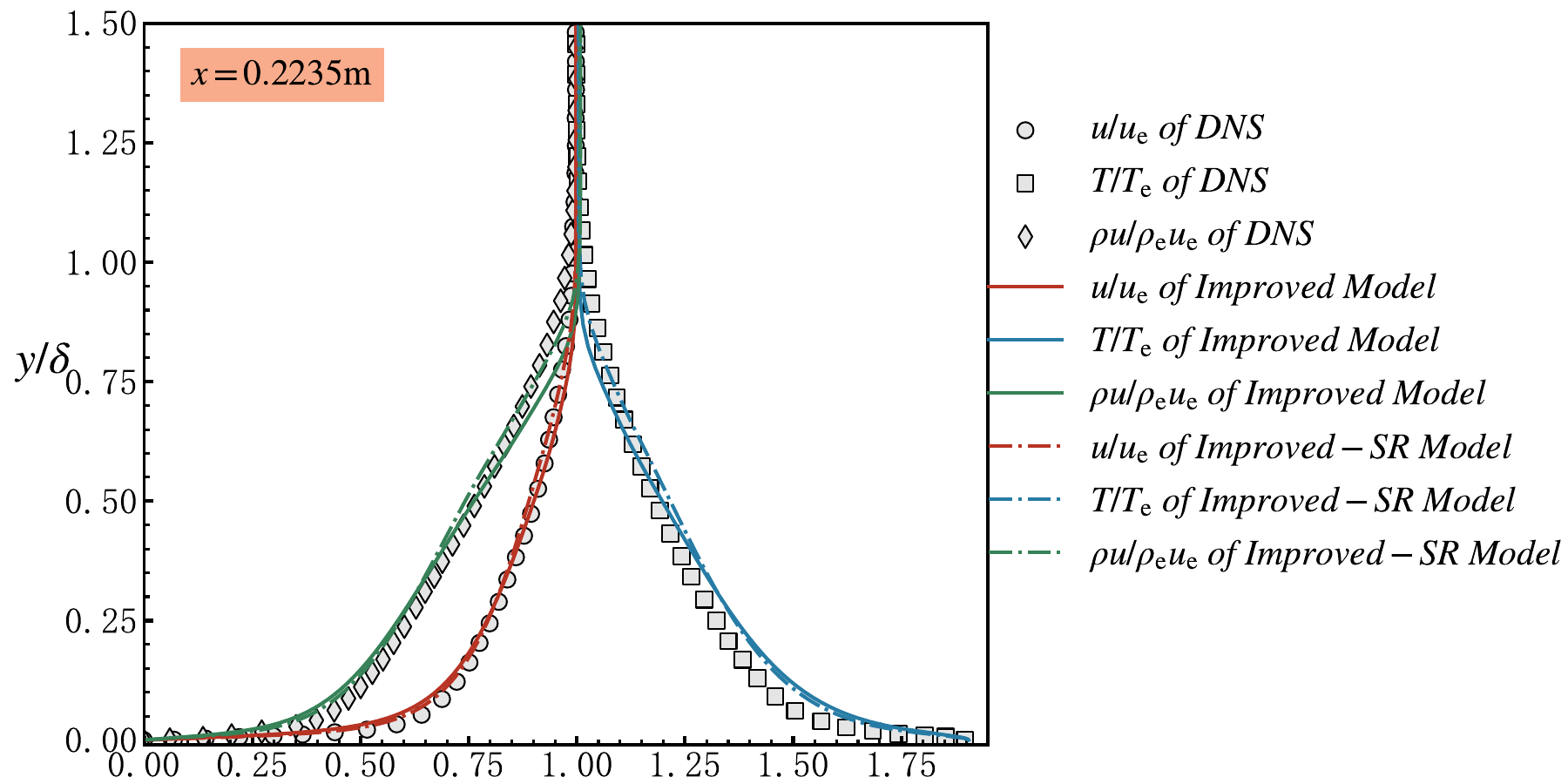}}
		\caption{Streamwise velocity and temperature profiles given by the improved model, SR-enhanced model, and CSS at $x=0.2235\rm{m}$ for case F1 (see Table~\ref{tab:Flat-plate-cases}). The vertical coordinate $y/\delta$ represents the dimensionless distance to the wall based on the boundary layer thickness $\delta$.}
		\label{fig:F1-uTrhou-profile-SST-G-DNS}
	\end{figure}

	Considering the availability of DNS data at $x=0.2235\rm{m}$, profiles of velocity, temperature, density-velocity are compared in Fig.~\ref{fig:F1-uTrhou-profile-SST-G-DNS}. As can be seen from Fig.~\ref{fig:F12-Cftau-Menter-LeiWu-Exp}, $x=0.2235\rm{m}$ corresponds to the late stage of transition process where still exhibits a discernible skin-friction overshoot. It is shown in Fig.~\ref{fig:F1-uTrhou-profile-SST-G-DNS} that the SR-modified model yields a near-wall velocity gradient that is slightly steeper than that of baseline model, bringing it into closer agreement with DNS result that captures the skin-friction overshoot. Similarly, both temperature and density-velocity profiles from SR model show improved fidelity to DNS compared to the unmodified model, which indicates that the symbolic regression correction improves not only the wall-shear stress but also the mean-flow structure of the boundary layer in the transitional zone.

\begin{figure}[h]
		\centerline{\includegraphics[scale=0.3]{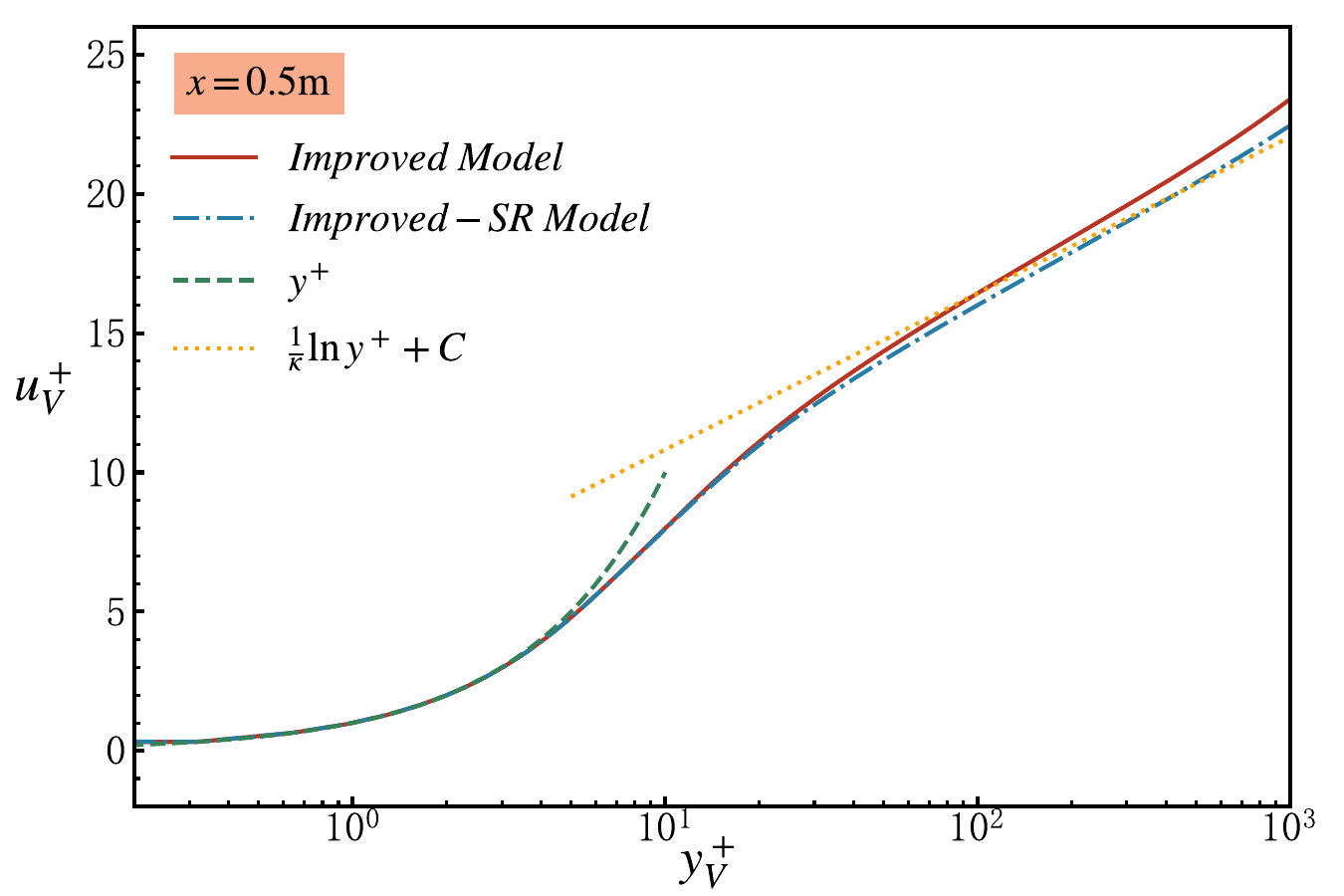}}
		\caption{Streamwise velocity profiles in wall units given by the improved model, SR-enhanced model at $x=0.5\rm{m}$ (turbulent state) for case F1 (see Table~\ref{tab:Flat-plate-cases}). Linear and logarithmic laws with $\kappa=0.41, C=5.2$ of the wall are plotted for comparison.}
		\label{fig:F1-yp-up}
	\end{figure}

	To assess the structure of turbulent compressible boundary layer in wall units, a so-called Volpiani transformation that designed to extend the concept of incompressible scaling to compressible flows is employed \cite{Volpiani-2020-PRF}. In this transformation, the wall-normal coordinate $y_{\rm{V}}^+$ and velocity $u_{\rm{V}}^+$ are defined as
	\begin{eqnarray}
		y_{\rm{V}}^+=\frac{y_{\rm{I}}}{\nu}, u_{\rm{V}}^+=\frac{u_{\rm{I}}}{u_{\tau}}.
		\label{equ:y_V-u_V}
	\end{eqnarray}
	Here, $y_{\rm{I}}=\int_{0}^{y}f_{\rm{I}}\mathrm{d}y$ and $u_{\rm{I}}=\int_{0}^{u}g_{\rm{I}}\mathrm{d}u$ are the integrated ``incompressible-like" coordinates with
	\begin{eqnarray}
		f_{\rm{I}}=\sqrt{\frac{R}{M^3}}, g_{\rm{I}}=\sqrt{\frac{R}{M}},
		\label{equ:fI-gI}
	\end{eqnarray}
	in which $R=\rho/\rho_{\rm{w}}, M=\mu/\mu_{\rm{w}}$ are the local-to-wall ratios of density and dynamic viscosity, respectively. Under this transformation, velocity profiles of compressible turbulent boundary layers under different Mach numbers and wall conditions will effectively collapse onto the incompressible profile, thereby facilitating direct comparison with the linear and logarithmic laws of the wall. Fig.~\ref{fig:F1-yp-up} presents the velocity profile of $x=0.5\rm{m}$ in wall units. In viscous sublayer ($y_{\rm{V}}^+ < 5$), both the improved model and SR-augmented model exhibit agreement with the linear law. For logarithmic region ($y_{\rm{V}}^+ > 30$), however, the baseline improved model begins to deviate from the theoretical log-law when shifting toward the outer edge of the log-layer. In contrast, the SR model remains closely aligned with the log-law across the entire overlap region, demonstrating enhanced accuracy in capturing the canonical structure of the turbulent boundary layer.

\begin{figure}[h]
		\centerline{\includegraphics[scale=0.2]{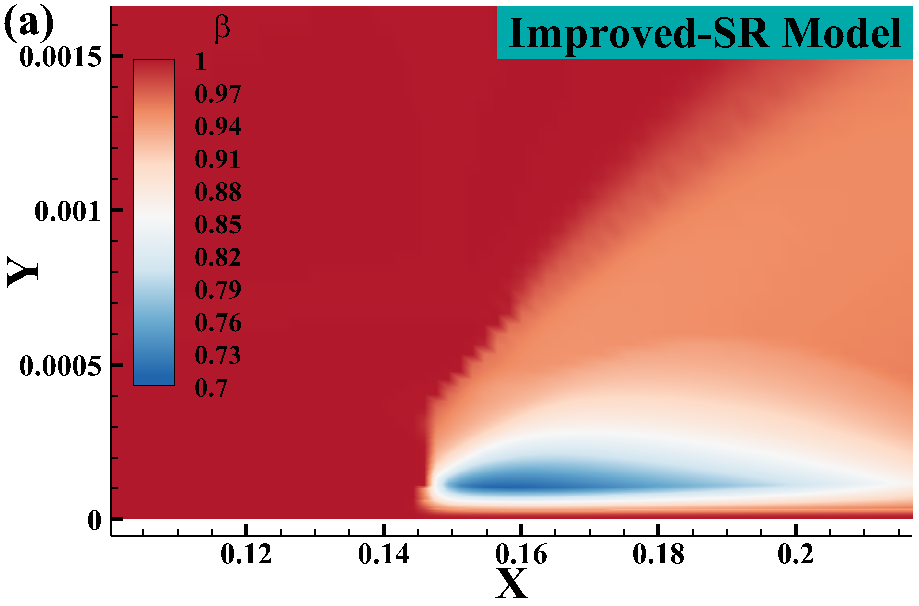}~\includegraphics[scale=0.2]{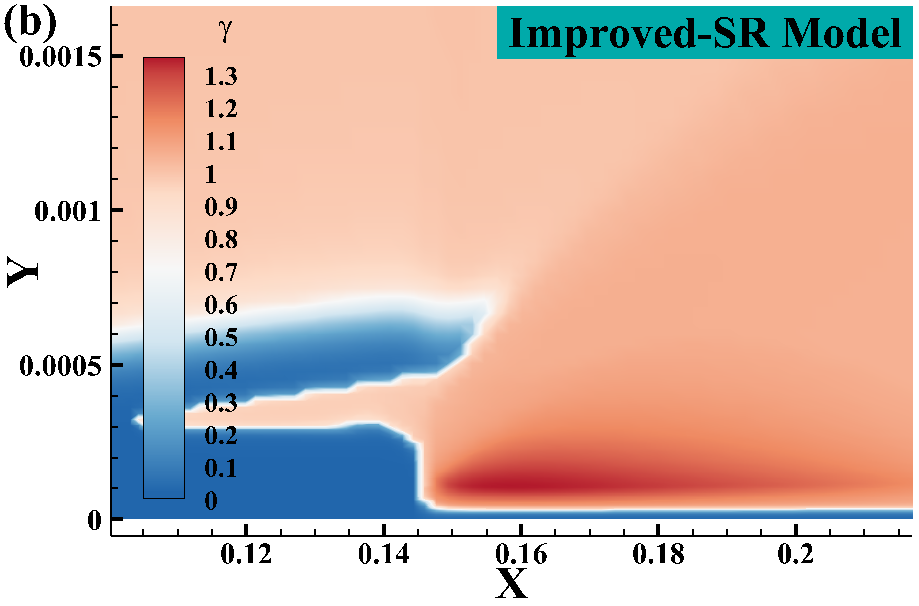}}
		\caption{Zoom-in view of (a) $\beta(\mathbf{x})$ and (b) intermittency factor $\gamma$ contours of SR-corrected model for case F1 in Table~\ref{tab:Flat-plate-cases}.}
		\label{fig:F1-SR-beta}
	\end{figure}

\begin{figure}[h]
		\centerline{\includegraphics[scale=0.2]{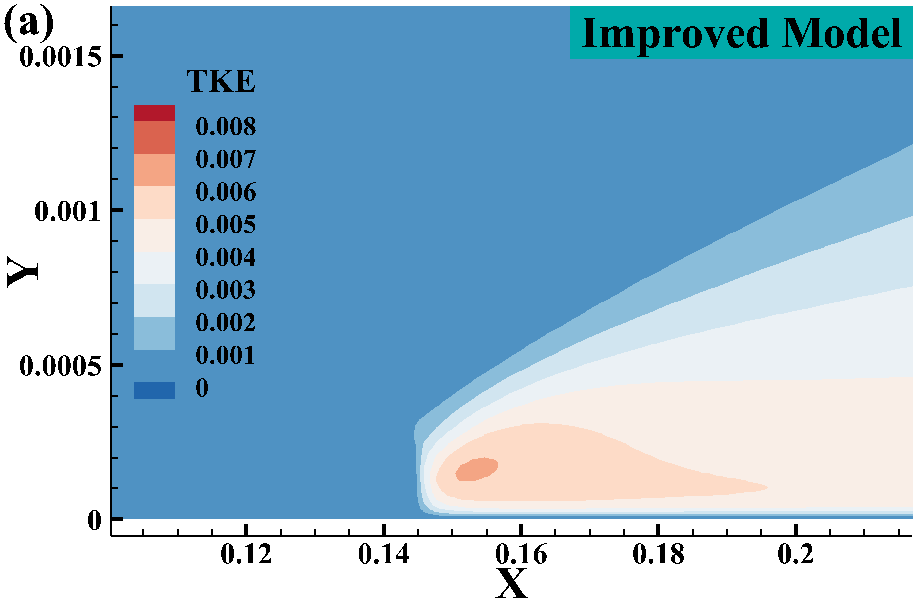}~\includegraphics[scale=0.2]{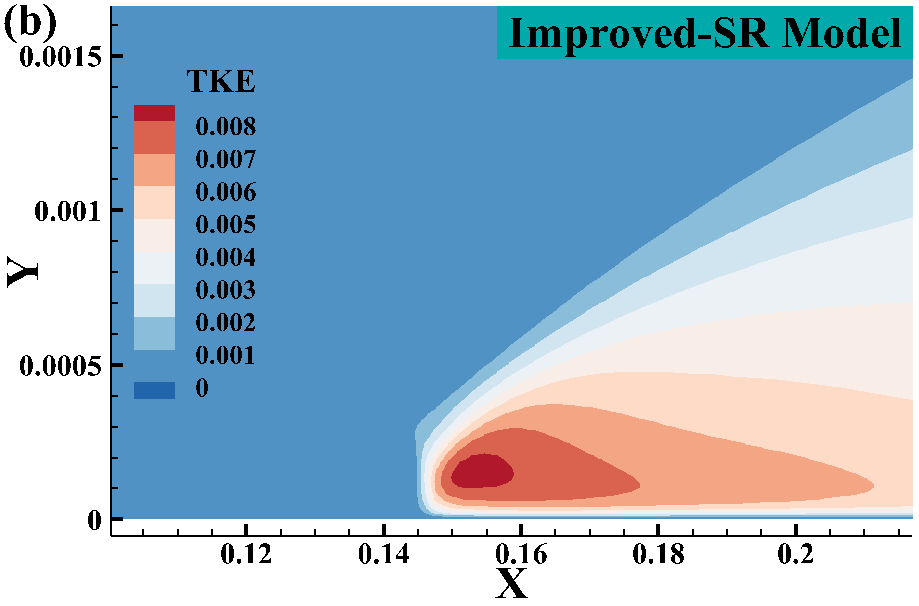}}
		\caption{Zoom-in view of turbulent kinetic energy contours from (a) high-speed baseline model and (b) SR-corrected model for case F1 (see Table~\ref{tab:Flat-plate-cases}).}
		\label{fig:F1-Improved-SR-TKE}
	\end{figure}

\begin{figure}[h]
		\centerline{\includegraphics[scale=0.3]{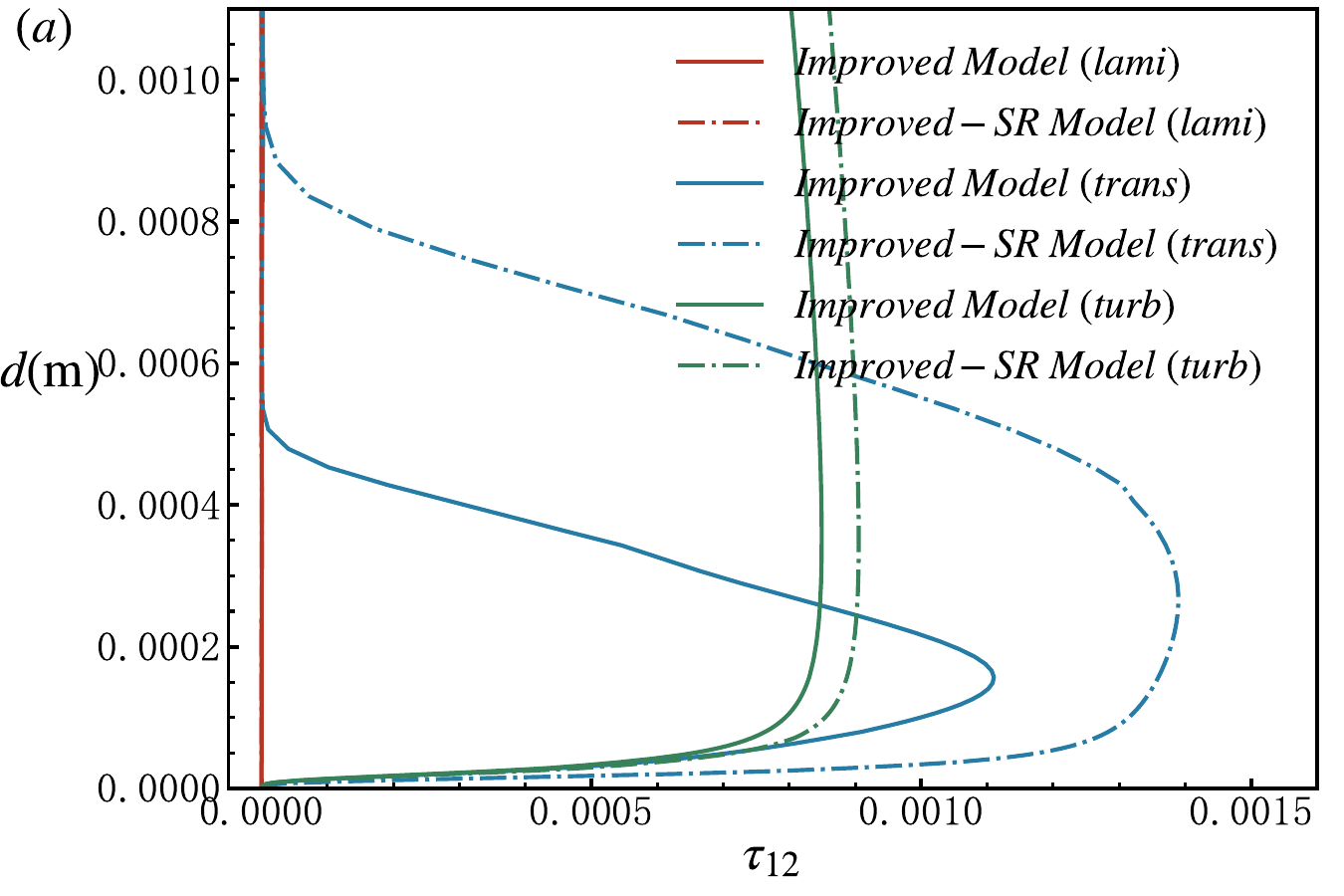}~\includegraphics[scale=0.3]{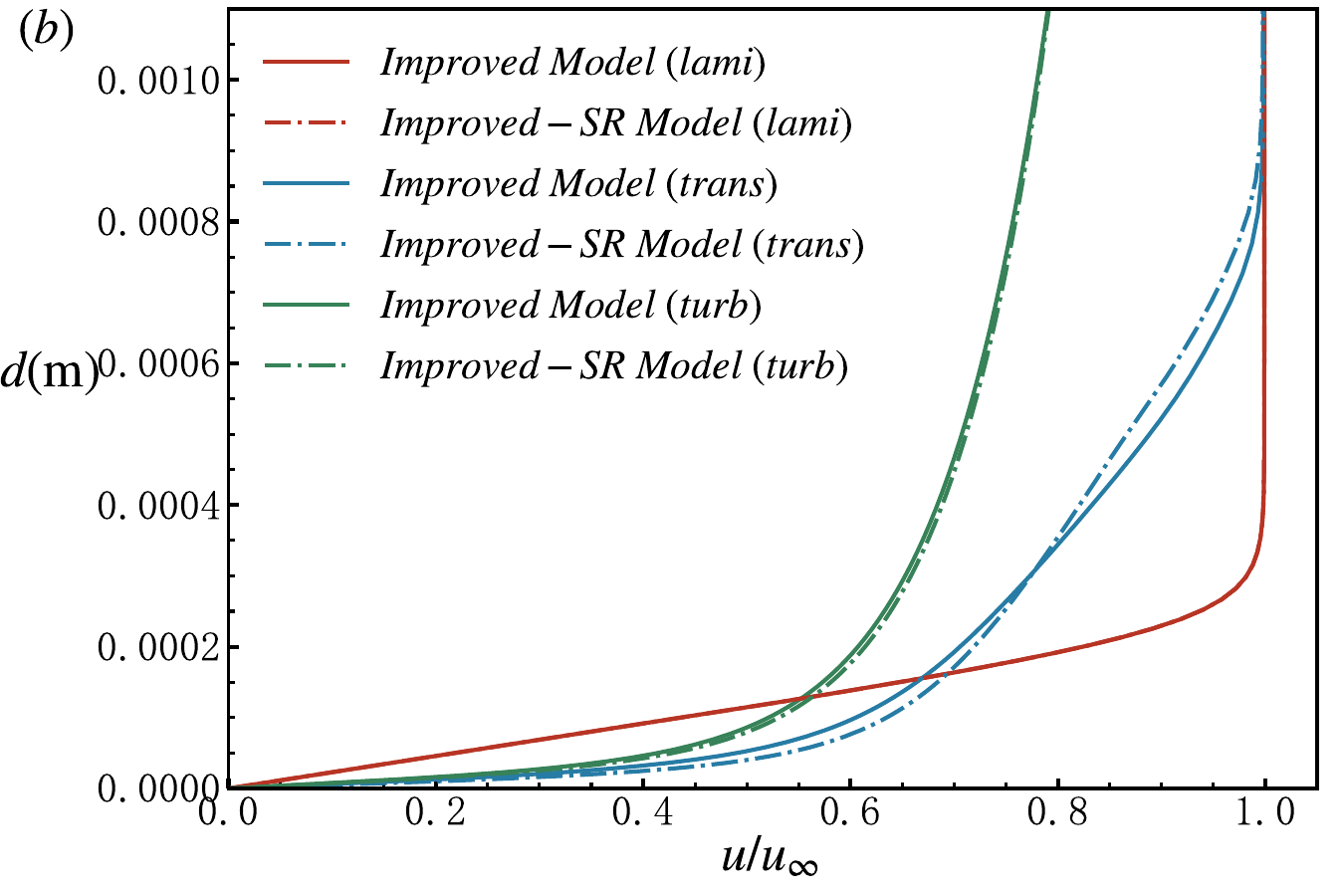}}
		\centerline{\includegraphics[scale=0.3]{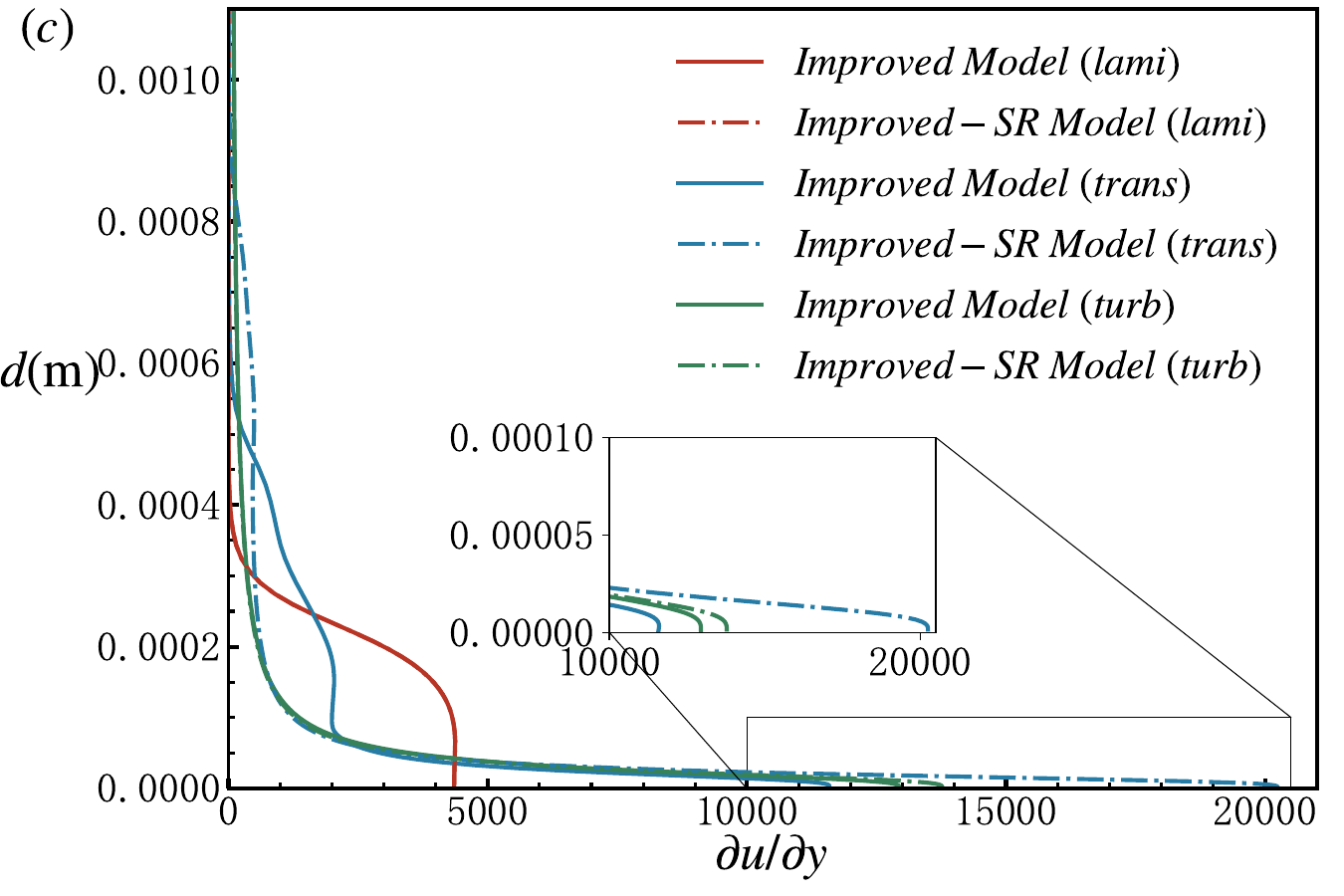}}
		\caption{Profiles of several crucial flow variables from baseline and SR-corrected models at representative laminar ($x=0.05\rm{m}$), transitional ($x=0.15\rm{m}$), and turbulent ($x=0.5\rm{m}$) locations for case F1 (see Table~\ref{tab:Flat-plate-cases}).}
		\label{fig:F1-t12-u-uy}
	\end{figure}

	Physical mechanism underlying the overshoot phenomenon is further elucidated. As shown in Fig.~\ref{fig:F1-SR-beta}, $\beta(\mathbf{x})$ from the SR-enhanced model decreases significantly in the transitional region, which contributes to a higher intermittency factor $\gamma$ than that predicted by the baseline model (see Fig.~\ref{fig:F1-Yt-contour} (a)). From the TKE equation in Eq.~\ref{equ:k-equ}, this elevated $\gamma$ directly enhances the production term of TKE as can be compared in Fig.~\ref{fig:F1-Improved-SR-TKE}, thereby intensifying the Reynolds shear stress $\tau_{12}$ in Fig.~\ref{fig:F1-t12-u-uy} (a). This amplified shear stress leads to a more pronounced distortion of the mean velocity profile, manifested as a steeper growth of streamwise velocity near the wall in Fig.~\ref{fig:F1-t12-u-uy} (b). Consequently, the SR-augmented model exhibits a larger velocity gradient (see Fig.~\ref{fig:F1-t12-u-uy} (c)) in the transitional region compared to the baseline model, thus resulting in a significant skin-friction overshoot in Fig.~\ref{fig:F12-Cftau-Menter-LeiWu-Exp} (a).

\begin{figure}[h]
		\centerline{\includegraphics[scale=0.3]{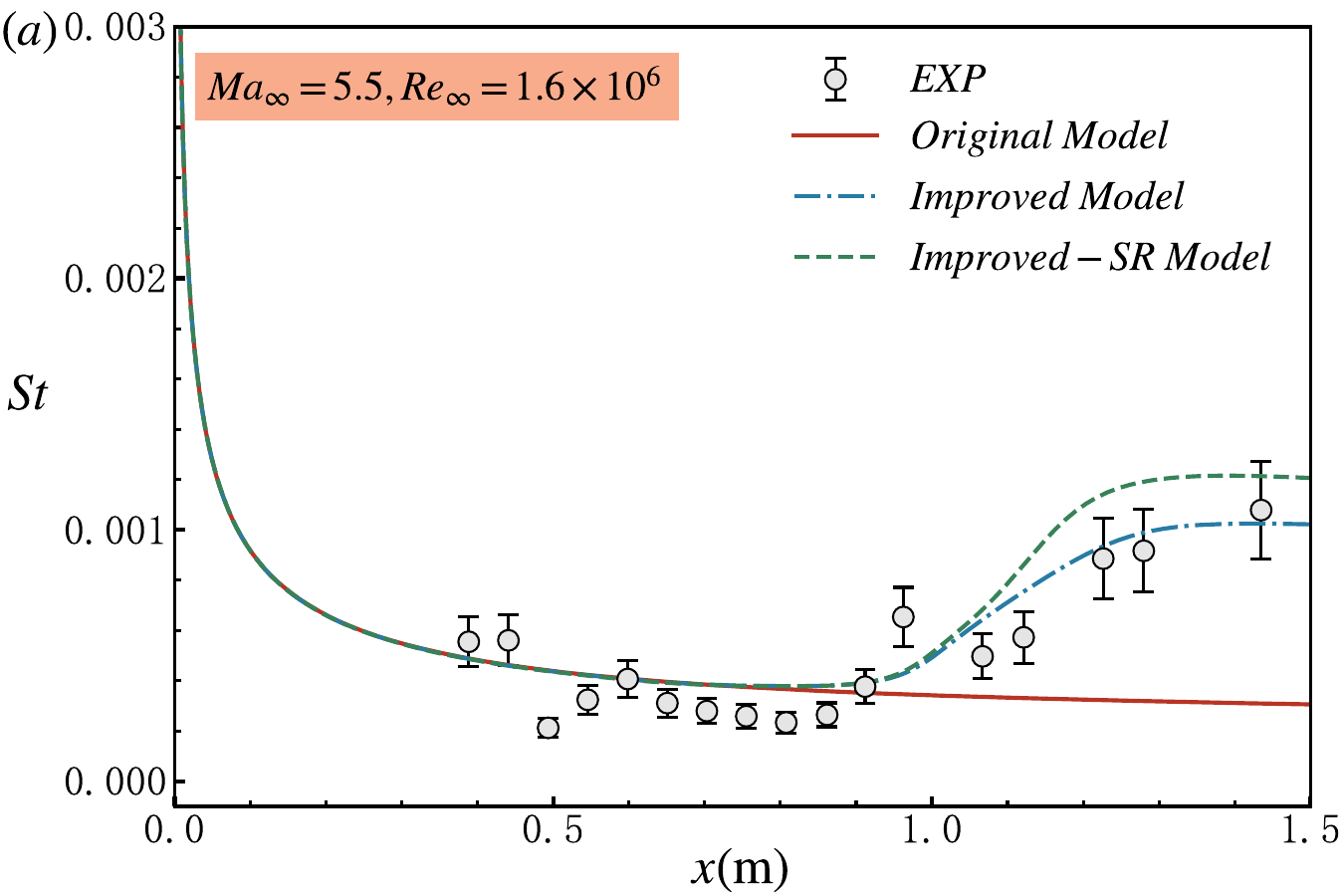}~\includegraphics[scale=0.3]{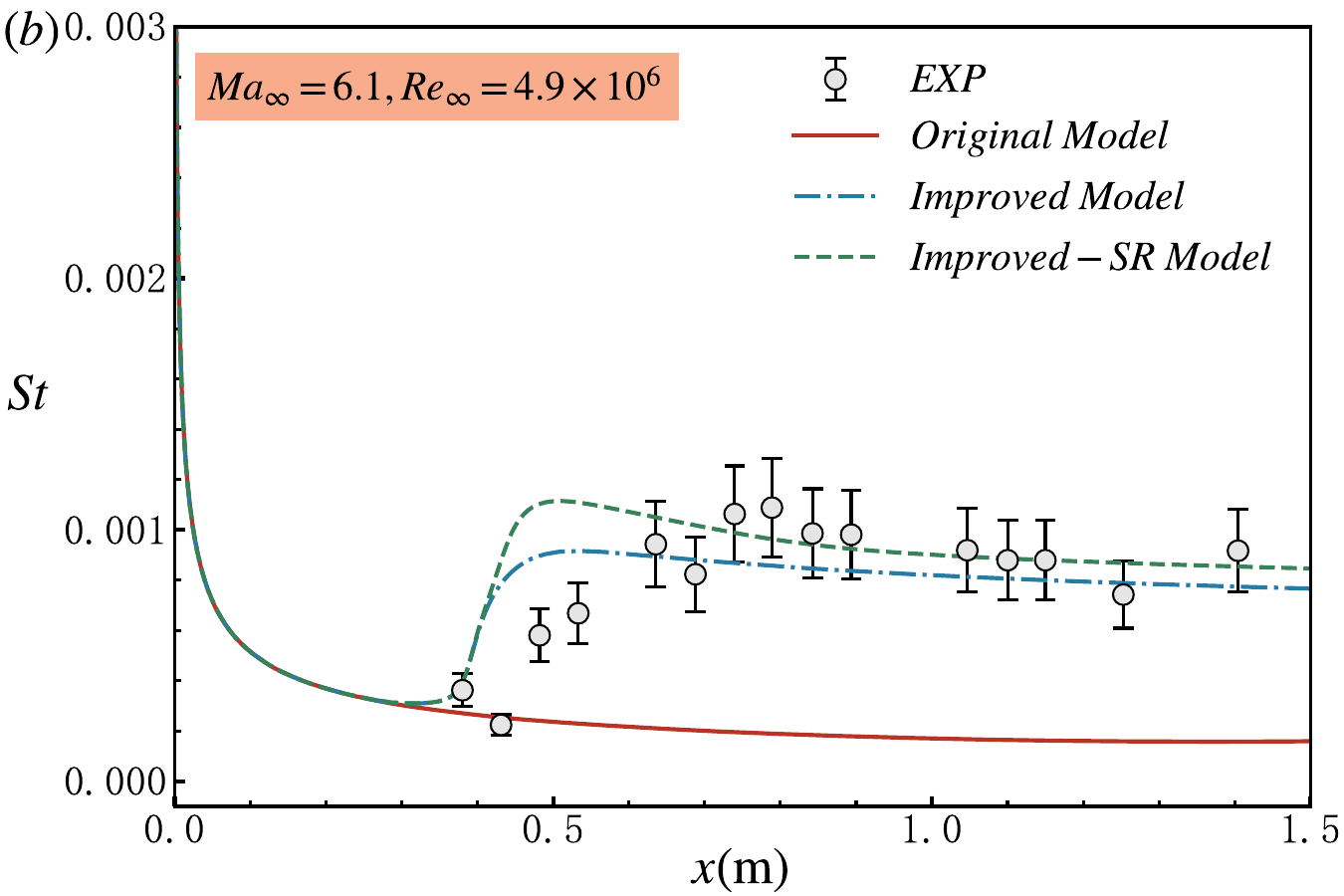}}
		\centerline{\includegraphics[scale=0.3]{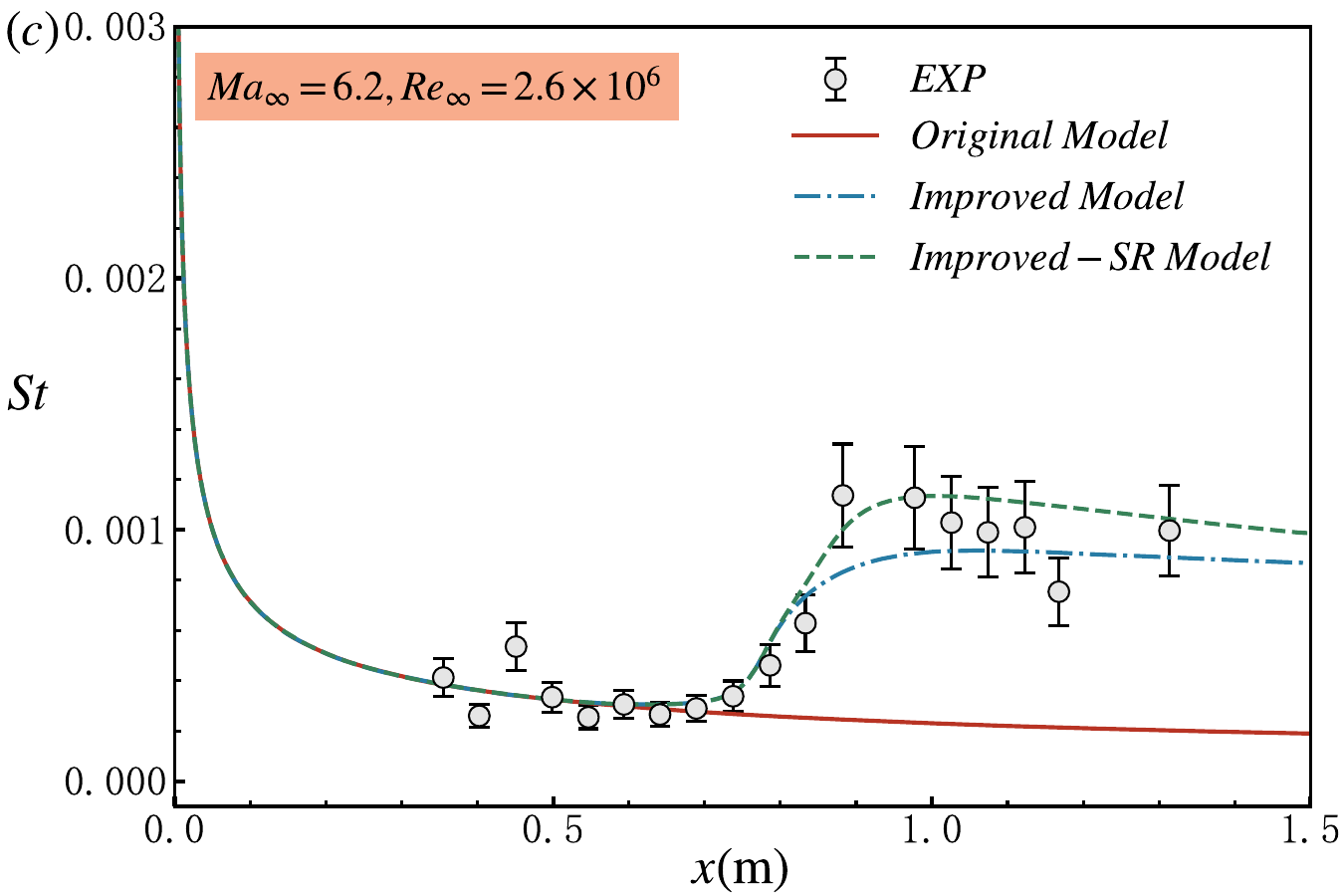}~\includegraphics[scale=0.3]{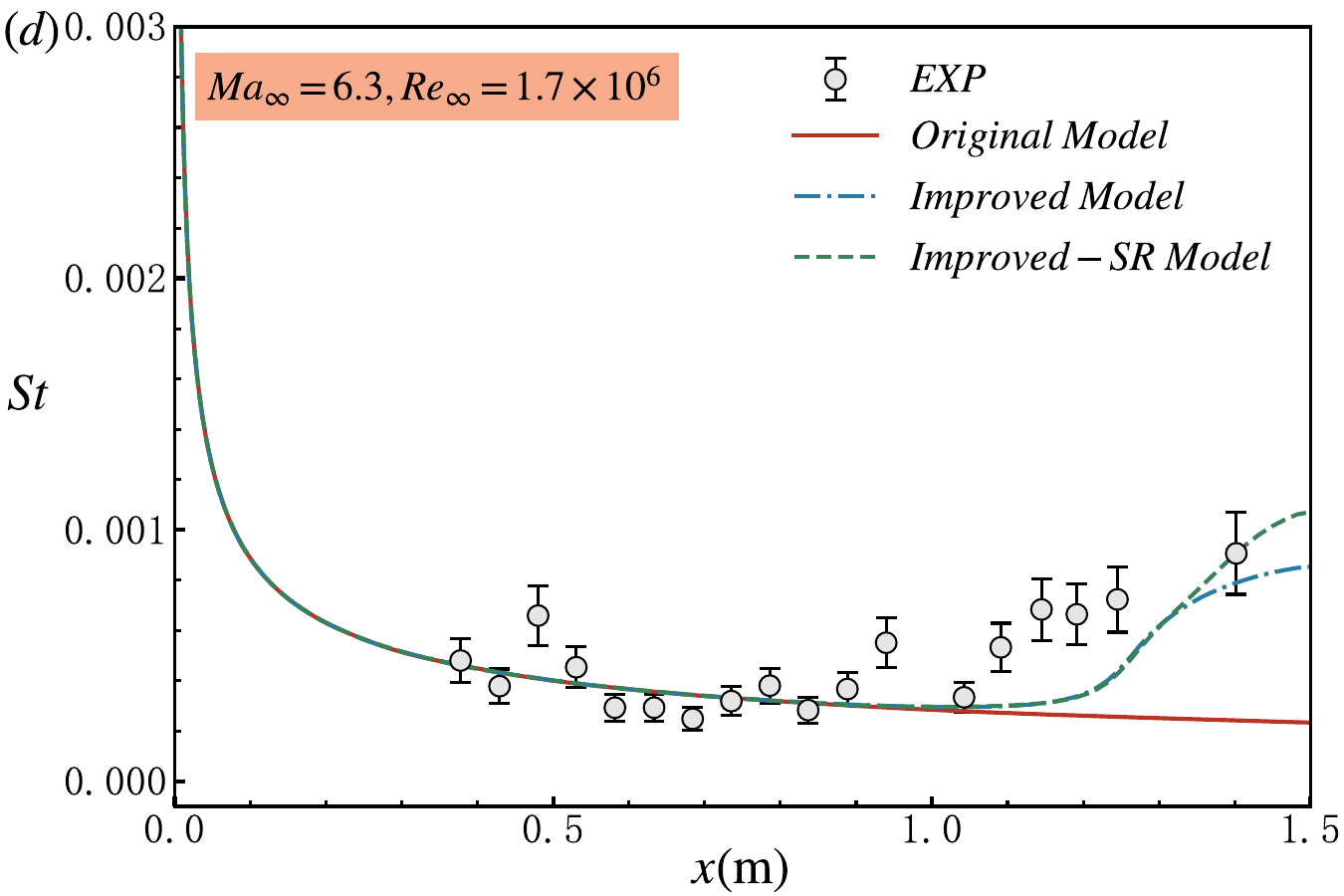}}
		\caption{Distributions of Stanton number $St$ for isothermal flat plate predicted by different models: (a) case F3, (b) case F4, (c) case F5, and (d) case F6 as listed in Table~\ref{tab:Flat-plate-cases}. The experimental results with $\pm 18\%$ uncertainty from \cite{Mee-Exp} are also depicted for comparison.}
		\label{fig:F3456-St-Menter-LeiWu-Exp}
	\end{figure}

	For the isothermal flat plate cases in Fig.~\ref{fig:F3456-St-Menter-LeiWu-Exp}, experimental results do not exhibit a pronounced heat-transfer overshoot. Consistent with this observation, the SR-augmented model also does not generate overshoot phenomenon. This behavior demonstrates that the SR-derived expression adjusts the model response based on local flow features rather than imposing a universal correction that forces an overshoot regardless of flow conditions. In addition, compared to the baseline model, a slight elevation in Stanton number $St$ is predicted within the fully turbulent region. Despite the improvement over baseline model is modest in Fig.~\ref{fig:F3456-St-Menter-LeiWu-Exp} (a), the SR-modified model demonstrates better agreement with experimental data for cases F4-F6.

	\subsection{Sharp straight cones}\label{subsec:SC}
	Flow over sharp straight cones at a $5$-degree half-angle and zero angle of attack is assessed in this section. Validation is performed against two experimental datasets \cite{Horvath, Chen-SharpCone-Exp}. The first dataset (cases S1-S4) corresponds to an isothermal cone with a length of $0.635\rm{m}$ and a nose bluntness of $0.00254\rm{mm}$ \cite{Horvath}. The second (cases S5-S8) involves an adiabatic cone measuring $0.381\rm{m}$ in length with a nose bluntness of $0.025\rm{mm}$ \cite{Chen-SharpCone-Exp}. Computational grids of $361\times181\times21$ and $481\times211\times21$ (streamwise $\times$ wall-normal $\times$ circumferential) are adopted for the simulations, respectively.
	\begin{table*}[width=.9\textwidth,cols=4,pos=h]
		\caption{\label{tab:Sharp-Cones-cases}Input parameters for flows past sharp straight cones.}
		\begin{tabular*}{\tblwidth}{@{} LLLLLLLLL@{}}
			\toprule
			Case & $R_{\rm{n}}(\rm{mm})$ & $Ma_\infty$ & $Re_\infty(\rm{m}^{-1})$ & $T_{\rm{w}}(\rm{K})$ & $T_\infty(\rm{K})$ & $Tu_\infty(\%)$ & $p_0(\rm{Mpa})$ & Reference  \\
			\midrule
			S1   & $0.00254$             & $6.0$       & $1.05\times10^7$         & $298.17$             & $61.630$           & $0.40$          & $1.24$          & EXP, Ref.~\cite{Horvath} \\
			S2   & $0.00254$             & $6.0$       & $1.41\times10^7$         & $298.17$             & $61.630$           & $0.40$          & $1.72$          & EXP, Ref.~\cite{Horvath} \\
			S3   & $0.00254$             & $6.0$       & $1.77\times10^7$         & $306.36$             & $63.324$           & $0.40$          & $2.24$          & EXP, Ref.~\cite{Horvath} \\
			S4   & $0.00254$             & $6.0$       & $2.03\times10^7$         & $306.36$             & $63.324$           & $0.40$          & $2.52$          & EXP, Ref.~\cite{Horvath} \\
			S5   & $0.025$               & $3.5$       & $4.90\times10^7$         & $T_{\rm{aw}}$        & $92.281$           & $0.02$          & $0.87$          & EXP, Ref.~\cite{Chen-SharpCone-Exp} \\
			S6   & $0.025$               & $3.5$       & $5.89\times10^7$         & $T_{\rm{aw}}$        & $92.539$           & $0.02$          & $1.05$          & EXP, Ref.~\cite{Chen-SharpCone-Exp} \\
			S7   & $0.025$               & $3.5$       & $6.88\times10^7$         & $T_{\rm{aw}}$        & $92.481$           & $0.02$          & $1.23$          & EXP, Ref.~\cite{Chen-SharpCone-Exp} \\
			S8   & $0.025$               & $3.5$       & $7.80\times10^7$         & $T_{\rm{aw}}$        & $92.330$           & $0.02$          & $1.39$          & EXP, Ref.~\cite{Chen-SharpCone-Exp} \\
			\bottomrule
		\end{tabular*}
	\end{table*}

\begin{figure}[h]
		\centerline{\includegraphics[scale=0.3]{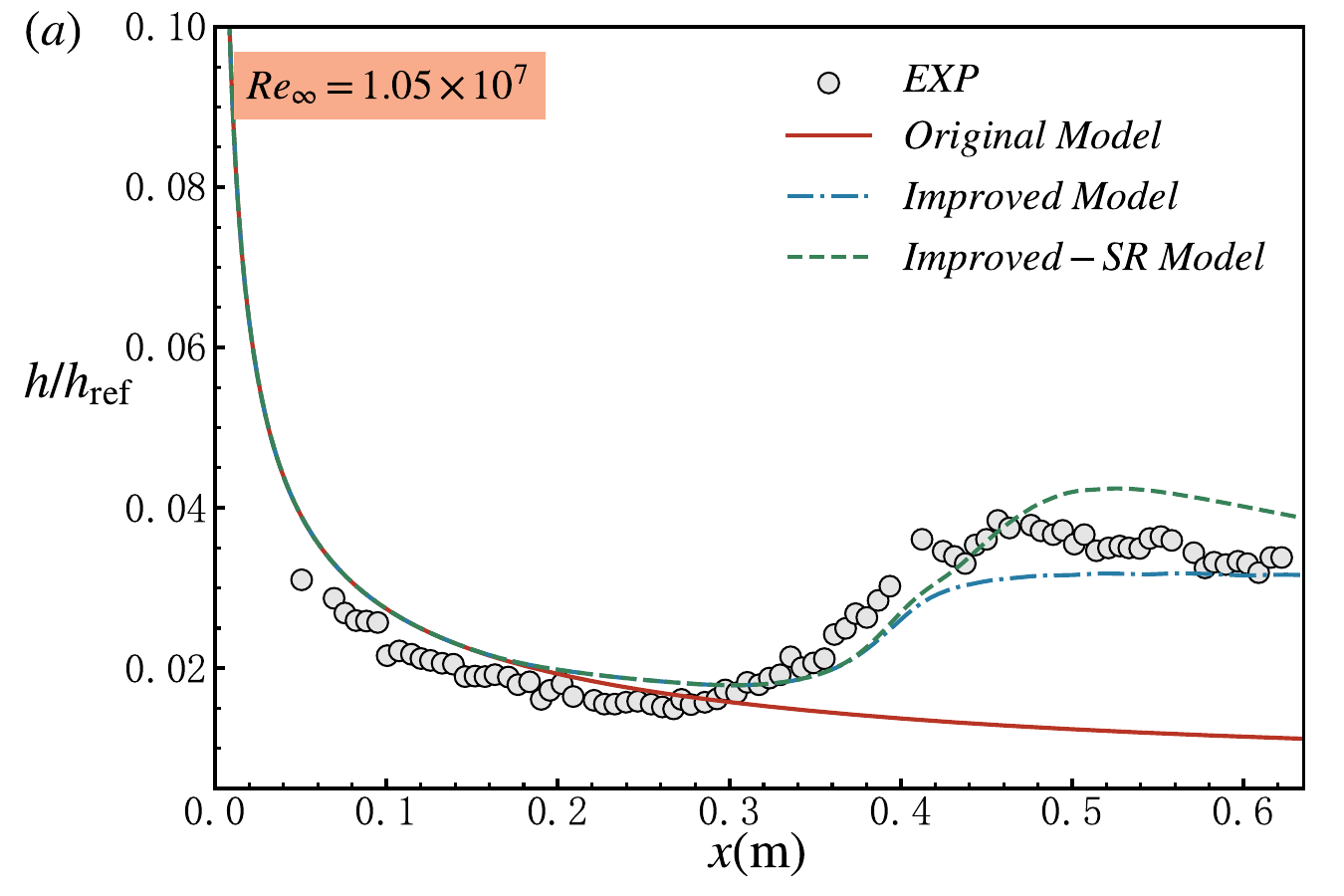}~\includegraphics[scale=0.3]{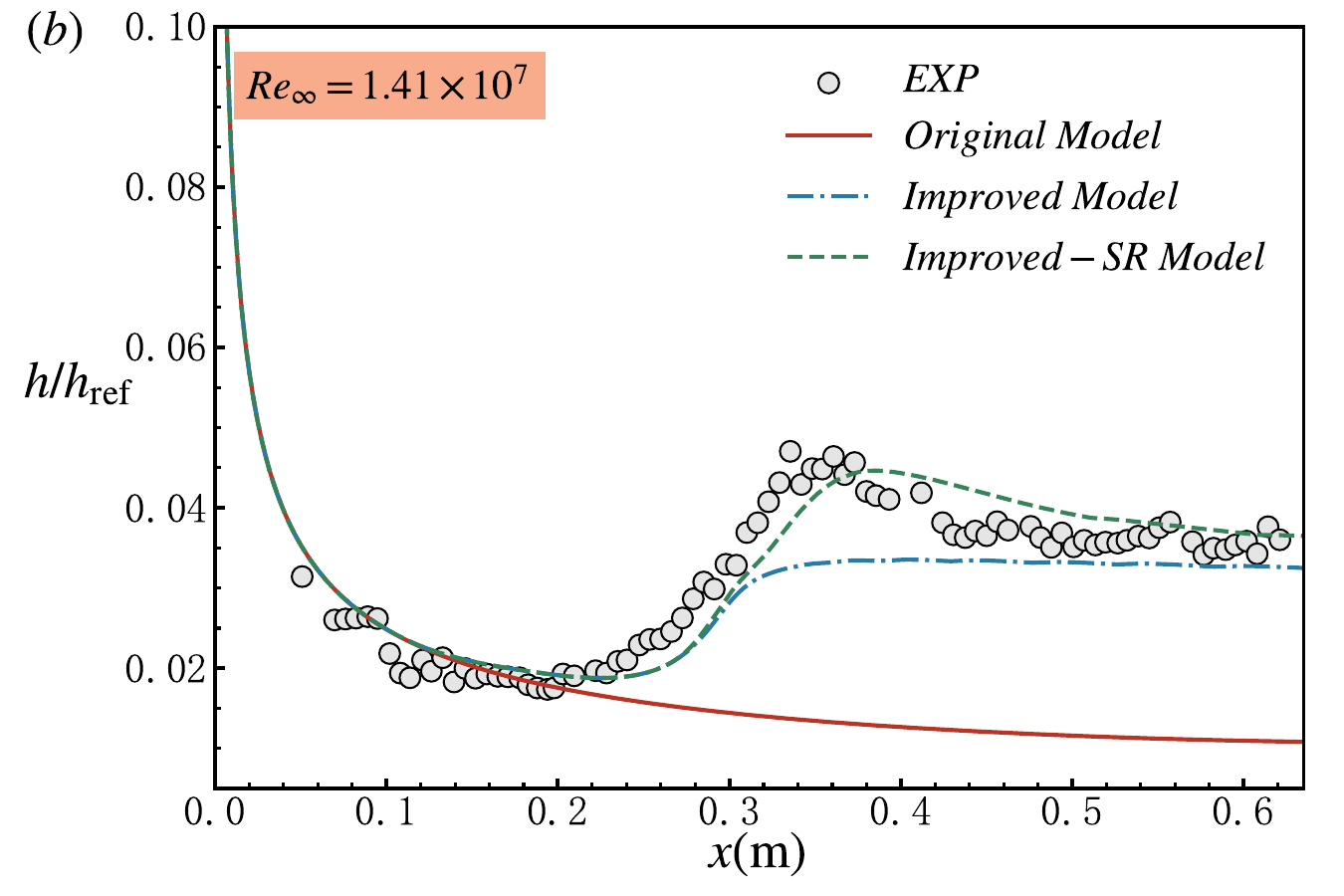}}
		\centerline{\includegraphics[scale=0.3]{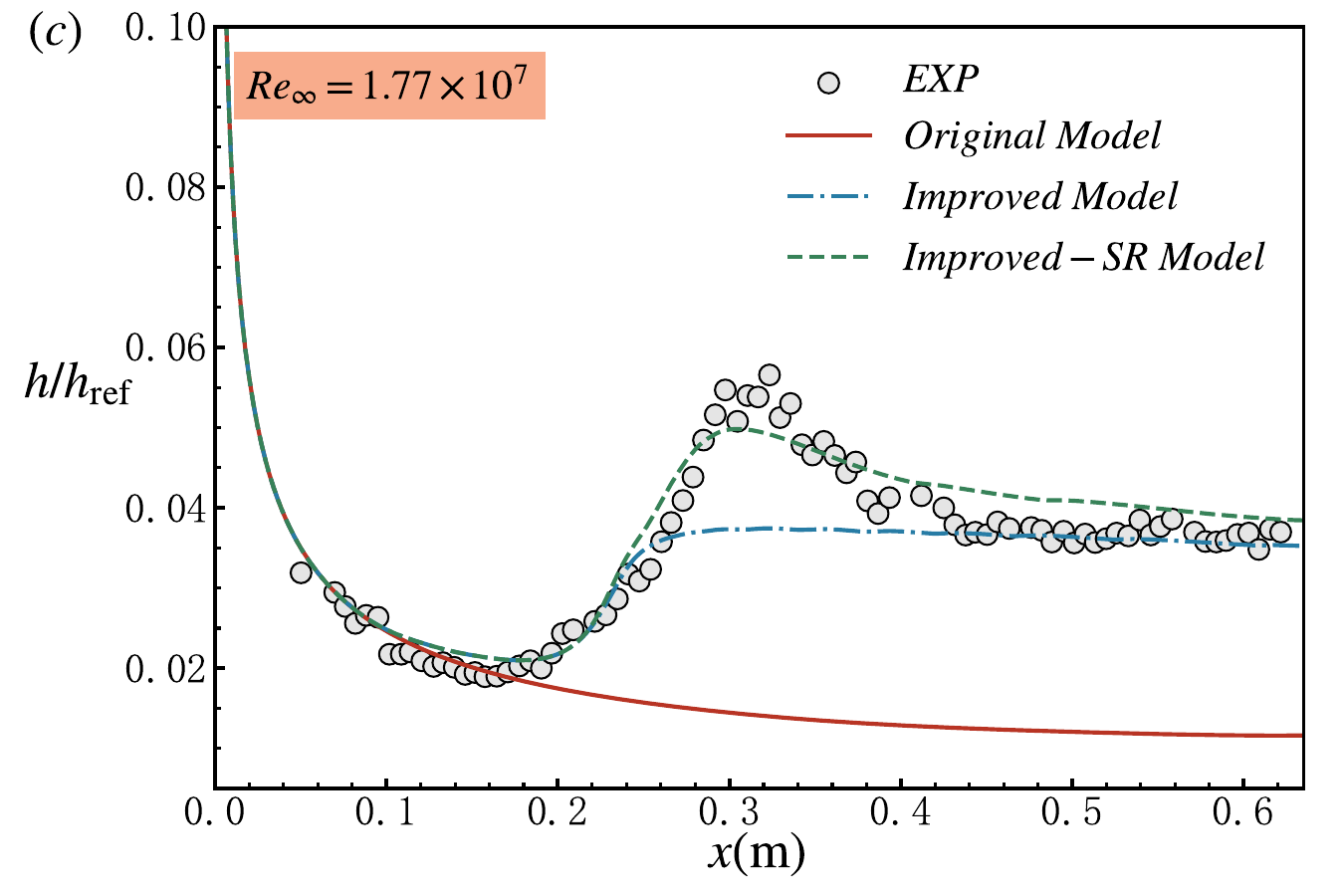}~\includegraphics[scale=0.3]{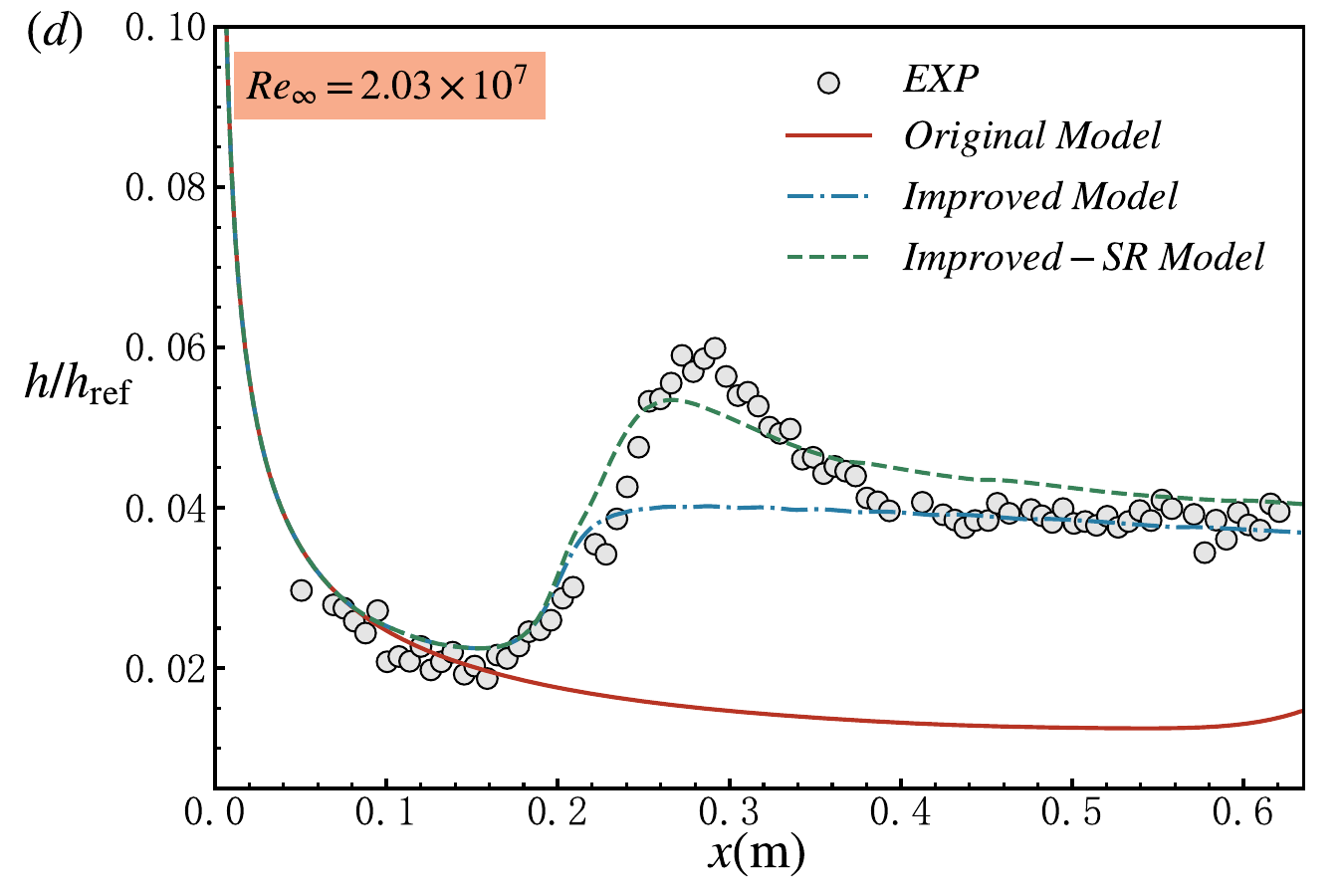}}
		\caption{Distributions of heat transfer coefficient $h/h_{\rm{ref}}$ for isothermal sharp straight cone predicted by different models in comparison with experimental data \cite{Horvath}: (a) case S1, (b) case S2, (c) case S3, and (d) case S4 as listed in Table~\ref{tab:Sharp-Cones-cases}.}
		\label{fig:S1234-h-Menter-LeiWu-Exp}
	\end{figure}

	Fig.~\ref{fig:S1234-h-Menter-LeiWu-Exp} presents the predicted heat-transfer coefficient $h/h_{\rm{ref}}$ distributions for the first isothermal sharp cone under different Reynolds numbers. The original low-speed model fails entirely to predict any transition, producing a smooth, laminar-like decay in heat-transfer across all cases. In contrast, the high-speed improved model accurately captures the trend of transition onset and length as $Re_\infty$ increases, yet consistently underpredicts the peak heat-transfer in the late transition region, \ie, the heat-transfer overshoot. The SR-augmented model, however, builds upon the accurate prediction of improved model for transition location and length and introduces a physically consistent correction that elevates the heat-transfer in the late transition zone, thus reproducing the overshoot behavior observed in experiments. Although the peak magnitude of overshoot still deviates from experimental values, the SR model demonstrates a substantial improvement over both the original and improved models, particularly in capturing the qualitative shape and spatial extent of overshoot.

\begin{figure}[h]
		\centerline{\includegraphics[scale=0.3]{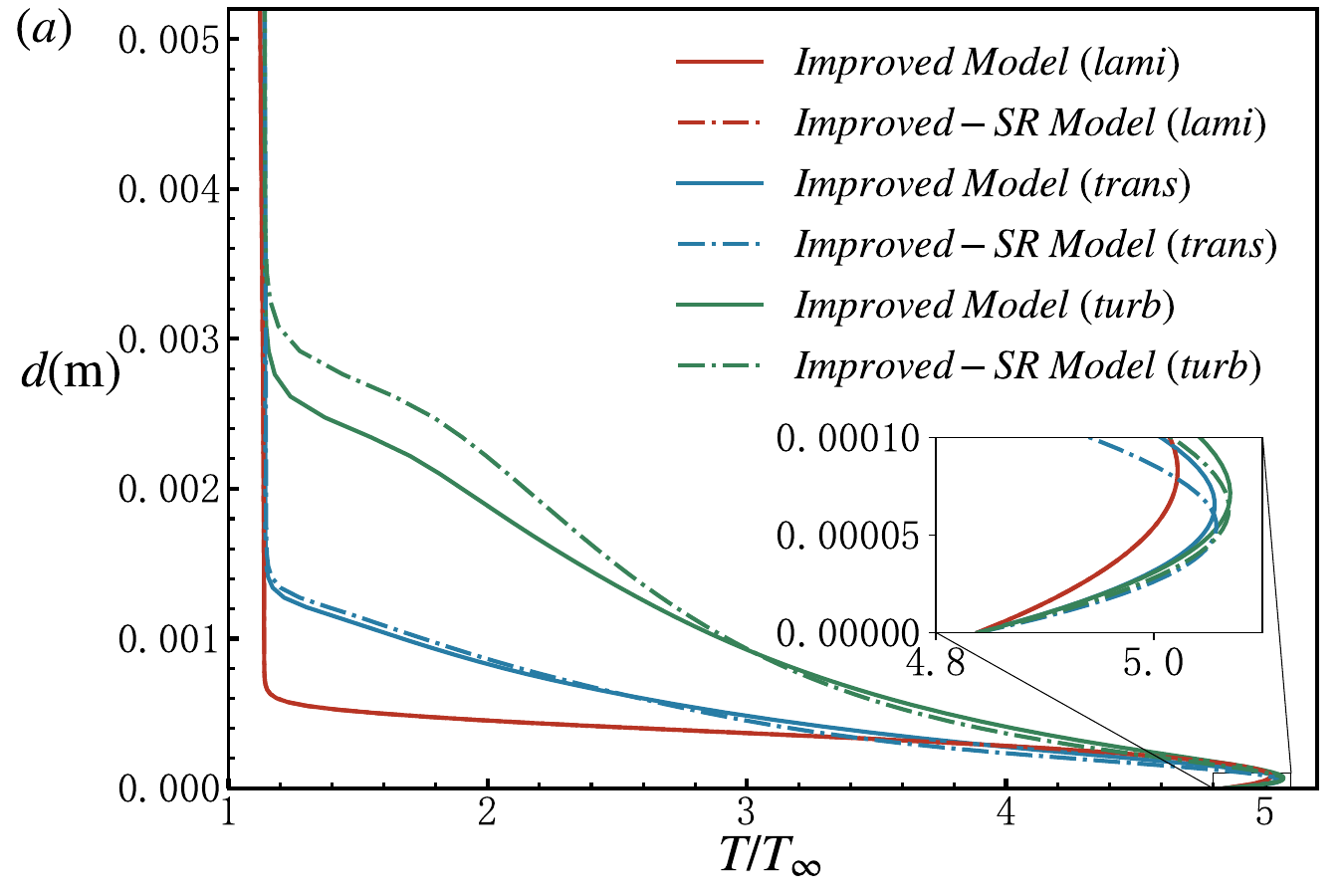}~\includegraphics[scale=0.3]{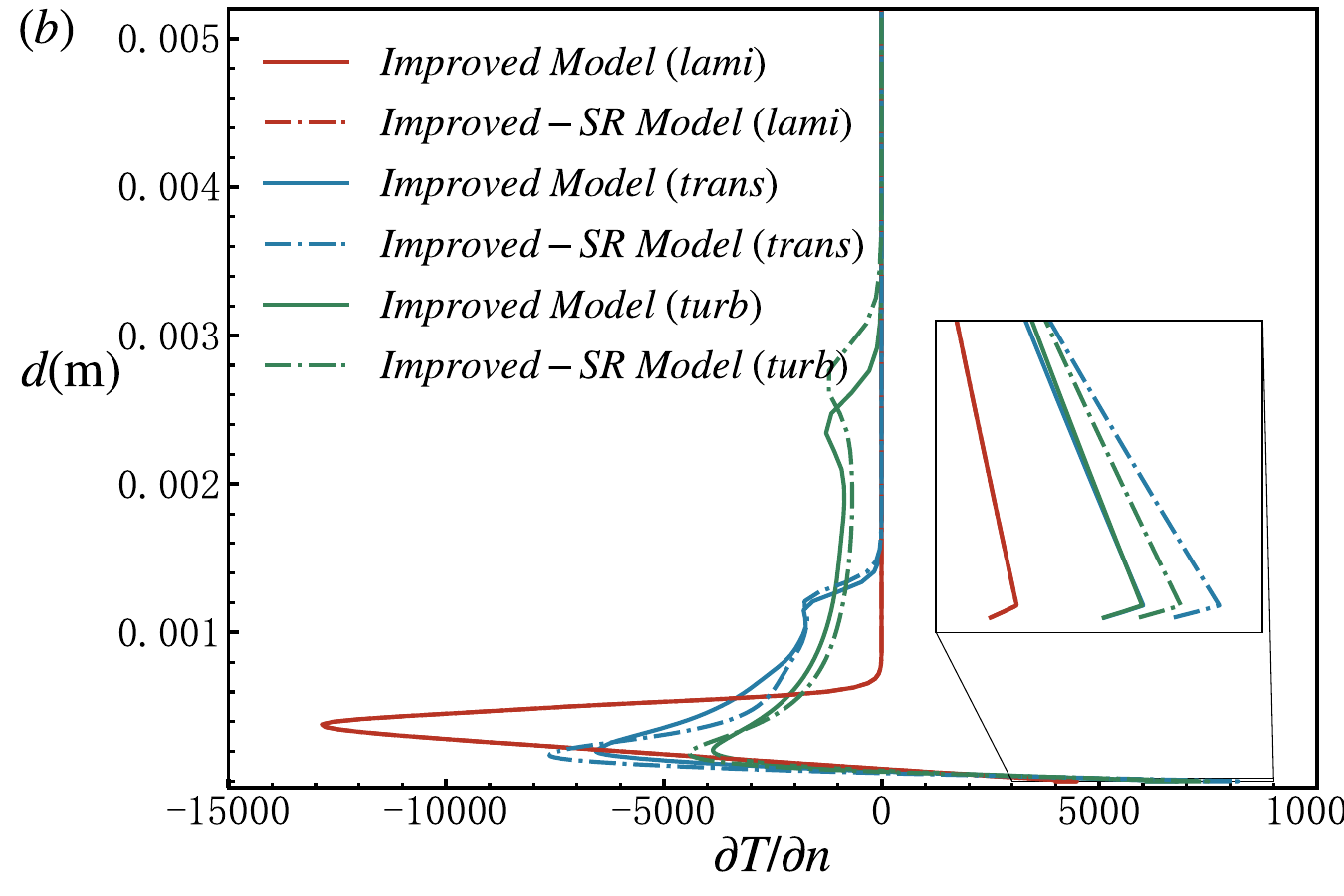}}
		\caption{Profiles of (a) temperature and (b) its wall-normal gradient from baseline and SR-corrected models at representative laminar ($x=0.1\rm{m}$), transitional ($x=0.26\rm{m}$), and turbulent ($x=0.5\rm{m}$) locations for case S4 (see Table~\ref{tab:Sharp-Cones-cases}).}
		\label{fig:S4-T-Ty}
	\end{figure}

	Depicted in Fig.~\ref{fig:S4-T-Ty} is the temperature and their wall-normal gradients profiles of baseline and SR-augmented models across laminar, transitional, and turbulent regions. As can be seen, both models yield nearly identical temperature distributions in the laminar region, confirming that the SR augmentation does not perturb the pre-transitional state. In the transitional zone, the SR-augmented model exhibits the steepest temperature gradient near the wall, significantly exceeding that of the baseline model. This enhanced gradient directly contributes to the increased heat-transfer (\ie, the overshoot phenomenon) in the late transition region.
	
\begin{figure}[h]
		\centerline{\includegraphics[scale=0.3]{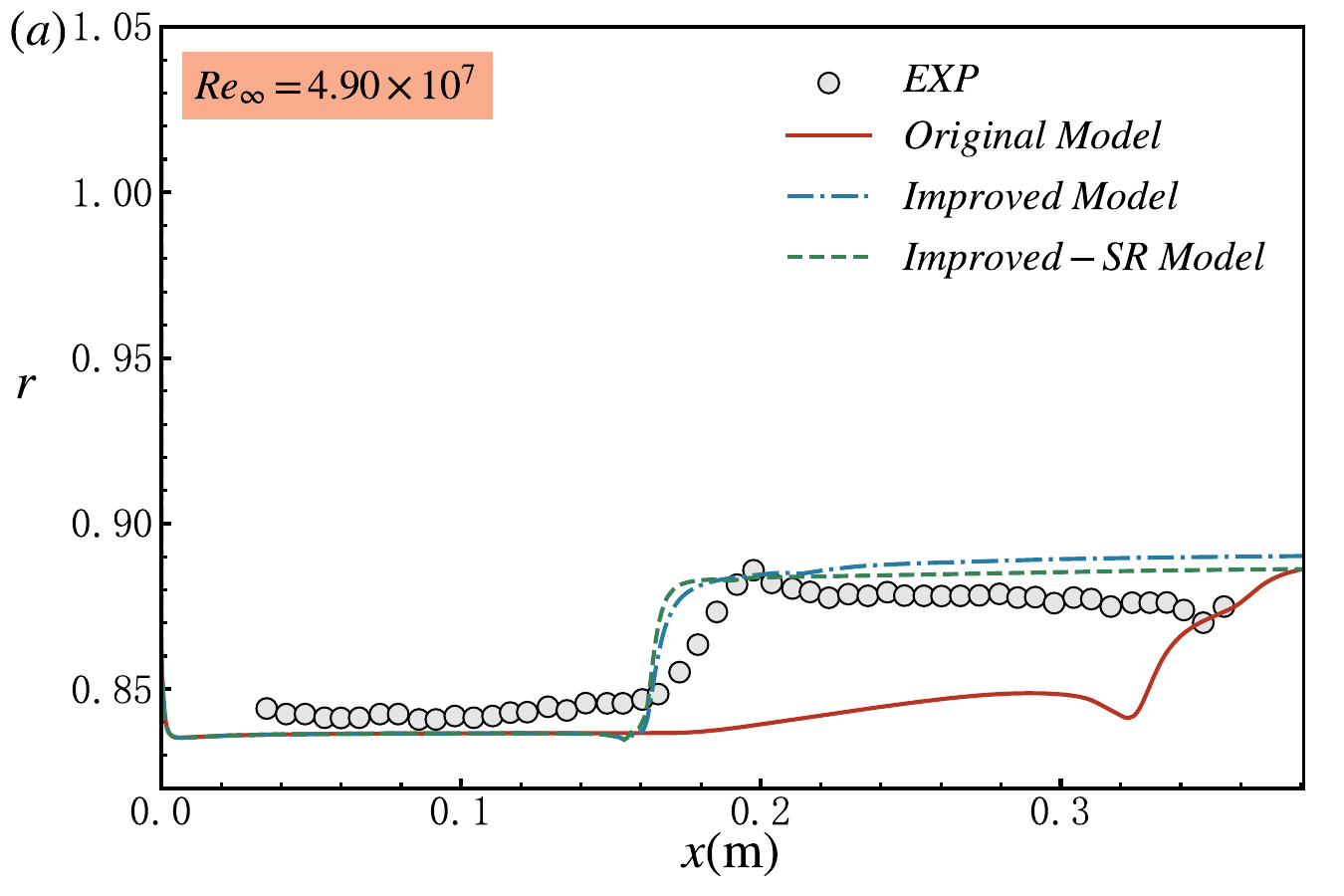}~\includegraphics[scale=0.3]{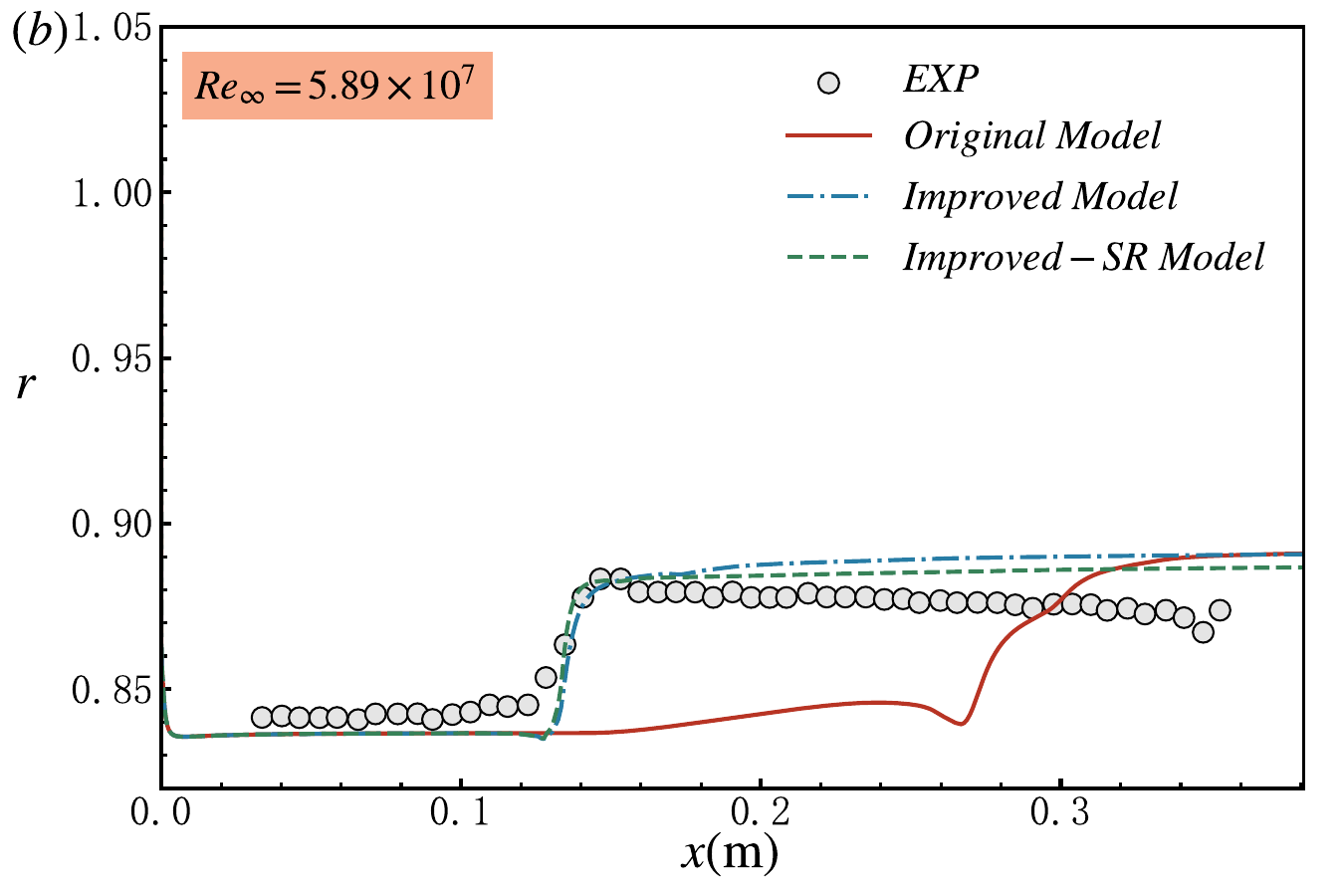}}
		\centerline{\includegraphics[scale=0.3]{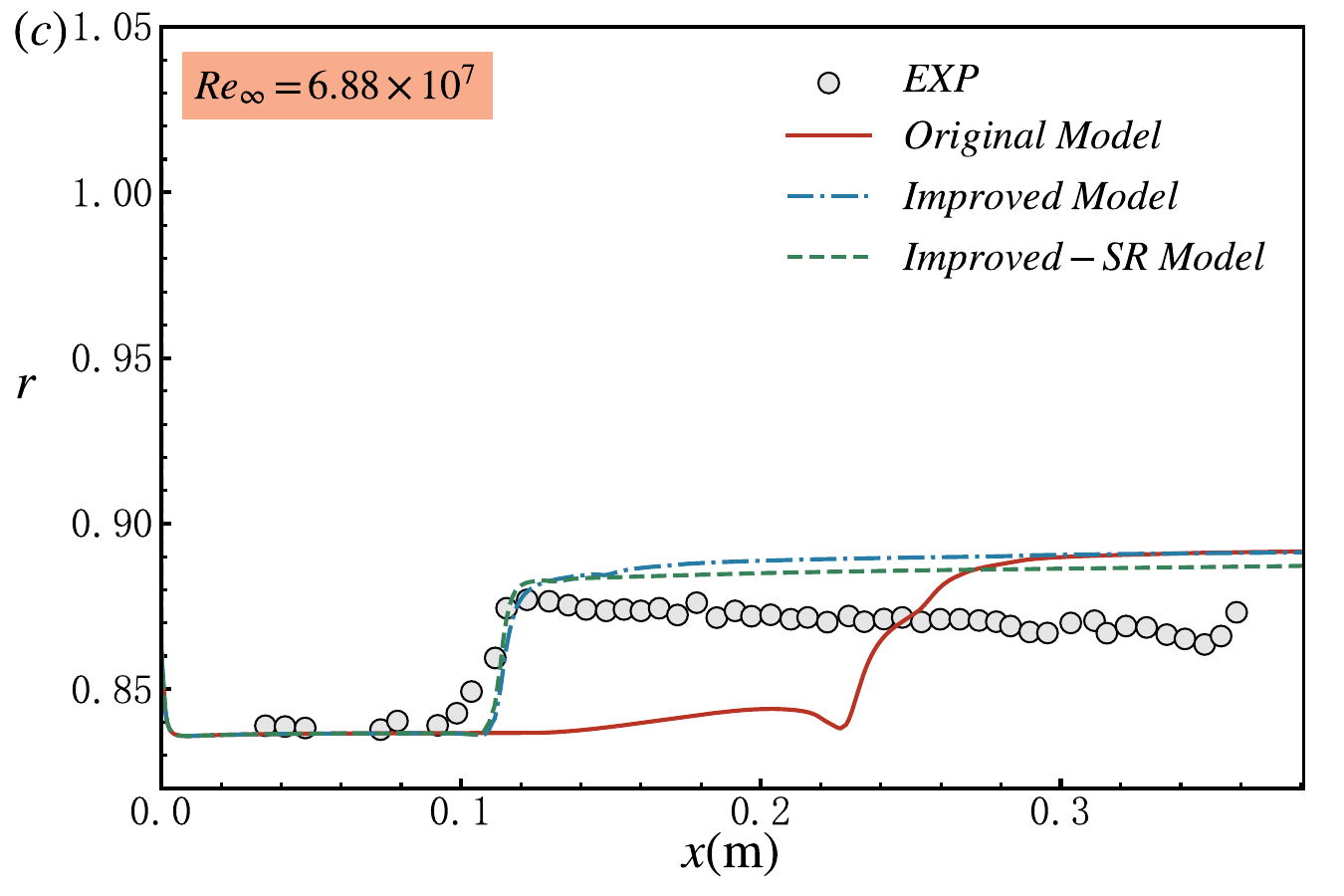}~\includegraphics[scale=0.3]{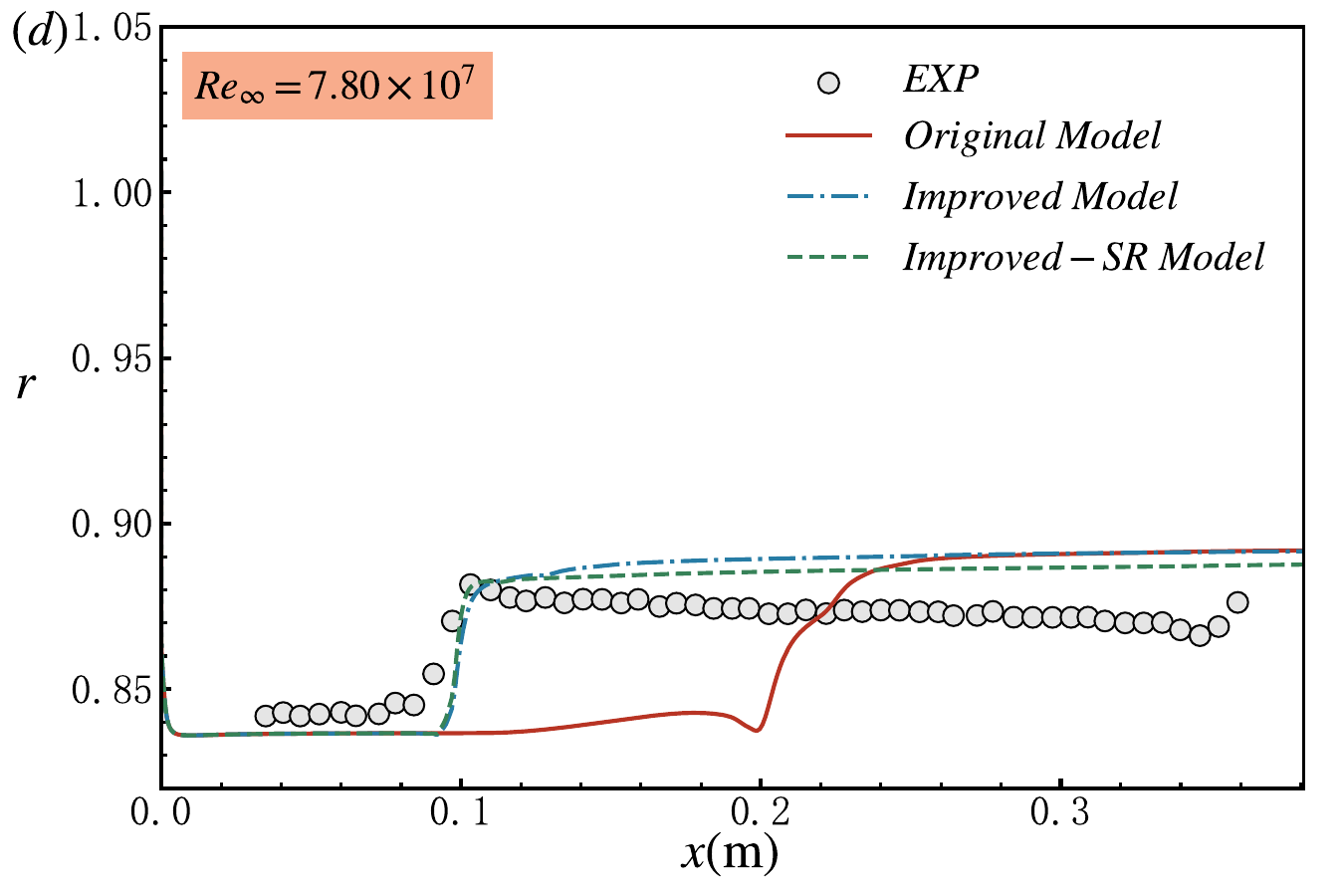}}
		\caption{Distributions of temperature recovery factor $r$ for adiabatic sharp straight cone predicted by different models in comparison with experimental data \cite{Chen-SharpCone-Exp}: (a) case S5, (b) case S6, (c) case S7, and (d) case S8 as listed in Table~\ref{tab:Sharp-Cones-cases}.}
		\label{fig:S5678-r-Menter-LeiWu-Exp}
	\end{figure}

\begin{figure}[h]
		\centerline{\includegraphics[scale=0.3]{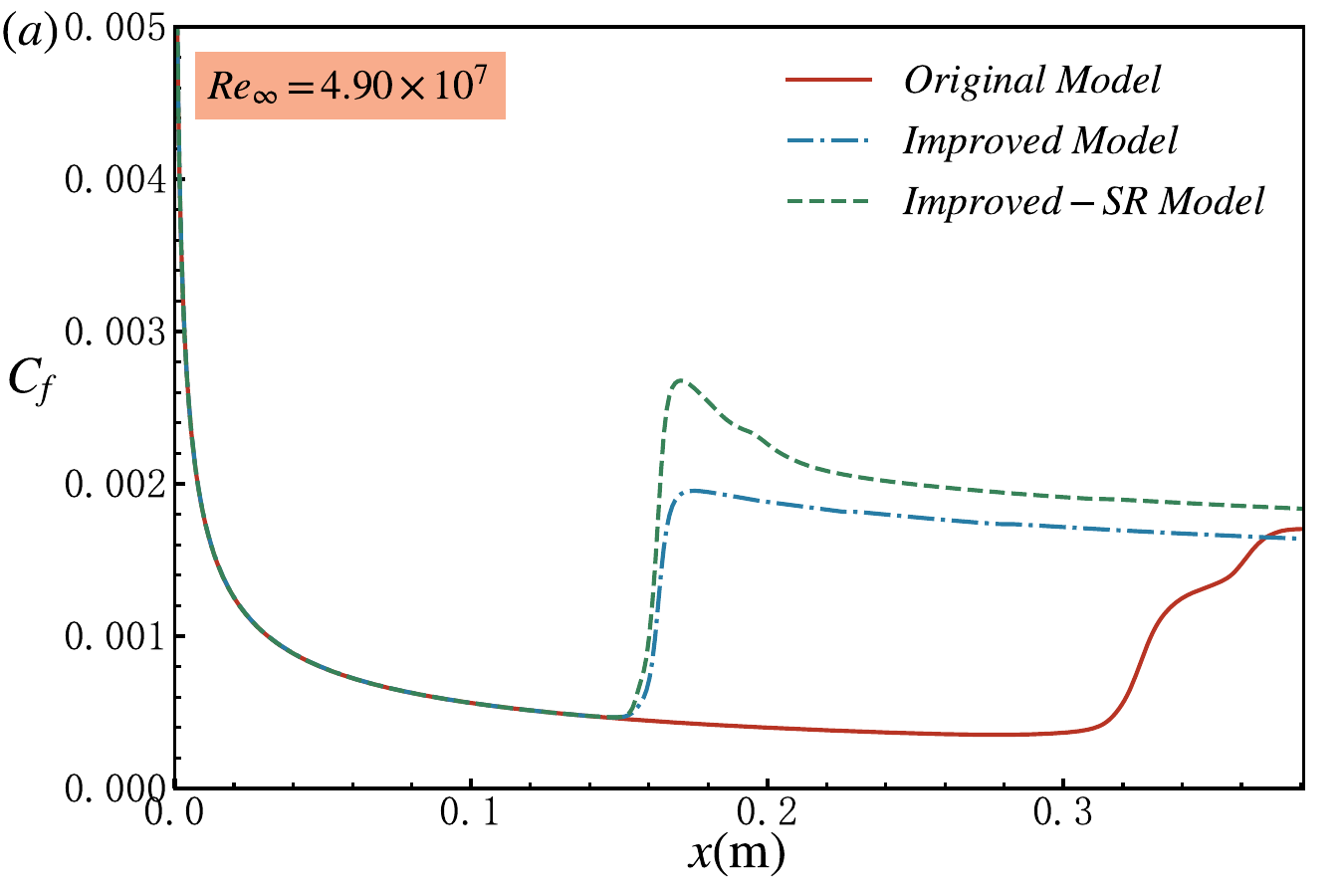}~\includegraphics[scale=0.3]{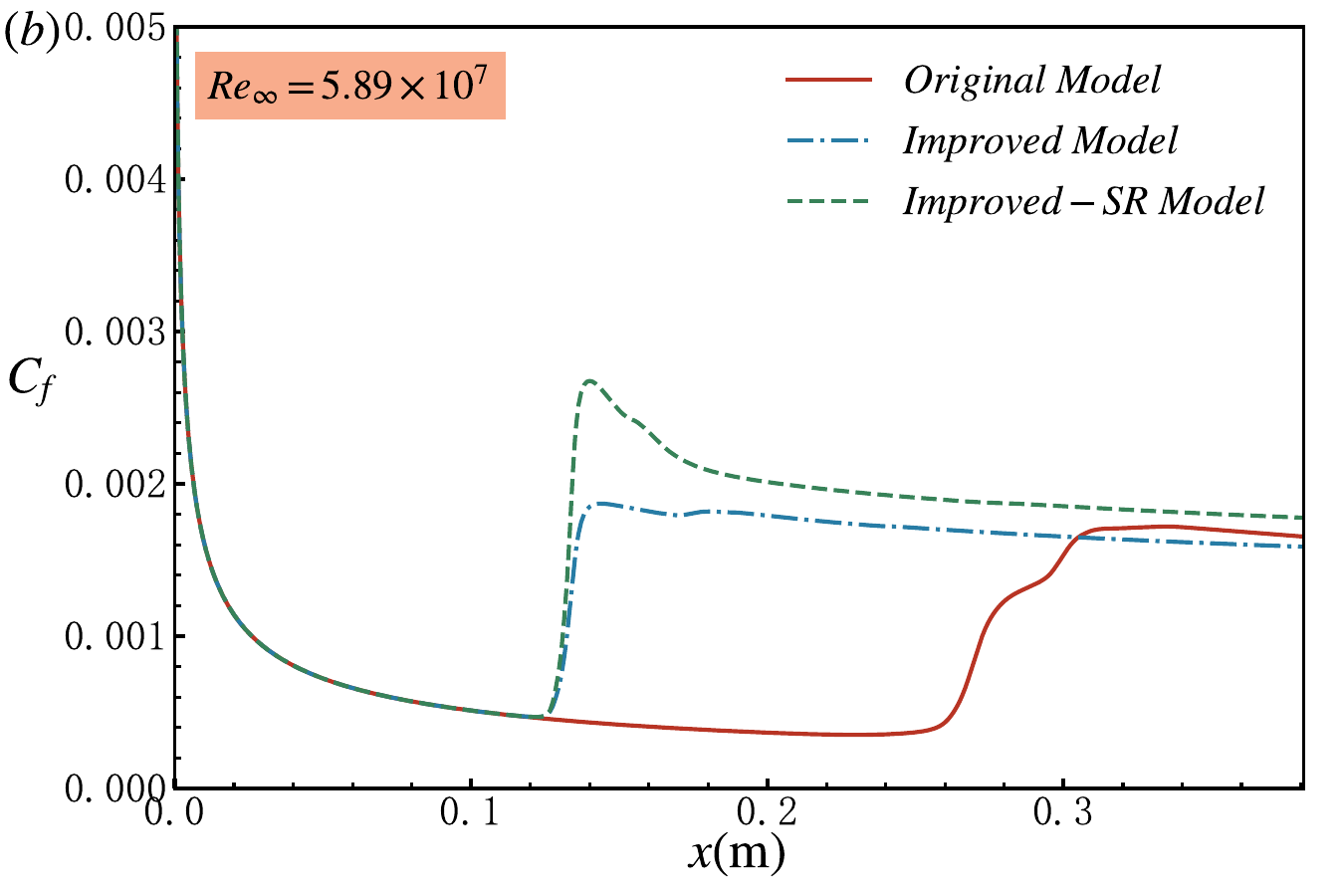}}
		\centerline{\includegraphics[scale=0.3]{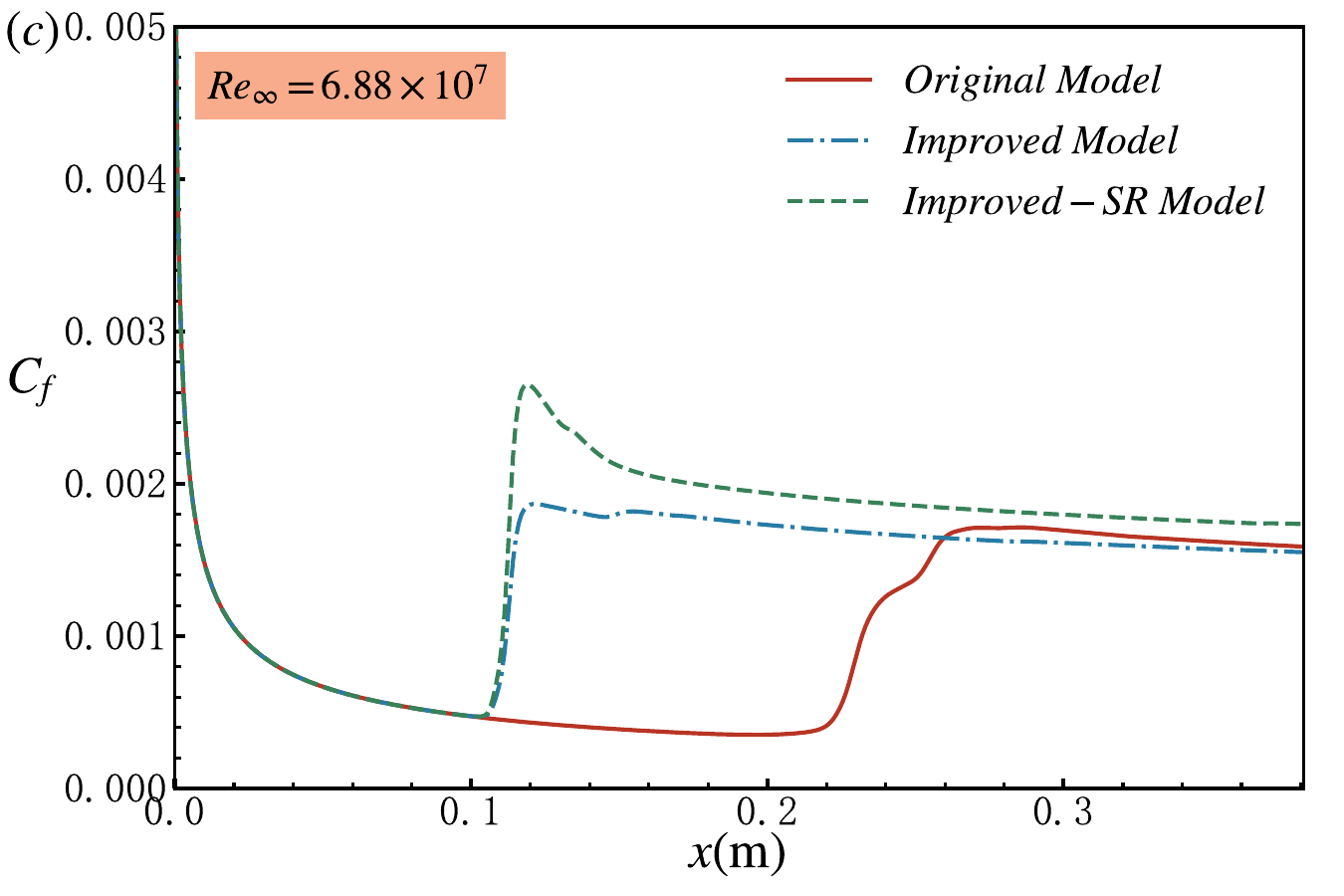}~\includegraphics[scale=0.3]{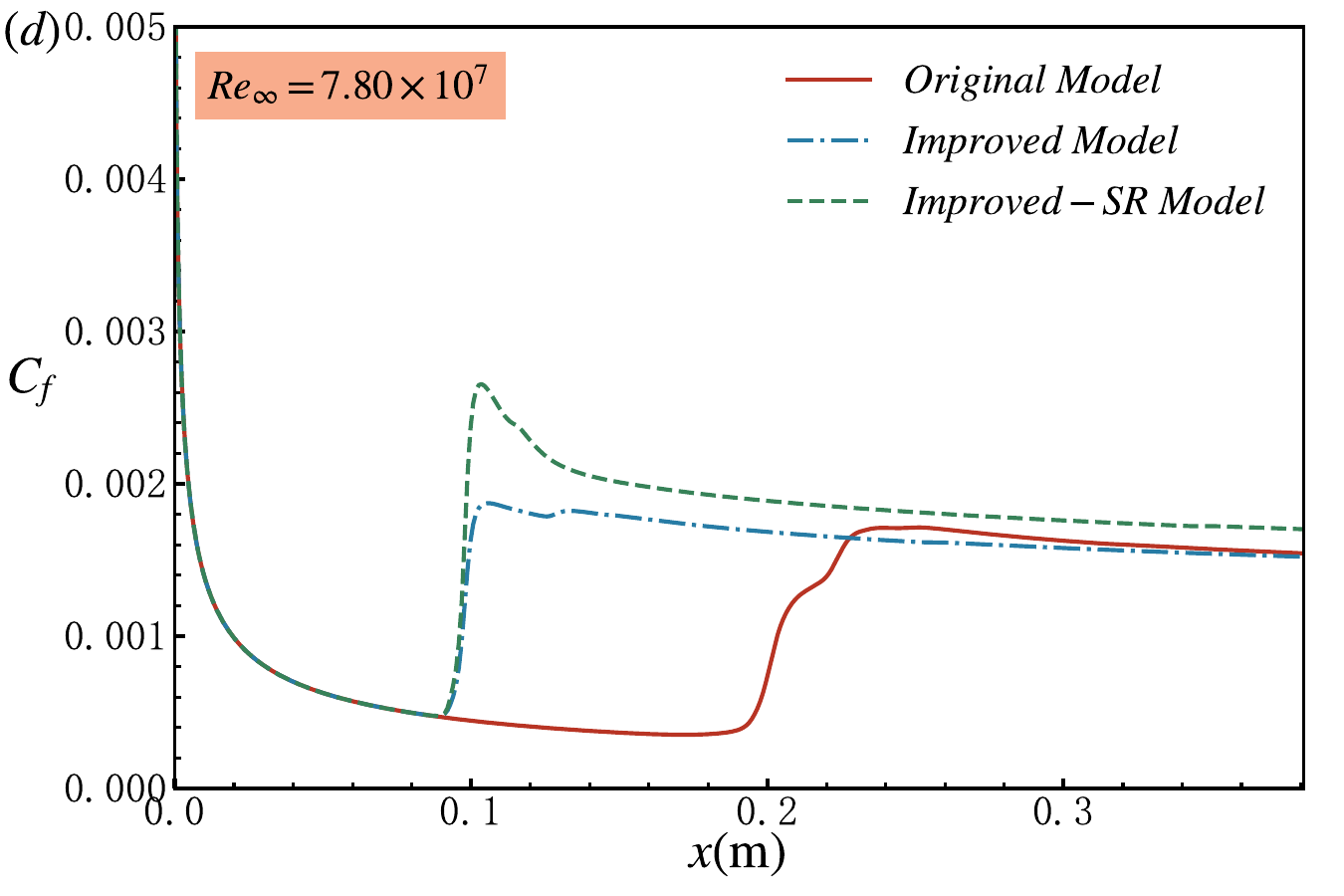}}
		\caption{Distributions of skin-friction coefficient $C_f$ for adiabatic sharp straight cone predicted by different models: (a) case S5, (b) case S6, (c) case S7, and (d) case S8 as listed in Table~\ref{tab:Sharp-Cones-cases}.}
		\label{fig:S5678-Cftau-Menter-LeiWu-Exp}
	\end{figure}

	Fig.~\ref{fig:S5678-r-Menter-LeiWu-Exp} shows the temperature recovery factor $r=(T_{\rm{aw}}-T_{\infty})/(T_0-T_\infty)$ profiles for adiabatic sharp straight cone. Compared to the low-speed transition model, both the baseline and SR-enhanced models exhibit similar morphology and agree well with experimental data. Regarding specific details, the SR model shows closer agreement with experiments in the fully turbulent region. Due to the adiabatic boundary condition, the adiabatic wall temperature $T_{\rm{aw}}$ adjusts passively to maintain energy conservation, so no overshoot is observed in $r$. When it comes to the skin-friction coefficient $C_f$ in Fig.~\ref{fig:S5678-Cftau-Menter-LeiWu-Exp}, the SR-corrected model clearly resolves the overshoot, whereas the baseline model fails to capture it.

	\subsection{Blunt straight cones}\label{subsec:BC}
	The final validation of present model is conducted in blunt straight cones with varying nose radii. Experimental measurements by \citet{Horvath} are taken as references (see cases B1-B3 listed in Table~\ref{tab:Blunt-Cones-cases}). Similar to our previous research \cite{WuLei-2026-IJHMT}, an additional case B4 with larger nose bluntness of $R_{\rm{n}}=2.38125\rm{mm}$ is introduced. Case B1 is analogous to the sharp cone configuration (S1-S4 in Sec.~\ref{subsec:SC}), cases B2-B4 share the same geometry but differ in nose radius. The grid resolutions are $361\times181\times21$, $457\times211\times21$, $553\times221\times21$, and $649\times231\times21$ (streamwise$\times$wall-normal$\times$circumferential) for cases B1 to B4, respectively.
	\begin{table*}[width=.9\textwidth,cols=4,pos=h]
		\caption{\label{tab:Blunt-Cones-cases}Input parameters for flows past blunt straight cones with various nose bluntnesses.}
		\begin{tabular*}{\tblwidth}{@{} LLLLLLLLL@{}}
			\toprule
			Case & $R_{\rm{n}}(\rm{mm})$ & $Ma_\infty$ & $Re_\infty(\rm{m}^{-1})$ & $T_{\rm{w}}(\rm{K})$ & $T_\infty(\rm{K})$ & $Tu_\infty(\%)$ & $p_0(\rm{Mpa})$ & Reference  \\
			\midrule
			B1   & $0.00254$             & $6.0$       & $2.56\times10^7$         & $306.36$             & $63.324$           & $0.40$          & $3.28$          & EXP, Ref.~\cite{Horvath} \\
			B2   & $0.79375$             & $6.0$       & $2.56\times10^7$         & $306.36$             & $63.324$           & $0.40$          & $3.28$          & EXP, Ref.~\cite{Horvath} \\
			B3   & $1.5875$              & $6.0$       & $2.56\times10^7$         & $306.36$             & $63.324$           & $0.40$          & $3.28$          & EXP, Ref.~\cite{Horvath} \\
			B4   & $2.38125$             & $6.0$       & $2.56\times10^7$         & $306.36$             & $63.324$           & $0.40$          & $3.28$          & None \\
			\bottomrule
		\end{tabular*}
	\end{table*}
	
\begin{figure}[h]
		\centerline{\includegraphics[scale=0.3]{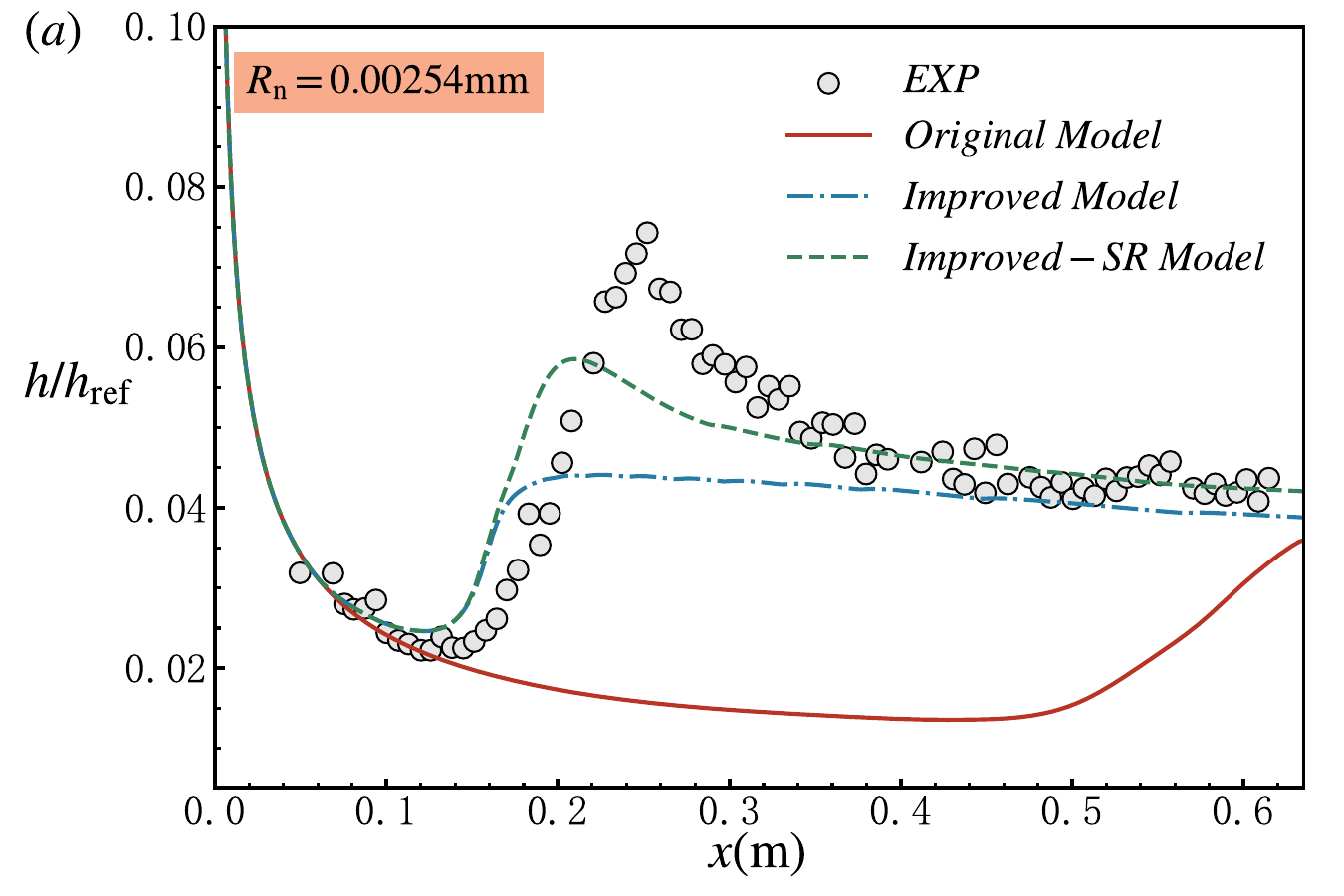}~\includegraphics[scale=0.3]{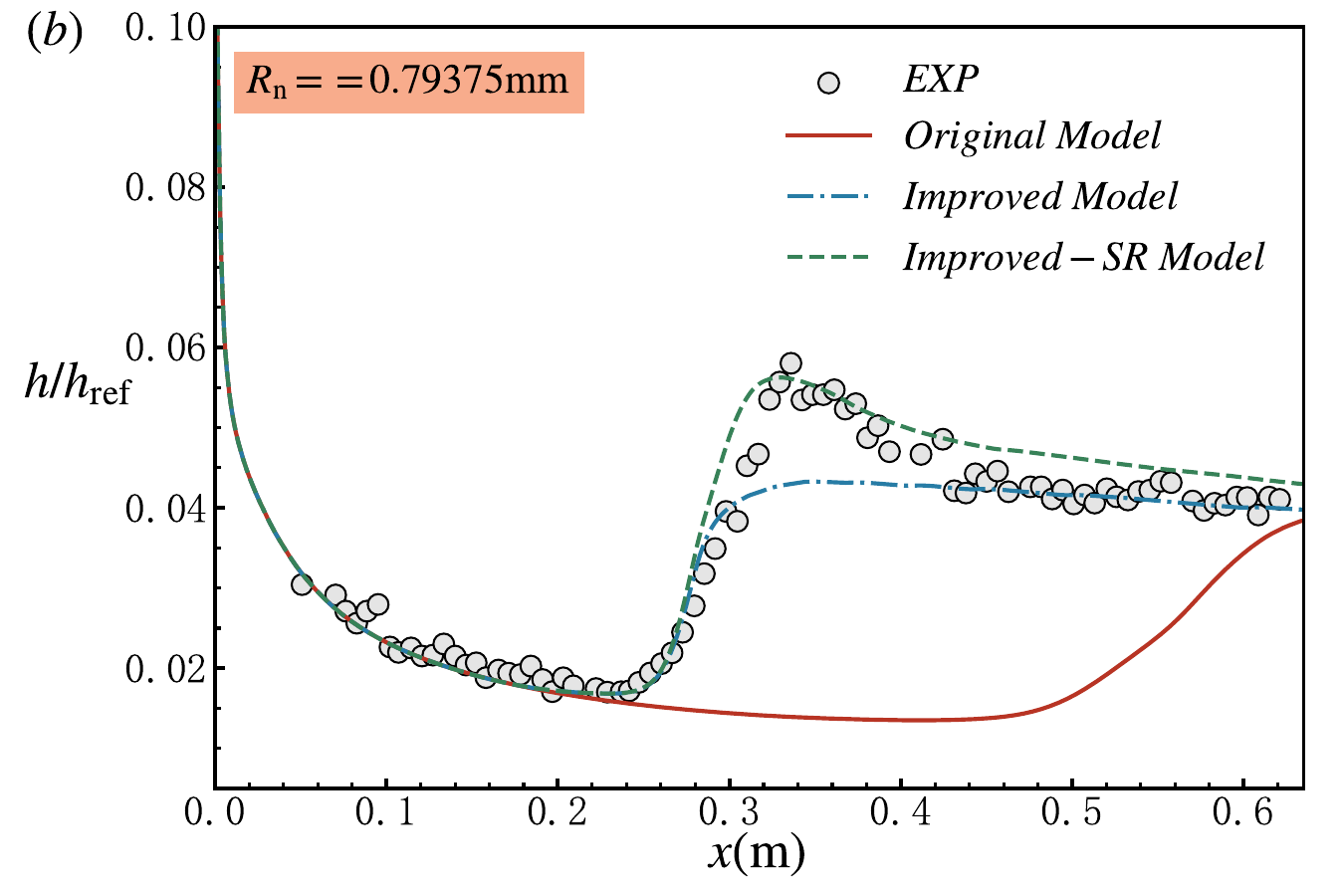}}
		\centerline{\includegraphics[scale=0.3]{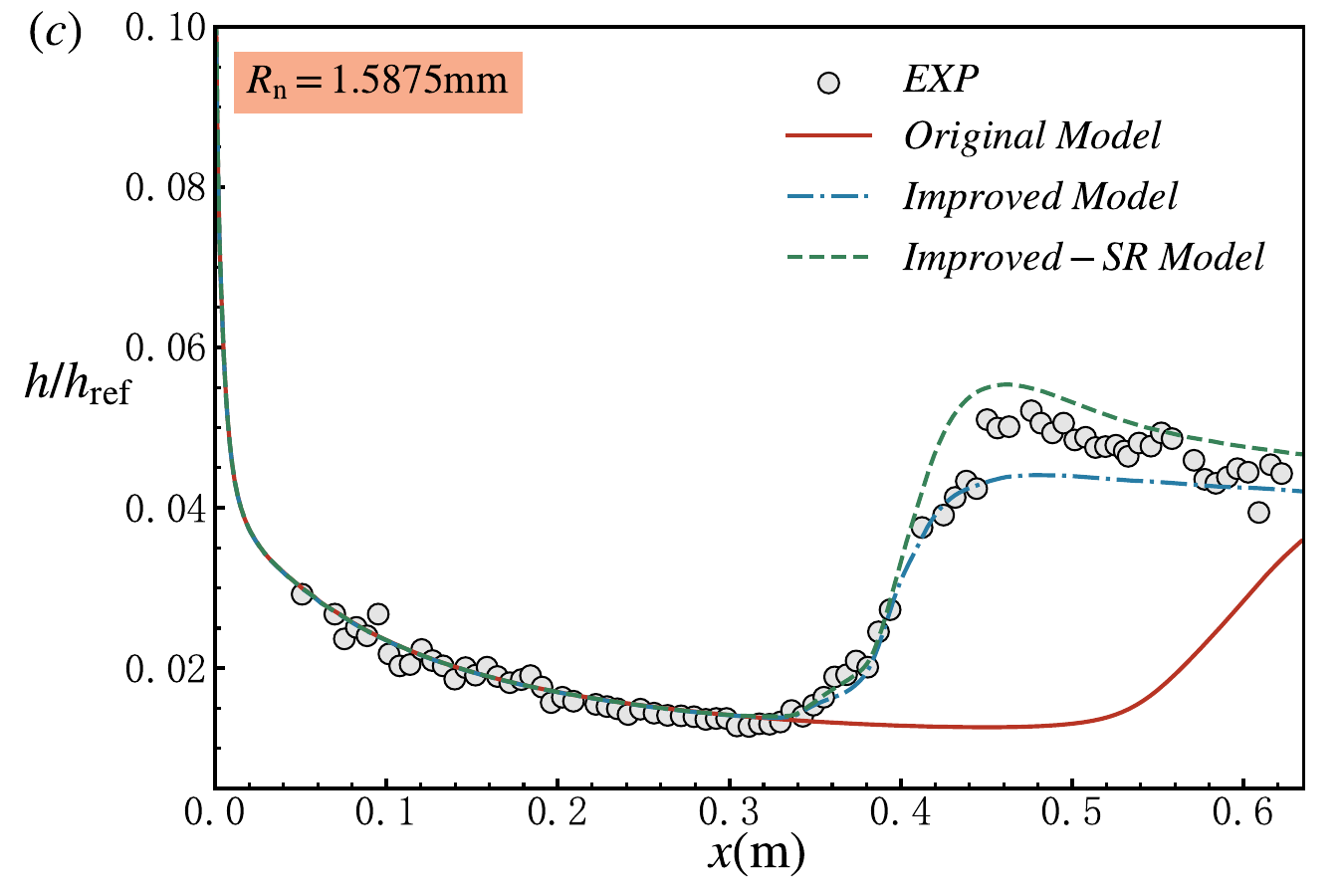}~\includegraphics[scale=0.3]{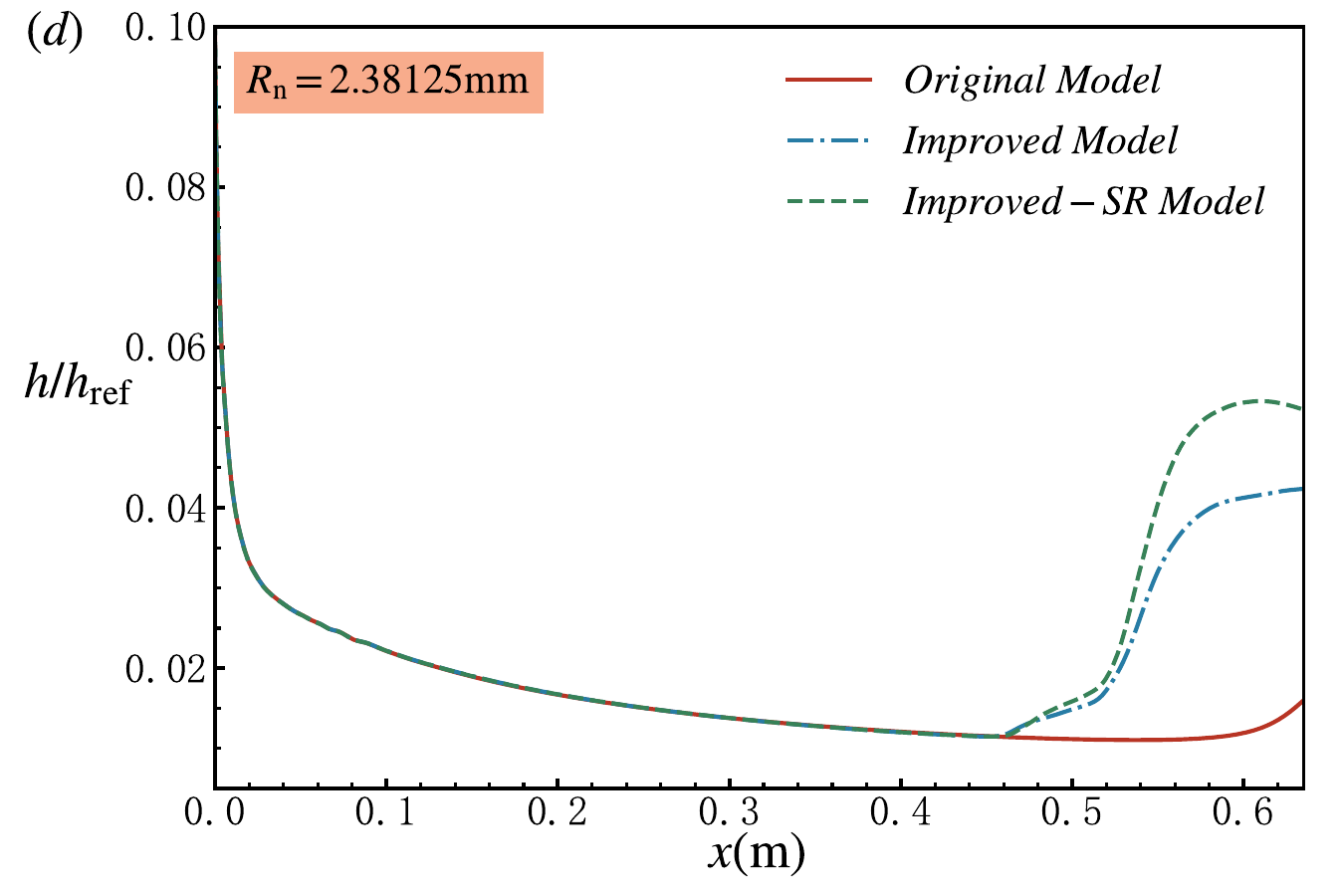}}
		\caption{Distributions of heat transfer coefficient $h/h_{\rm{ref}}$ for isothermal blunt straight cones predicted by different models in comparison with experimental data \cite{Horvath}: (a) case S1, (b) case S2, (c) case S3, and (d) case S4 as listed in Table~\ref{tab:Blunt-Cones-cases}.}
		\label{fig:B1234-h-Menter-LeiWu-Exp}
	\end{figure}

	As can be seen in Fig.~\ref{fig:B1234-h-Menter-LeiWu-Exp}, as $R_{\rm{n}}$ increases, both the baseline model and SR-augmented model exhibit a delayed transition onset, which agrees well with experimental observations and the well-known bluntness effect to transition. Similar to previous flat plate and sharp straight cone cases, the baseline model fails to resolve any overshoot in heat-transfer, while SR-augmented model successfully reproduces the overshoot phenomenon, aligning qualitatively with experimental trends. Although the peak magnitude of overshoot in case B1 still deviates noticeably from experimental values, the SR model demonstrates a substantial improvement over both original low-speed and high-speed improved model.
	
	As of now, the current SR-corrected model has undergone comprehensive validation across a wide range of configurations and has successfully resolved the previously unattainable overshoot phenomenon. Remarkably, this SR-based correction is trained exclusively on two canonical cases (flat plate case F1 and sharp cone case S4), yet it demonstrates well generalization across a wide spectrum of high-speed transitional flows. It accurately captures the overshoot not only in DNS-based flat plate scenarios but also in experimental sharp and blunt cone configurations with varying nose bluntness, Reynolds numbers, and thermal boundary conditions. These results suggest that the obtained SR expression has discovered a relatively universal physical mechanism for capturing the overshoot phenomenon, which confers its predictive capability across diverse, unseen flow configurations.

	\subsection{Importance of protect function $s_{\rm{prot}}$}
\begin{figure}[h]
		\centerline{\includegraphics[scale=0.3]{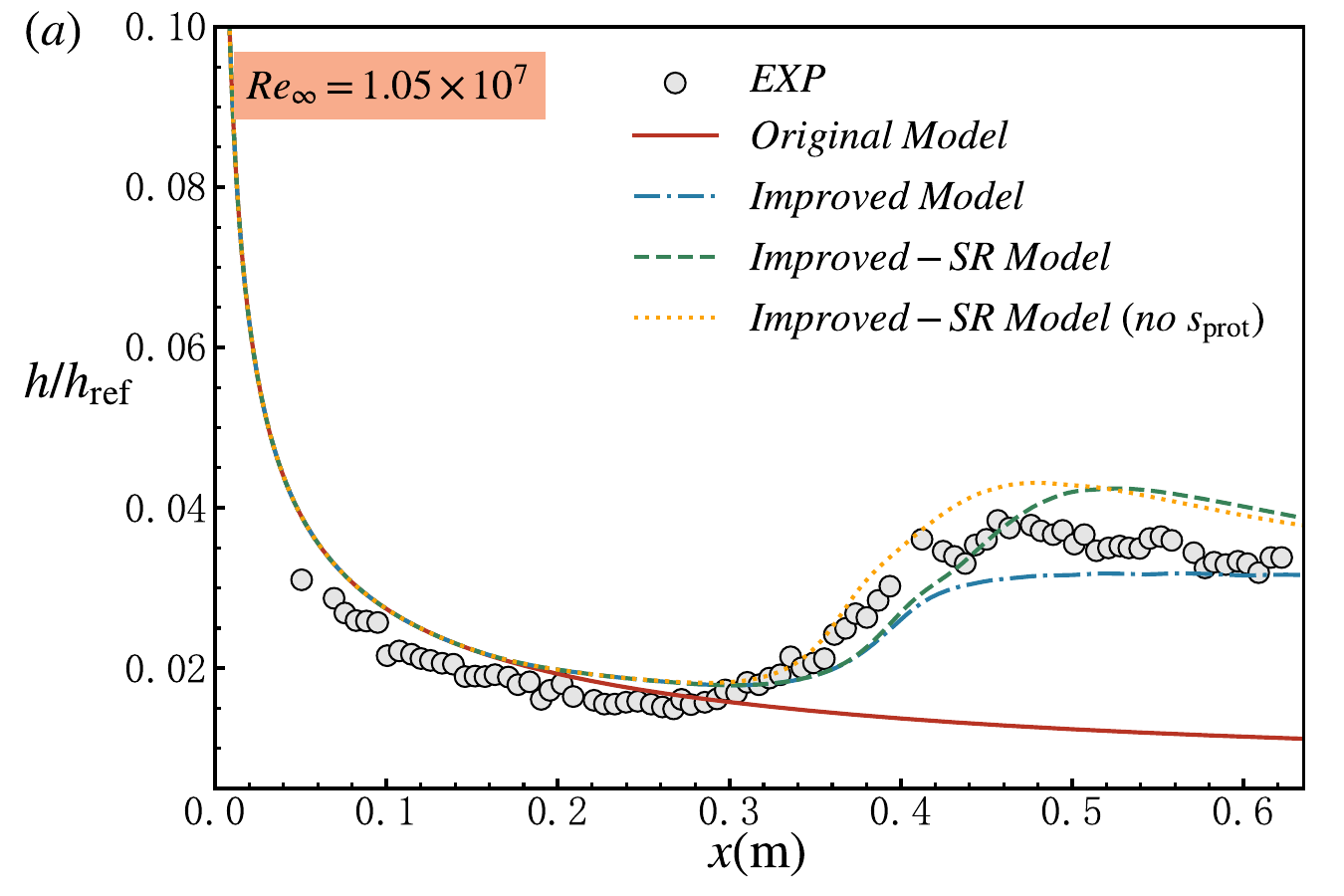}~\includegraphics[scale=0.3]{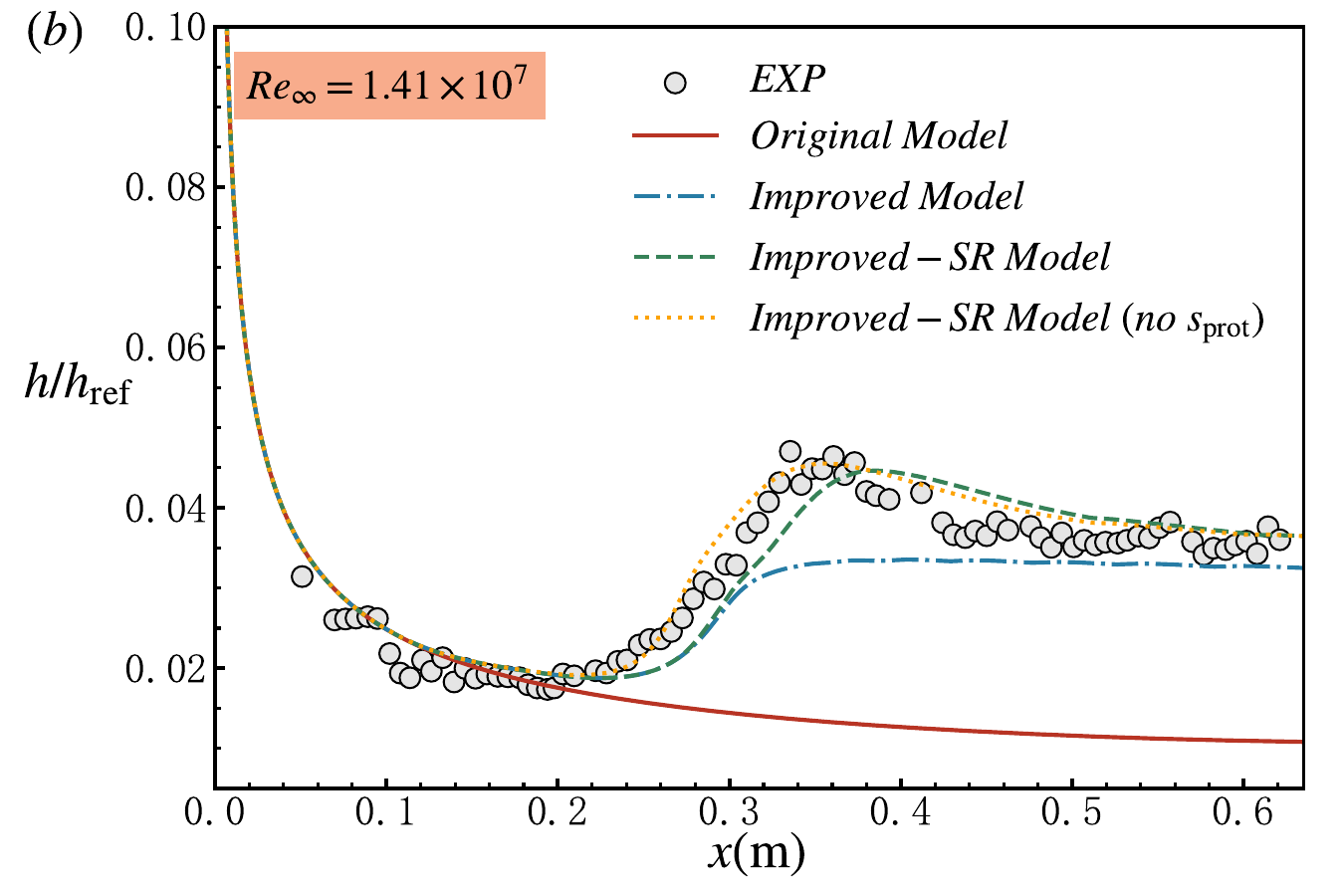}}
		\centerline{\includegraphics[scale=0.3]{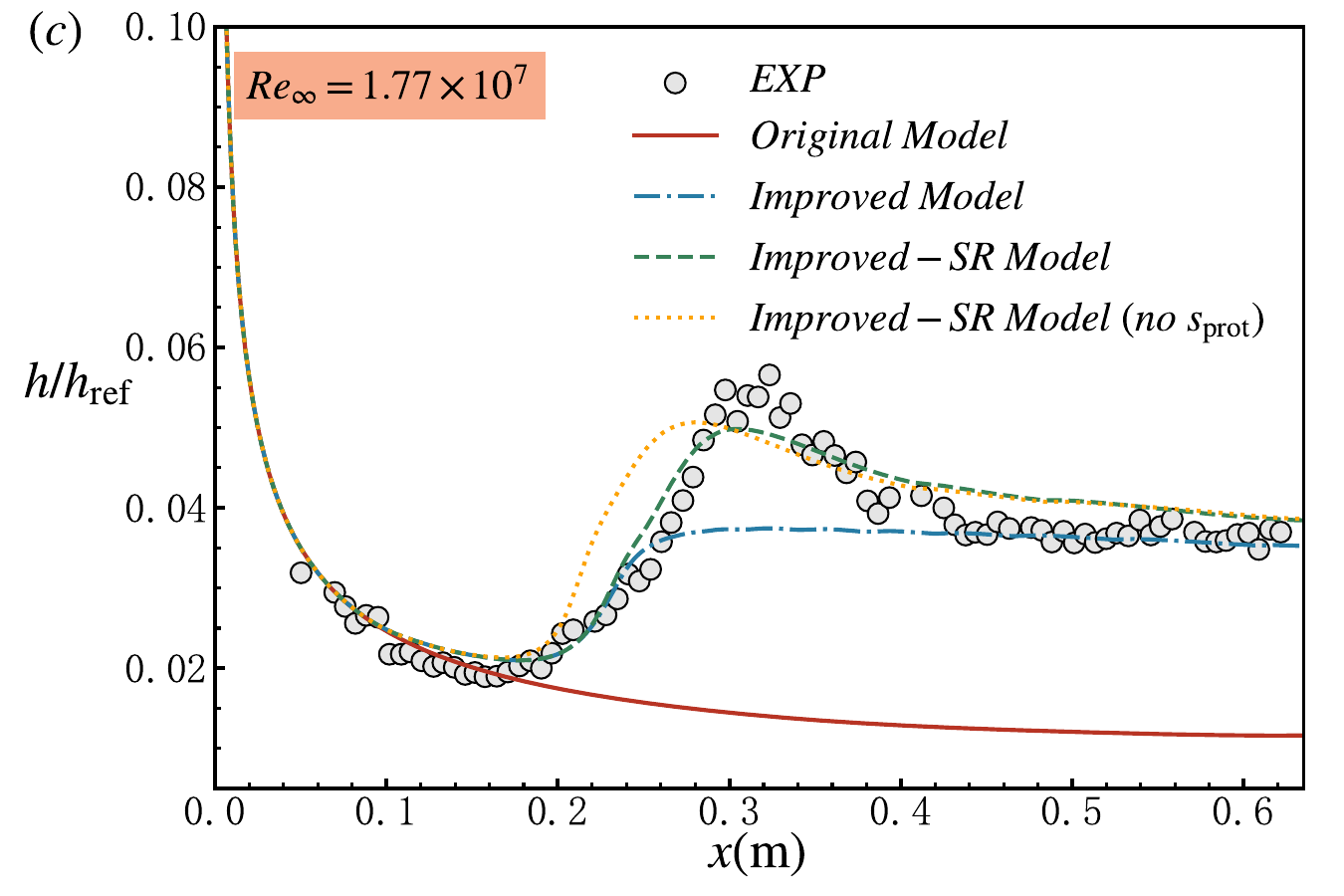}~\includegraphics[scale=0.3]{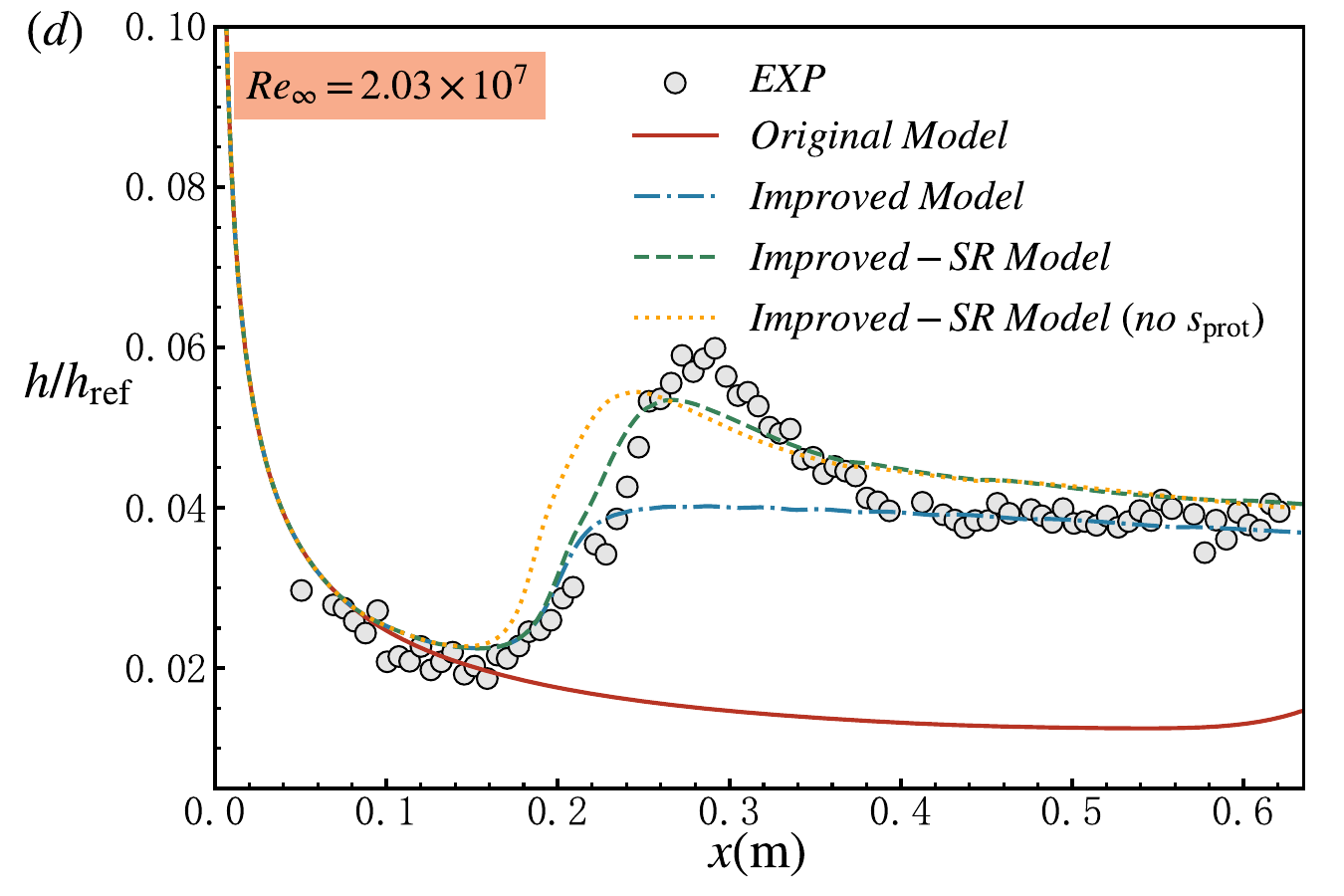}}
		\caption{Distributions of heat transfer coefficient $h/h_{\rm{ref}}$ for isothermal sharp straight cone predicted by different models in comparison with experimental data \cite{Horvath}: (a) case S1, (b) case S2, (c) case S3, and (d) case S4 as listed in Table~\ref{tab:Sharp-Cones-cases}.}
		\label{fig:S1234-h-Menter-LeiWu-Exp-no_s_prot}
	\end{figure}

	Take cases S1-S4 as example, the influence of protect function $s_{\rm{prot}}$ on SR-enhanced model is examined in Fig.~\ref{fig:S1234-h-Menter-LeiWu-Exp-no_s_prot}. While $s_{\rm{prot}}$ has minimal effect on the peak magnitude of heat-transfer overshoot and fully turbulent region, its absence leads to a premature transition onset, which is clearly undesirable. Our goal is not merely to enhance the model's ability to capture the overshoot, but to do so without compromising the fidelity of underlying transition location and length, which are considered to be well-predicted by the high-speed improved model. In this scenario, $s_{\rm{prot}}$ ensures that the correction is activated only after the flow commits to transition, thereby preserving the integrity of predictive capability for pre-transitional zone.
	
	Notably, for cases S1 and S2, SR model without $s_{\rm{prot}}$ appears to align more closely with experimental data. However, this improvement is coincidental: the baseline improved model itself exhibits a delayed transition in these cases, while removing $s_{\rm{prot}}$ artificially advances it, thus compensating for the delay. This behavior stems not from an improved physical mechanism, but from an unintended cancellation of errors. Therefore, the inclusion of $s_{\rm{prot}}$ is both necessary and physically justified. It enforces a disciplined application of the model-form uncertainty correction, ensuring that augmentation is activated only where it is physically warranted.

	\subsection{More concise expression for $\beta(\mathbf{x})$}
\begin{figure}[h]
		\centerline{\includegraphics[scale=0.3]{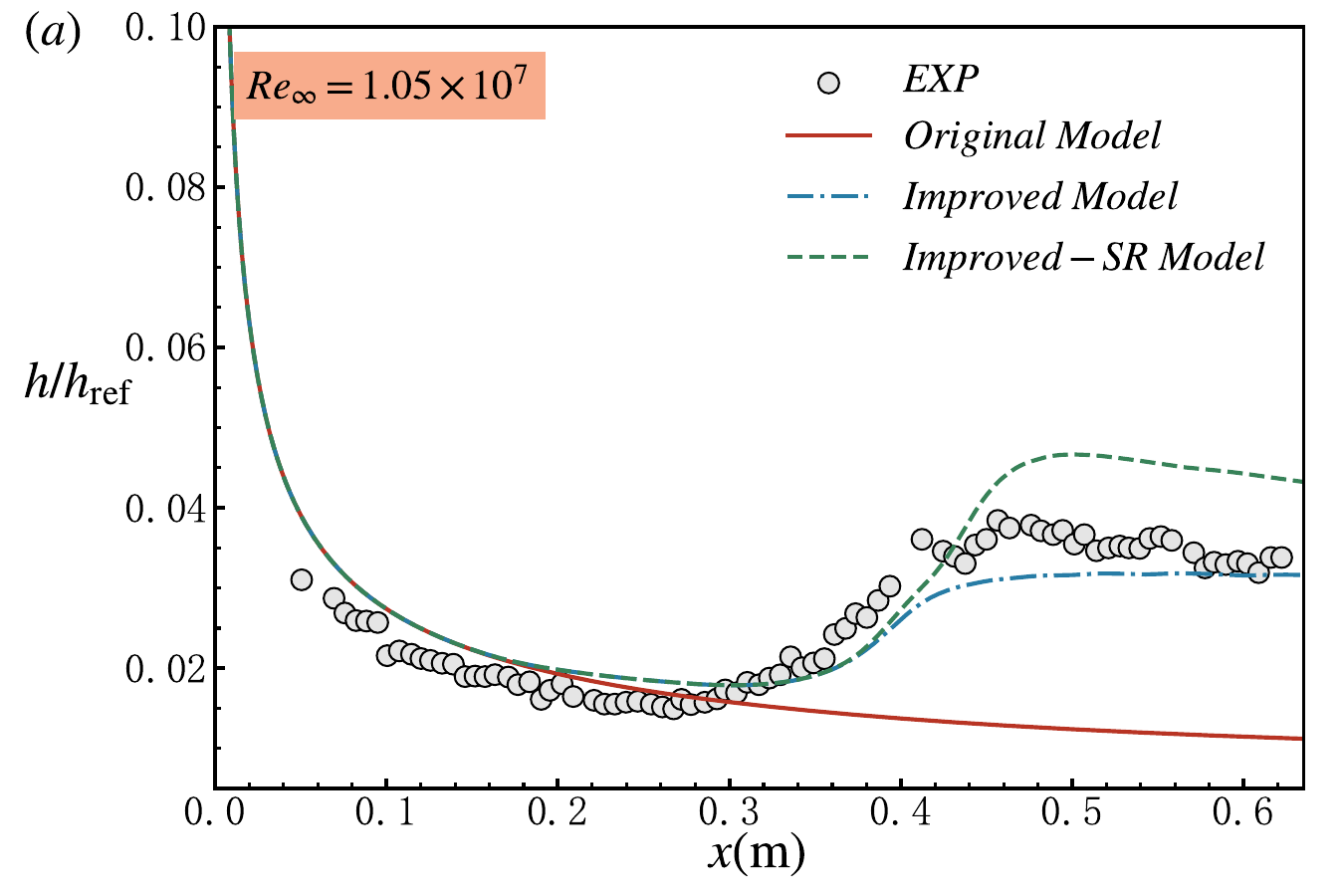}~\includegraphics[scale=0.3]{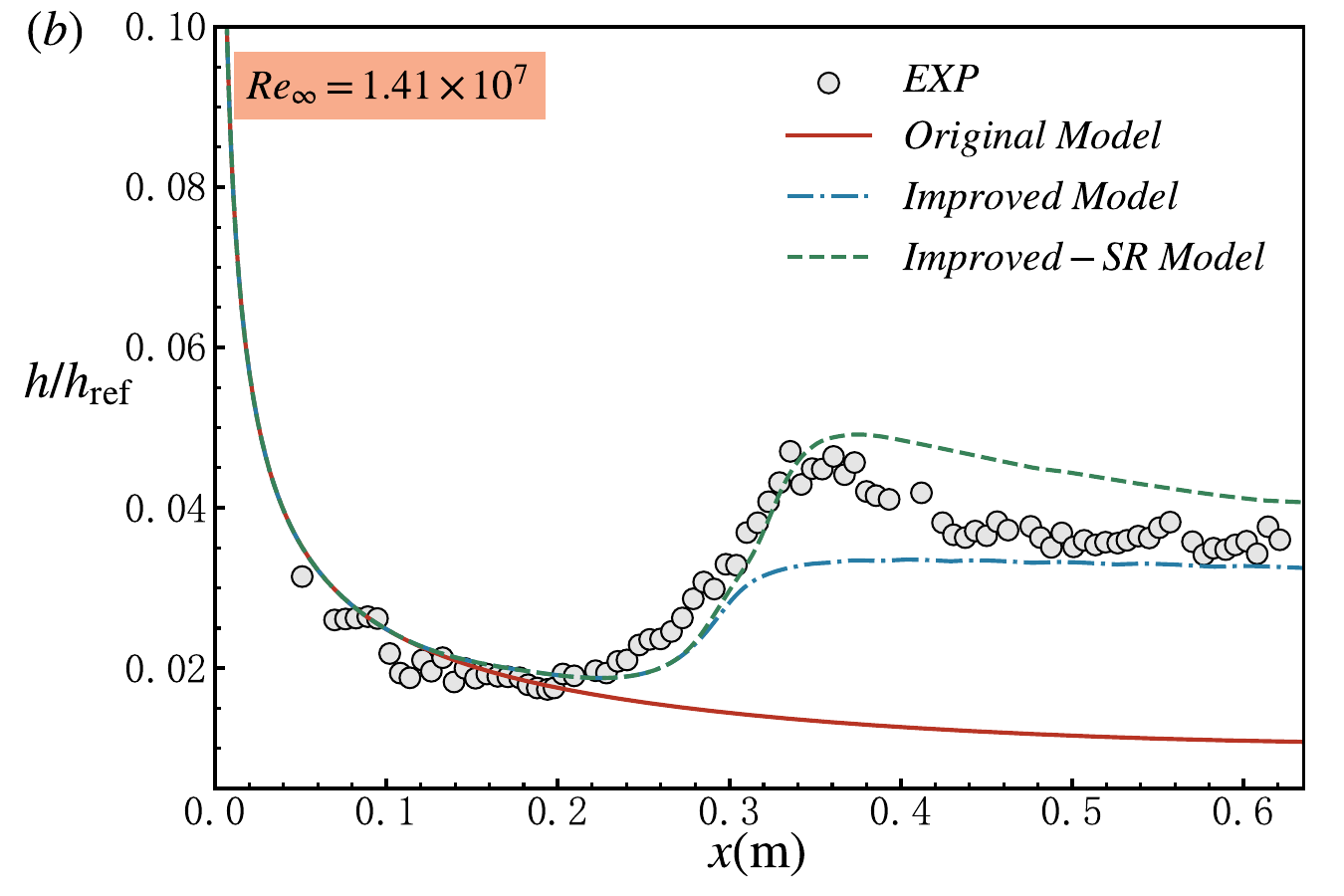}}
		\centerline{\includegraphics[scale=0.3]{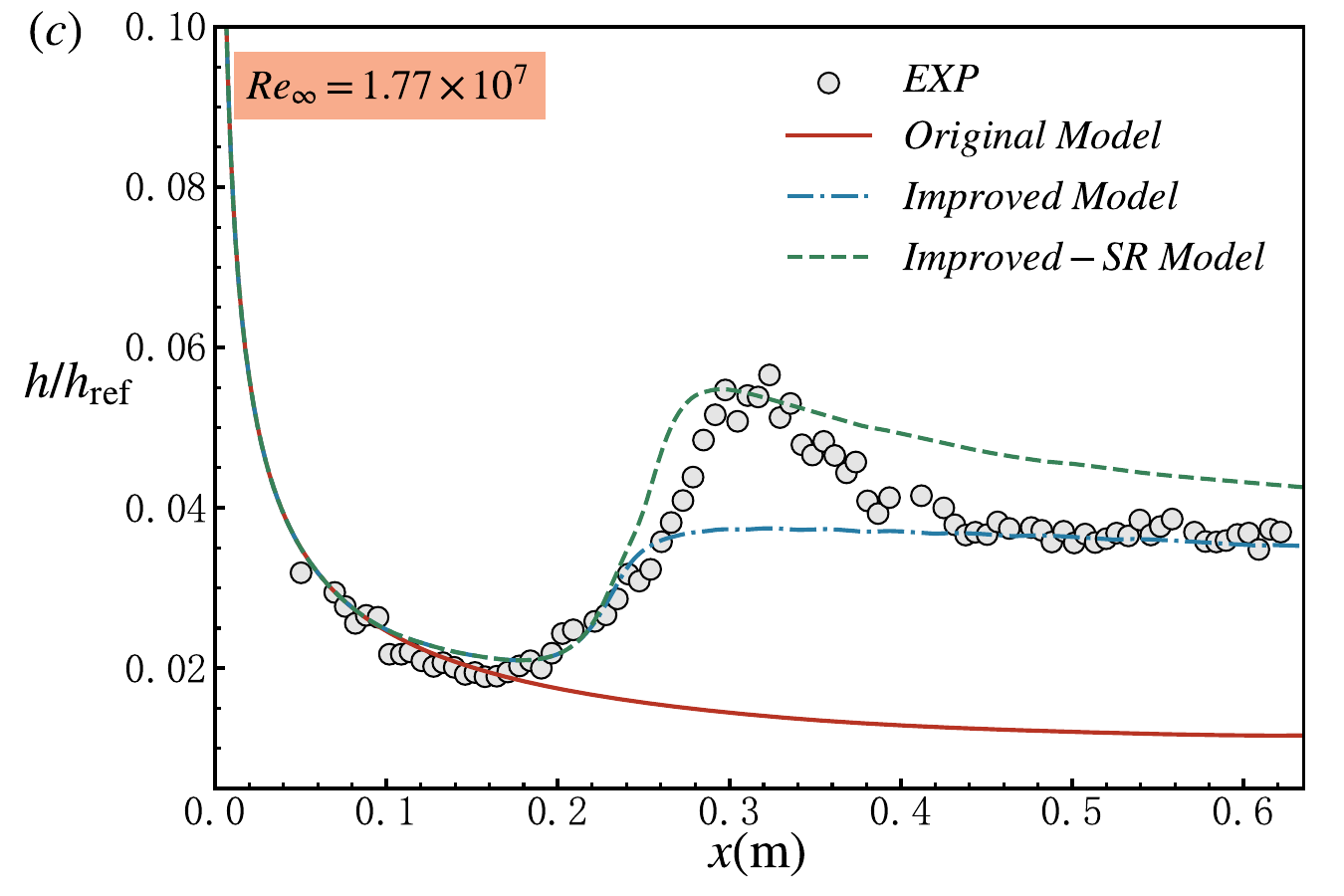}~\includegraphics[scale=0.3]{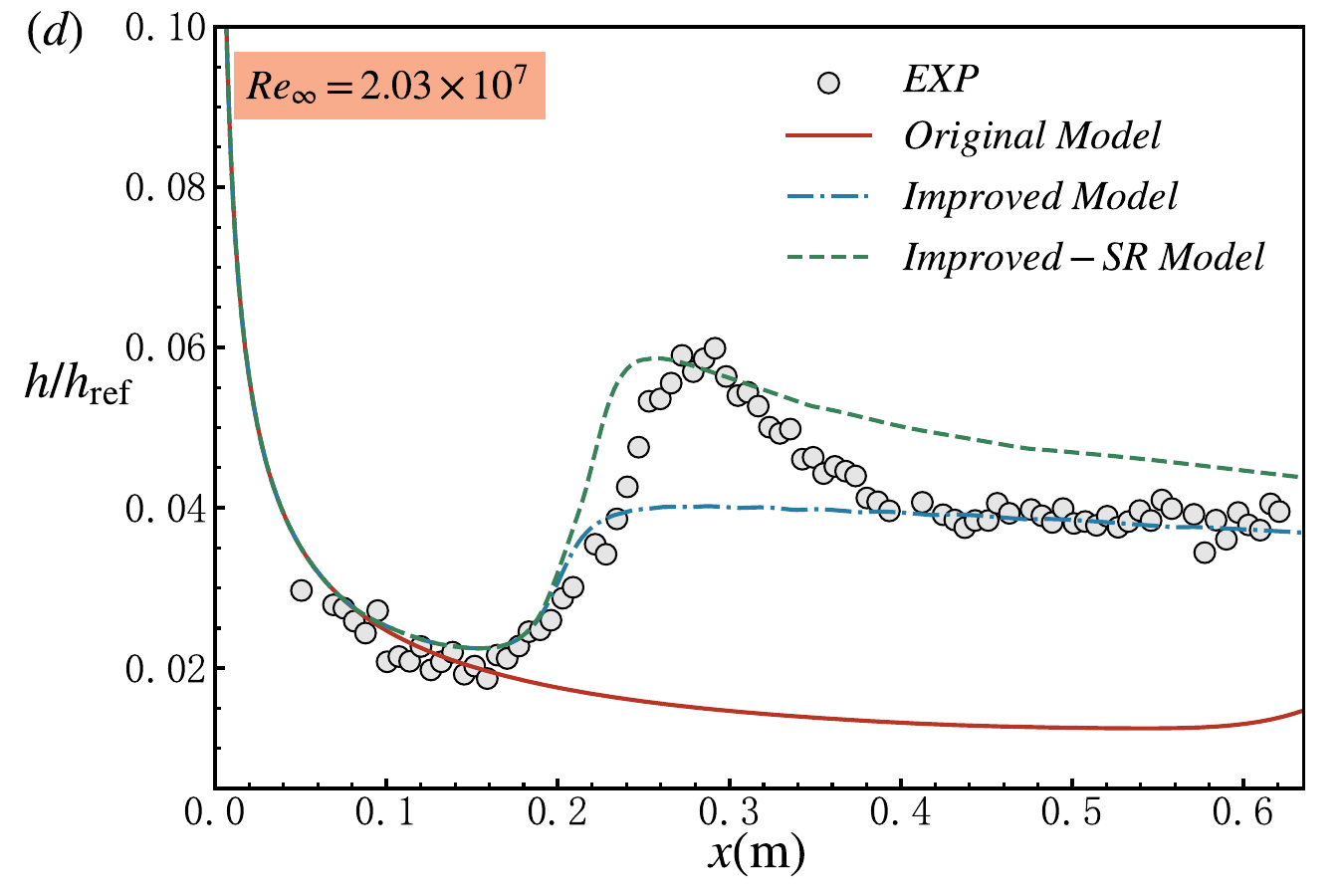}}
		\caption{Distributions of heat transfer coefficient $h/h_{\rm{ref}}$ for isothermal sharp straight cone predicted by different models in comparison with experimental data \cite{Horvath}: (a) case S1, (b) case S2, (c) case S3, and (d) case S4 as listed in Table~\ref{tab:Sharp-Cones-cases}.}
		\label{fig:S1234-h-Menter-LeiWu-Exp-concise}
	\end{figure}

	Despite the success of SR-derived expression $\beta(\mathbf{x})$ in Eq.~\eqref{equ:SR-expression2}, several terms are complex and dependent on our high-speed improved model (\eg, the compressibility correction correlation $f(Ma_{\rm{local}},T_{\rm{w}}/T_{\rm{local}})$). To explore a simpler and more fundamental form of overshoot-resolved correction, the PySR is re-employed using only features available in the original low-speed transition model of \citet{Langtry-2009-AIAA} (\ie, excluding any high-speed-specific modification functions), and thus yielding a significantly more compact expression
	\begin{eqnarray}
		\beta(\mathbf{x}) = 1 - \frac{s_{\rm{prot}} \cdot Re_k Ma_k} {1+\exp(F_{\rm{onset}})}.
		\label{equ:SR-expression3}
	\end{eqnarray}
	As shown in Fig.~\ref{fig:S1234-h-Menter-LeiWu-Exp-concise}, this simplified form still successfully captures the heat-transfer overshoot, while preserving the accurate prediction of transition onset and length inherited from the baseline model. This result demonstrates the possibility of a concise augmentation without high-speed-specific function. Despite the preliminary success, the pursuit of an even more precise and concise formulation that achieves an optimal balance among accuracy, simplicity, and interpretability remains a subject of ongoing research.

	\subsection{Generalizability of $\beta(\mathbf{x})$ for low-speed transitional flows}
\begin{figure}[h]
		\centerline{\includegraphics[scale=0.3]{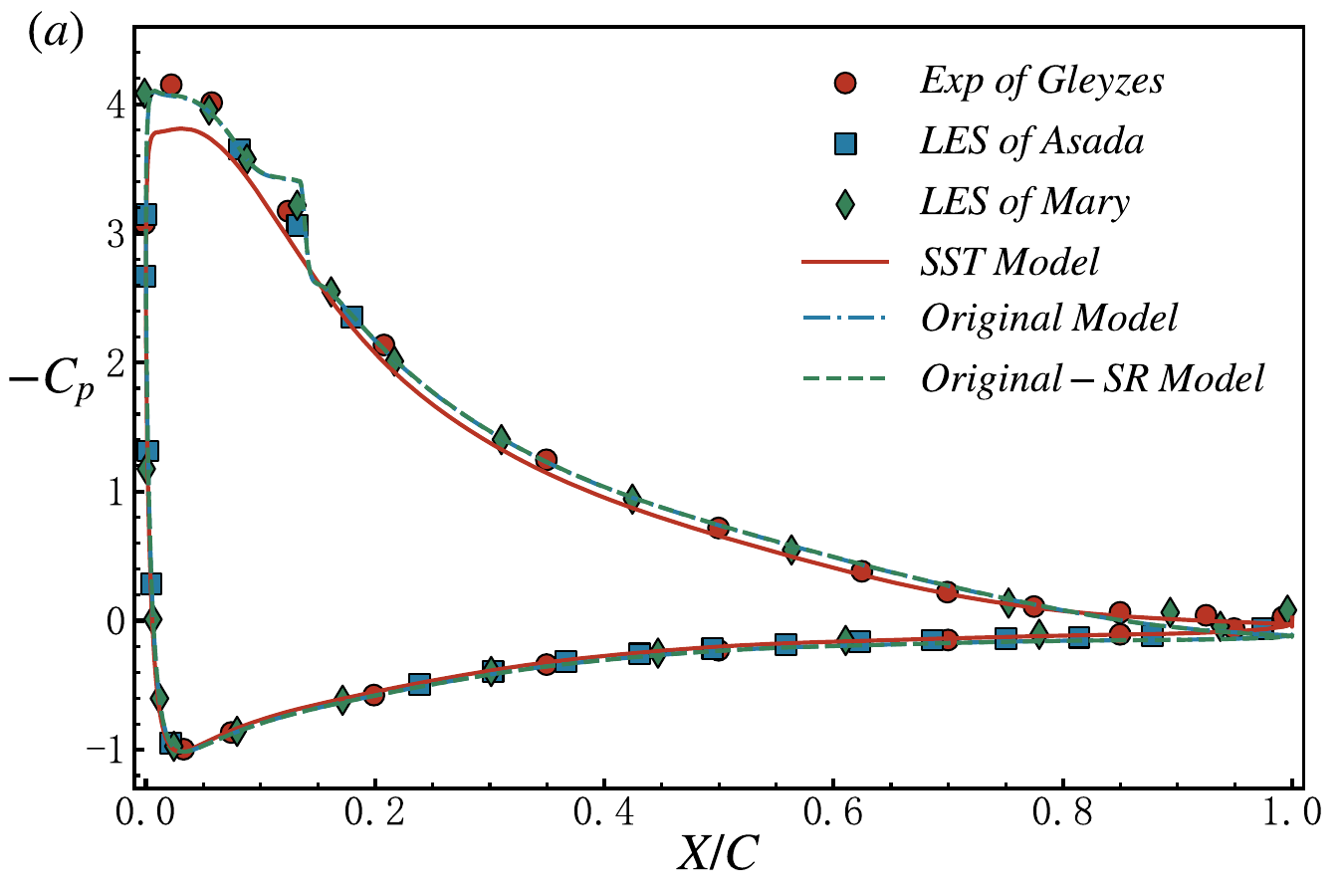}~\includegraphics[scale=0.3]{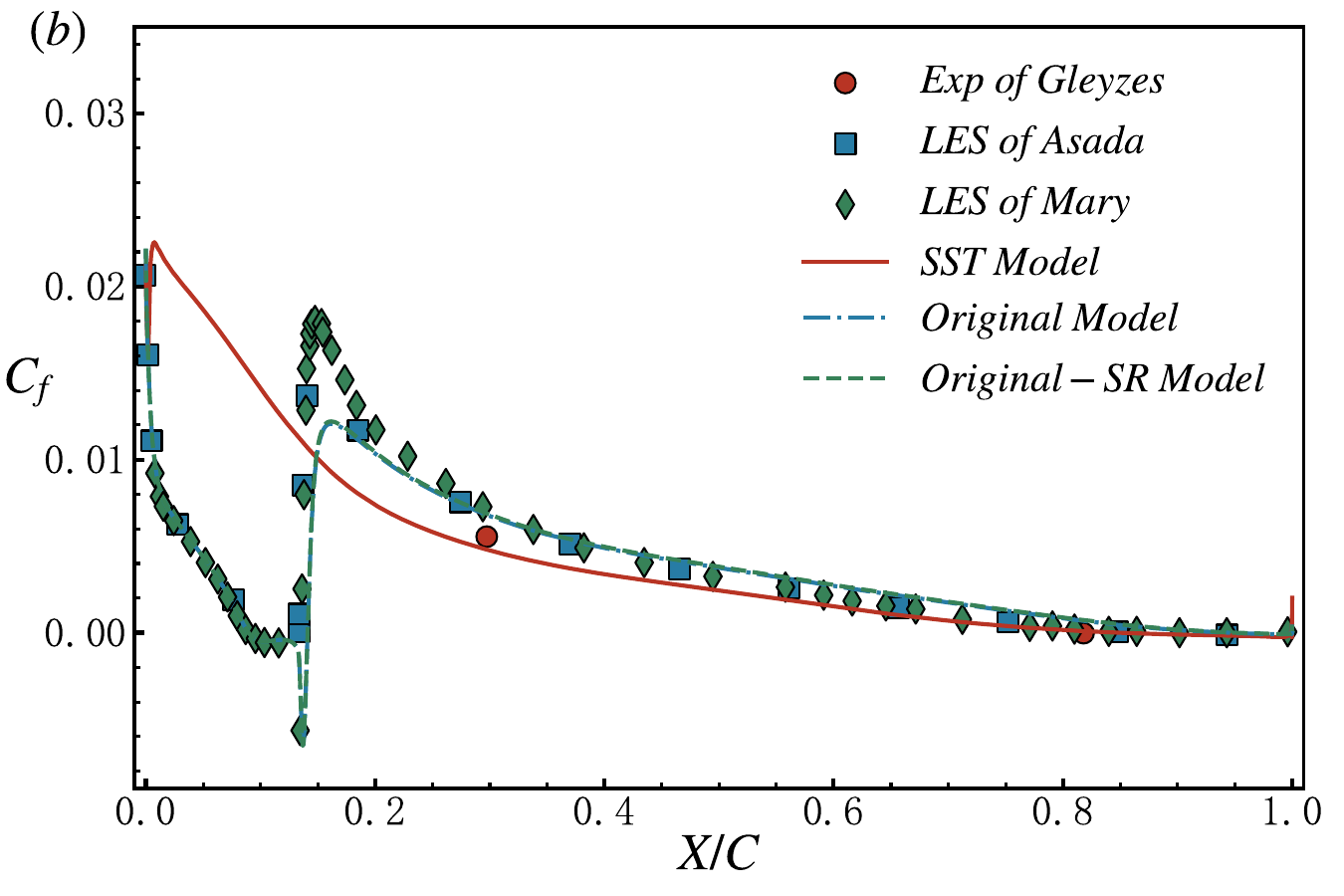}}
		\caption{Distributions of (a) mean pressure coefficient and (b) skin-friction coefficient on the surface of A-Airfoil at $AoA=13.3^{\circ}, Ma=0.15, Re=2.1\times10^6$. The experimental data from \citet{Gleyzes} and LES data from \citet{Asada-2018-POF, Mary-2002-AIAA} are presented for comparison.}
		\label{fig:A-Airfoil-CpCf}
	\end{figure}

\begin{figure}[h]
		\centerline{\includegraphics[scale=0.25]{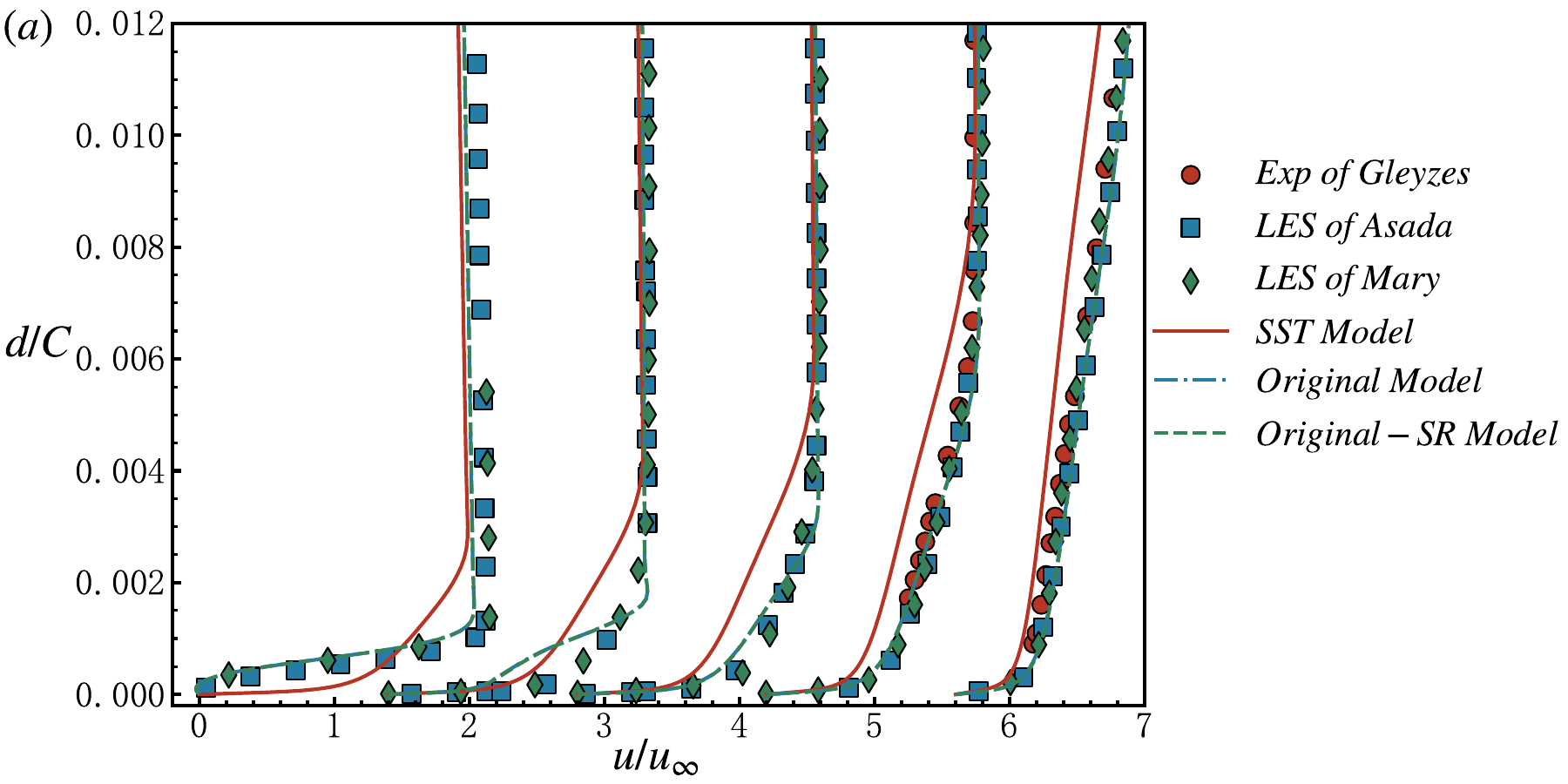}~\includegraphics[scale=0.25]{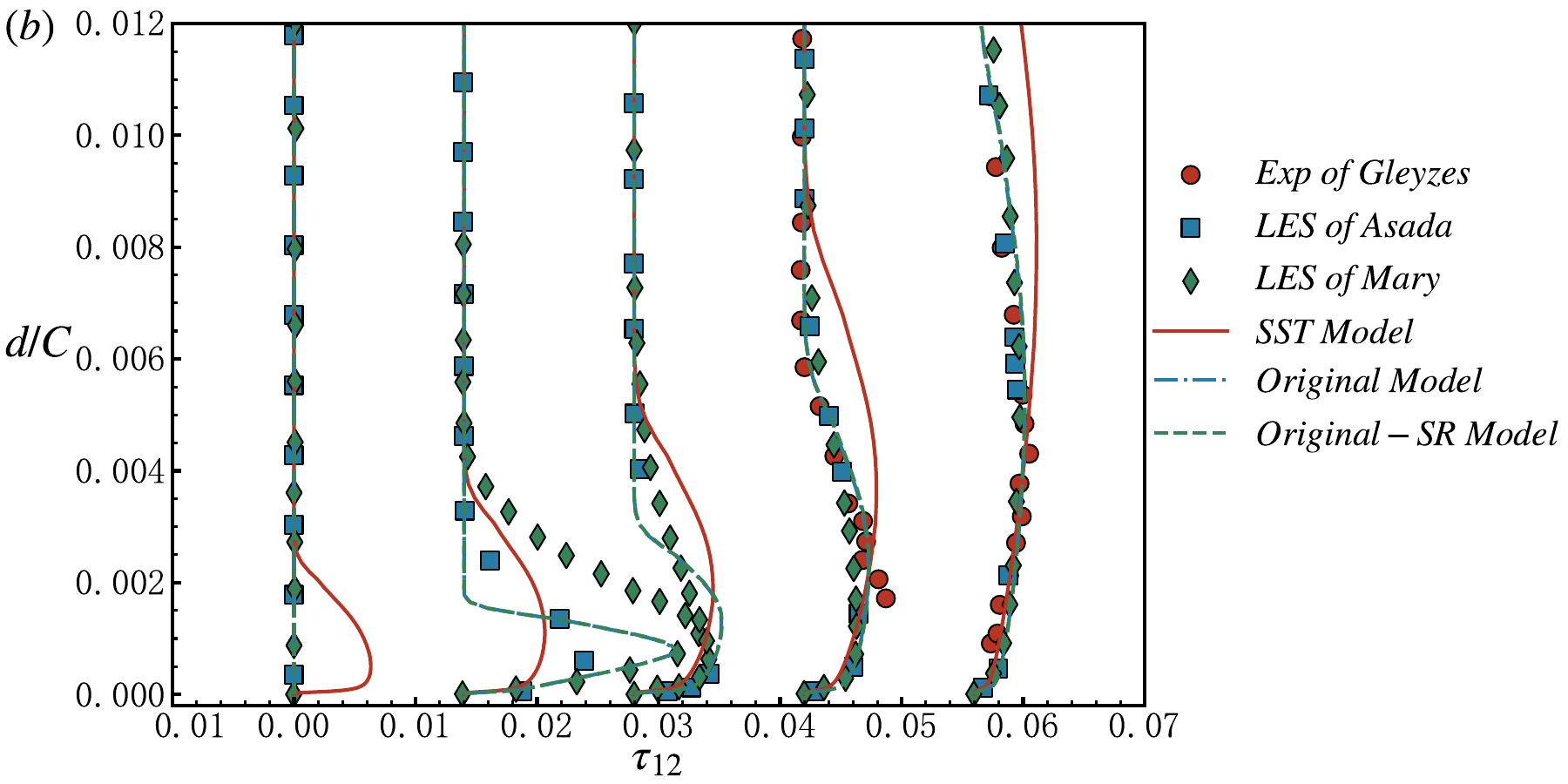}}
		\caption{Profiles of (a) mean streamwise velocity and (b) Reynolds shear stress along the wall-normal direction of A-Airfoil at $AoA=13.3^{\circ}, Ma=0.15, Re=2.1\times10^6$. The experimental data from \citet{Gleyzes} and LES data from \citet{Asada-2018-POF, Mary-2002-AIAA} are presented for comparison. All the curves at different positions are separated from each other by a constant distance.}
		\label{fig:A-Airfoil-u-t12}
	\end{figure}

	As discussed in Sec.~\ref{subsec:elucidation}, the overshoot phenomenon is not expected to occur in low-speed transitional flows. To assess the generalizability of SR-derived correction $\beta(\mathbf{x})$ for low-speed regimes, Fig.~\ref{fig:A-Airfoil-CpCf} depicts the distributions of pressure and skin-friction coefficients of A-Airfoil at $AoA=13.3^{\circ}, Ma=0.15, Re=2.1\times10^6$. As can be seen, low-speed model with $\beta(\mathbf{x})$ exhibits identical behavior to the original low-speed model. And both models significantly outperform the fully turbulent SST model. Similar conclusion can also be demonstrated in profiles of streamwise velocity and Reynolds shear stress along the wall-normal direction (see Fig.~\ref{fig:A-Airfoil-u-t12}). This outcome is not surprising. As elucidated by our previous analysis for mechanism of SR expression underlying the overshoot phenomenon (see Sec.~\ref{subsec:elucidation} and Sec.~\ref{subsec:FP}), the activation of model-form uncertainty $\beta(\mathbf{x})$ is fundamentally tied to a sufficient level of turbulent Mach number $\sqrt{2k}/{c}$. In low-speed flows, while the turbulent kinetic energy $k$ can be significant, the turbulent Mach number inherently remains a small quantity, thus preventing the spurious activation of the correction.

	\section{Conclusions and perspectives}\label{sec:conclusion}
	In this paper, a generalizable and interpretable RANS-based transition model capable of resolving the high-speed transitional overshoot phenomenon is developed through a field inversion and symbolic regression (FISR) framework.
	
	First, a comprehensive sensitivity analysis reveals that uncertainty encoded by $c_{e1}$ governs both the emergence and disappearance of the overshoot phenomenon. Leveraging this insight, ensemble Kalman filter (EnKF)-based field inversion is applied to two representative cases (\ie, adiabatic flat plate case F1 in Table~\ref{tab:Flat-plate-cases} and isothermal sharp cone case S4 in Table~\ref{tab:Sharp-Cones-cases}) to generate the overshoot-resolved RANS flow fields that faithfully match high-fidelity DNS and experimental data.
	
	Building upon these inversion fields, the spatially varying model-form uncertainty $\beta(\mathbf{x})$ is algebraically inverted. Six physically motivated input features are selected as candidate, thus the symbolic regression is employed to discover the first interpretable expression for the corrective field $\beta(\mathbf{x})$. Physical significance and functional role of each term are individually interpreted. To prevent spurious correction in the laminar regions, a dedicated protection function $s_{\rm{prot}}$ is proposed.
	
	The derived corrective expression $\beta(\mathbf{x})$ is then introduced into our previous high-speed-improved model.  Extensive validation across flat plates, sharp straight cones at different Reynolds numbers, and blunt straight cones with several nose bluntness confirms that the SR-enhanced model accurately captures the overshoot phenomenon, while preserving the fidelity of baseline model in transition location and length. Additional studies further demonstrate: (i) the necessity of $s_{\rm{prot}}$ to prevent premature transition; (ii) the feasibility of a more concise expression that eliminates dependence on high-speed transition model-specific correlations; and (iii) generalizability of $\beta(\mathbf{x})$ to low-speed transitional flows.
	
	In summary, the present work address, to some extent, the long-standing challenge of the missing overshoot phenomenon in traditional RANS-based transition models for high-speed transitional flows. Based the FISR framework, we have developed the first overshoot-resolved transition with clear physical significance and interpretability. Furthermore, to the best of our knowledge, this study identifies the first clear and physically well-founded pathway for realizing an overshoot-resolved transition model: a turbulent Mach number leads to an elevated intermittency factor, which in turn amplifies turbulent kinetic energy, ultimately resulting in distortions of velocity and temperature profiles that manifest as the overshoot. Nevertheless, the present work represents only an initial step toward an overshoot-capable transition model. Development of a more concise, accurate, and universally applicable white-box expression for overshoot phenomenon is open for further research.

	\section*{Declaration of competing interest}
	The authors do not have any known competitive economic interests or personal relationships that may affect the work reported in this article.
	\section*{Data availability}
	The data that support the findings of this study are available from the corresponding author upon reasonable request.
	\section*{Acknowledgment}
	Numerical simulations were carried out on the Polars computing platform of Peking University in Beijing, China. The authors acknowledge the financial supports provided by National Natural Science Foundation of China (grants No. 92152202 and No. 11988102).

	% Use the style of numbering in square brackets.
	% If nothing is used, default style will be taken.
	
	%\begin{enumerate}[a)]
	%\item 
	%\item 
	%\item 
	%\end{enumerate}  
	
	%\begin{itemize}
	%\item 
	%\item 
	%\item 
	%\end{itemize}  
	
	%\begin{description}
	%\item[]
	%\item[] 
	%\item[] 
	%\end{description}  
	
	% Figure
	% \begin{figure}[<options>]
	% 	\centering
	% 		\includegraphics[<options>]{}
	% 	  \caption{}\label{fig1}
	% \end{figure}

	% \begin{table}[<options>]
	% \begin{tabular*}{\tblwidth}{@{}LL@{}}
	% \toprule
	% \midrule
	%  & \\
	%  & \\
	%  & \\
	%  & \\
	% \bottomrule
	% \end{tabular*}
	% \end{table}
	
	% Uncomment and use as the case may be
	%\begin{theorem} 
	%\end{theorem}
	
	% Uncomment and use as the case may be
	%\begin{lemma} 
	%\end{lemma}
	
	
	
	\bibliographystyle{elsarticle-num-names}
	%\bibliographystyle{cas-model2-names}
	%\bibliographystyle{abbrvnat} % You can also try other styles such as plainnat or abbrvnat
	\bibliography{refs}

	
	
\end{document}